\documentclass[a4paper,fleqn,usenatbib]{mnras}
\usepackage{newtxtext,newtxmath}
\usepackage[dvipsnames]{xcolor}
\usepackage{ae,aecompl}
\usepackage{graphicx}	
\usepackage{amsmath}	
\usepackage{mathtools}
\usepackage{array}
\usepackage{booktabs}
\usepackage{longtable}

\usepackage{hyperref}
\hypersetup{draft=false}
\title[DEVILS: Data Release 1]{Deep Extragalactic VIsible Legacy Survey (DEVILS):  First Data Release Covering The D10 (COSMOS) Region}
\author[L. J. M. Davies]{L. J. M. Davies$^{1}$\thanks{E-mail: luke.j.davies@uwa.edu.au}, M. Bravo$^{2}$, R. H. W. Cook$^{1}$, A. Hashemizadeh$^{1}$, J. E. Thorne$^{1}$, S. Bellstedt$^{1}$,  S. P. Driver$^{1}$,  \newauthor A. S. G. Robotham$^{1}$, S. Koushan$^{1}$, N. Adams$^{3}$, S. Huynh$^{4}$, E. J. A. Mannering$^{4}$, J. Tocknell$^{4}$,   M. J. I. Brown$^{5}$, \newauthor J. Bland-Hawthorn$^{6}$, L. Cortese$^{1}$, B. Catinella$^{1}$, M. Meyer$^{1}$,  S. Phillipps$^{7}$,    M. Siudek$^{8,9}$, C. Wolf$^{10}$ \\
$^{1}$ ICRAR, The University of Western Australia, 35 Stirling Highway, Crawley, WA 6009, Australia \\
$^{2}$ Department of Physics \& Astronomy, McMaster University, 1280 Main Street W, Hamilton, ON, L8S 4M1, Canada \\
$^{3}$ Jodrell Bank Centre for Astrophysics, University of Manchester, Oxford Road, Manchester M13 9PL, UK \\
$^{4}$ Australian Astronomical Optics, Macquarie University, Sydney, NSW, Australia \\
$^{5}$ School of Physics \& Astronomy, Monash University, Clayton, VIC, Australia \\
$^{6}$ Sydney Institute for Astronomy, School of Physics, A28, The University of Sydney, NSW 2006, Australia\\
$^{7}$ Astrophysics Group, School of Physics, University of Bristol, Bristol BS8 1TL, UK \\
$^{8}$ Instituto de Astrof\'{\i}sica de Canarias, V\'{\i}a L\'actea, 38205 La Laguna, Tenerife, Spain\\
$^{9}$ Instituto de Astrof\'isica de Canarias (IAC); Departamento de Astrof\'isica, Universidad de La Laguna (ULL),  38200, La Laguna, Tenerife, Spain \\
$^{10}$ Research School of Astronomy and Astrophysics, Australian National University, Canberra, ACT 2611, Australia\\
}

\date{Accepted XXX. Received YYY; in original form ZZZ}

\pubyear{2025}

\begin{document}
\label{firstpage}
\pagerange{\pageref{firstpage}--\pageref{lastpage}}
\maketitle

\begin{abstract}

The Deep Extragalactic VIsible Legacy Survey (DEVILS) is a deep, high-completeness multi-wavelength survey based around spectroscopic observations using the Anglo-Australian Telescope's AAOmega spectrograph. The survey covers $\sim4.5$\,deg$^{2}$ over three extragalactic fields to Y$_{AB}<21.2$\,mag and probes sources at $0<z<1.2$, with a median redshift of $z=0.53$. Here we describe the DEVILS spectroscopic observations, data reduction and redshift analysis. We then describe and release to the community all DEVILS data in the 10h (D10, COSMOS) region including: i) catalogues of redshifts, photometry, SED fitting for physical properties, visual morphologies, structural decompositions and group environments/halo masses, ii) matched imaging in 28 bands from x-rays to radio continuum, and iii) reduced 1D spectra. All data are made publicly available through \texttt{Data Central}\footnotemark. Within D10 we obtain 5,442 new high-quality spectroscopic redshifts. When combined with existing, lower-quality, redshift information ($i.e.$ photometric redshifts) this is increased to 7,946. Of these, 3,122 have a spectroscopic redshift from another source (many that was not available at the time of the DEVILS observations). As such, DEVILS provides new unique high-quality spectroscopic redshifts for 4,824 faint sources in COSMOS. This increases the spectroscopic completeness at Y-mag$\sim$21 from $\sim$50\% in other samples to $\sim$90\% in DEVILS.  Finally, we show the power of this dataset by exploring the suppression of star formation in over-dense environments, split by morphology and stellar mass, and highlighting the ubiquitous nature of environmental quenching.

\end{abstract}


\footnotetext{\href{https://datacentral.org.au}{\texttt{datacentral.org.au}}}

\section{Introduction}

Over the last few decades large multi-object spectroscopic galaxy evolution-focused surveys have revolutionised our understanding of the processes which shape the Universe we live in today. These surveys have now allowed us to explore the properties of over a million galaxies covering a range of epochs and perform statistical analyses to identify the most significant drivers in the galaxy evolution process \citep[$e.g.$][]{Kauffmann03b, Lewis02b, Peng10, Taylor11}. These studies have uncovered a wealth of information, suggesting that galaxy evolution is a complex and multi-faceted process, being driven by many co-dependant and overlapping factors, such as age, stellar mass, metallicity, available gas supply,  star-formation rate, morphology, active galactic nuclei, mergers, interactions, and larger-scale environmental effects, to mention but a few. 

Given the complex nature of the galaxy evolution process, larger and most robustly parameterised galaxy samples allow us to better constrain this complex web of processes by both improving number statistics and allowing us to more-finely subdivide galaxies based on their characteristics, while still retaining adequate statistically-significant sample sizes \citep[$e.g.$ see][]{Driver22a}. However, care must be taken as sample biases can plague our understanding of the core mechanisms which shape galaxy evolution. Primarily three biases can lead misinterpretation of results: i) sample incompleteness in any one of the properties being studied can lead to interpretation biases \citep[$e.g.$ see][]{Cameron07}, particularly when controlling for one observable to explore another ($e.g.$ controlling for stellar mass, to explore another property, when the sample is incomplete in stellar mass), ii) observational accuracy limitations (for example, fundamental limitations in a particular survey's ability to measure galaxy properties due to limited signal-to-noise spectra or photometric bands) and iii) progenitor bias, where it is problematic to compare similar galaxy populations over a range of epochs \citep[$e.g.$ see][]{vanDokkum96, Belli15}. The latter can lead to significant biases when exploring evolutionary trends. Many surveys aim to correct for one or more of these three effects to better constrain observational galaxy evolution studies, with varying degrees of success.    

Historically, galaxy evolution-focused surveys have fallen into two broad classes, largely driven by observational limitations. The first targets large volumes of the local ($z$$<$0.2) Universe to high spatial and spectroscopic completeness, aiming to reduce sample incompleteness by targeting a significant fraction of the galaxy population over a fixed volume and down to a limiting observational brightness/magnitude. Surveys such as the Sloan Digital Sky Survey \citep[SDSS, $e.g.$][]{Abazajian09}, the Two Degree Field Galaxy Redshift Survey \citep[2dFGRS, ][]{Colless01},  Galaxy And Mass Assembly Survey \citep[GAMA, ][]{Driver11, Liske15} and more recently the Dark Energy Spectroscopic Survey Bright Galaxy Sample \citep[DESI-BGS, ][]{Hahn23} have revolutionised our view of the local Universe. 

To mention just a few key areas, they have transformed our understanding the dark matter distribution on scales of a few kpc to $>$1\,Mpc \citep[ $e.g.$][]{Peacock01,Robotham11,Yang07, Tempel14a, Alam17, Driver22b}, the effect of both large-scale environment \citep[$e.g.$][]{Lewis02, Peng10, Alpaslan15} and local environment \citep{Patton11, Knobel15, Davies15b, Davies16a, Davies19b} on galaxy evolution, allowed an exploration of the low stellar mass Universe \citep[$e.g.$][]{Baldry10, Wright17}, explored fundamental relations linking galaxy stellar mass, star-formation and metallically \cite[$e.g.$][]{Kauffmann03a, Tremonti04, Brinchmann04, Gunawardhana11, Davies16b}, determined the impact of active galactic nuclei on galaxy properties \citep[$e.g.$][]{Kauffmann03b, Gadotti09b}, and linked morphological and structural parameters with key evolutionary mechanisms \citep[$e.g.$][]{Gadotti09a, Kelvin12, Lange15, Driver22a}.  While these surveys have proved invaluable in parameterising the local galaxy population, they are limited in measuring the \textcolor{black}{astrophysics} that led to their formation - $i.e.$ using these surveys we are witnessing the in-situ processes that are driving galaxy evolution right now, but are limited in our ability to understand the \textcolor{black}{mechanisms} that have led to the $z\sim0$ galaxy population over the last $\sim10$\,Gyrs of Universal history.      

The second class of galaxy evolution-focused surveys have aimed to address this by targeting smaller areas but fainter sources, extending out into the distant Universe. Surveys such as zCOSMOS \citep{Lilly07}, the Very Large Telescope (VLT) VIsible Multi-Object Spectrograph (VIMOS) Deep Survey \citep[VVDS, ][]{LeFevre13}, VIMOS Ultra-Deep Survey \citep[VUDS, ][]{LeFevre15}, VIMOS Public Extragalactic Redshift Survey \citep[VIPERS,][]{Garilli14} and DEEP2/3 \citep{Cooper12, Newman13} have explored earlier epochs ($z>0.5$), probing \textcolor{black}{earlier stages} of galaxy evolution. However, due to observational limitations in parameterising faint sources, these surveys typically trade off spectral completeness for sample size, sparse-sampling the galaxy population at a given epoch. While this approach does not inhibit these surveys' specific science goals, it does lead to increased sample bias, which then needs to be corrected for when exploring the evolution of galaxy properties. This is particularly true when studying the environmental impact on galaxy properties, where environmental metrics on group- and merger-scales ($i.e.$ sub-Mpc) are very sensitive to small-scale spatial and spectral incompleteness \citep{Robotham11,Tempel14b, Tempel17}. On these scales, dark matter haloes virialize and merge, and gas collapses to form galaxies. Thus, this regime is paramount to our understanding of baryon physics, and the interplay between dark matter and directly-observable galaxy components. To study sub-Mpc scales spectroscopic completeness is key, as even the most high fidelity photometric redshifts are not precise enough to identify these structures.

Within the Deep Extragalactic VIsible Legacy Survey \citep[DEVILS][]{Davies18, Davies21} we aim to combine the two approaches detailed above, producing a deep magnitude-limited (Y$<$21.2\,mag), but also high spectral and spatial completeness ($>85\%$) spectroscopic survey of three well-established legacy fields: XMM-Newton Large-Scale Structure field \textcolor{black}{(XMM-LSS)}, Extended Chandra Deep Field-South \textcolor{black}{(ECDFS)}, and Cosmological Evolution Survey field \textcolor{black}{(COSMOS)}. DEVILS is designed to detect down to the stellar masses of M$^{*}_{z=0}$ galaxies to $z$\,=\,1 \citep[$10^{10.8}$\,M$_{\odot}$ - the typical galaxy in the local Universe in terms of mass-density budget,][]{Wright17}, major merger pairs of M$^{*}_{z=0}$ galaxies to $z$\,$\sim$\,0.5, and groups down to $10^{13}$\,M$_{\odot}$ to $z$\,=\,0.5. The approach, science goals and target selection for DEVILS are described extensively in \cite{Davies18}. However, we note here that the key focus of the DEVILS survey is to explore the group-scale environment and its impact on galaxy evolution at intermediate redshift. 

The survey is also specifically designed to match the GAMA sample in a number of key respects, but to extend out to the more distant Universe - allowing us to define comparable samples of galaxies across a broad evolutionary range with minimal selection and measurement biases. Briefly, within GAMA and DEVILS we define input catalogues using similar imaging \citep[see][]{Bellstedt20a, Davies21} and the same photometric codes \citep{Robotham18}, observe spectra using the same instrument+facility (the AAOmega Spectrograph on the Anglo-Australian Telescope, AAT)  and observational/reduction set-up, measure redshifts using the same methodology \citep{Baldry14}, measure galaxy properties such as stellar mass, star formation rate (SFR) and star-formation histories (SFH) using the same methodology \citep{Robotham20, Bellstedt20b, Thorne21}, and define environmental metrics such as nearest-neighbour density \citep{Davies25d} and group/cluster properties \citep{Robotham11, Bravo25} using the same approach. In combination this minimises selection and measurement biases across the two surveys allowing us to directly compare galaxies and environments across a large evolutionary baseline (also aiming to minimise progenitor bias).  DEVILS also serves as a precursor to the upcoming 4m Multi-Object Spectrograph Telescope \citep[4MOST,][]{deJong19} Wide Area Vista Extragalactic Survey \citep[WAVES,][]{Driver19} deep program, which will target similar galaxy populations but covering a much larger area. As such, the DEVILS data can be used to plan data collection and science exploitation for WAVES. What's more, in the near future there will be deluge of surveys probing high-completeness samples of galaxies outside of the local Universe from facilities such a William Hershel Telescope's Enhanced Area Velocity Explorer \citep[WEAVE,][]{Jin04}, Subaru's Prime Focus Spectrograph \citep[PFS,][]{Greene22}, the Very Large Telescope's Mulit-Object Optical and Near-IR Spectrograph \citep[MOONS,][]{Cirasuolo20} and ultimately with the European Southern Observatory's proposed Wide-field Survey Telescope \citep[WST,][]{Mainieri24}. Thus, the DEVILS dataset represents a timely comparison point for all of these upcoming surveys.

The DEVILS team have already published a number of key science and technical papers exploring the extragalactic background light \citep{Koushan21}, the evolution of galaxy morphology \citep{Hashemizadeh21} and structure \citep{Hashemizadeh22, Cook25}, the evolution of star-formation and stellar mass \citep{Thorne21, Davies22}, the evolution of the SFR-mass-metallicity relation \citep{Thorne22}, the evolution of cosmic star-formation and AGN activity \citep{D'Silva23}, SFR timescales using deep radio continuum imaging \citep{Cook25}, robust galaxy merger rates at intermediate redshift \citep{Fuenblackba25}, the evolution of the star-forming sequence using recent SFHs \citep{Davies25a}, the impact of AGN feedback in galaxy quenching \citep{Davies25b}, the environmental quenching of satellites at intermediate redshifts \citep{Davies25c} and the evolution of the morphology-density relation \citep{Davies25d}. In this work we now present the first public data release of the DEVILS sample used in the majority of these works. This includes all observed and derived properties in the D10/COSMOS region, including catalogues of photometric measurements, redshifts, SED fit parameters, visual morphological classifications, structural decomposition parameters, and group environmental diagnostics, standardised imaging covering 28 bands in the D10 regions, and standardised 1D spectroscopic data in the D10 region from our DEVILS observations at the AAT, zCOMSOS and DESI for DEVILS sources. All data is made available to the community via the \texttt{Data Central} web portal: \href{https://datacentral.org.au}{\texttt{datacentral.org.au}} to allow further scientific exploration outside of the core DEVILS team. In this paper, we detail the DEVILS target catalogue (Section \ref{sec:targets}), spectroscopic observations and data reduction at the AAT (Section \ref{sec:obs}), sample diagnostics in the D10 region (including comparisons to existing samples, Section \ref{sec:sample}), the compilation of the master D10 redshift catalogue (Section \ref{sec:redshiftcomp}), and details of all data products provided in this release (Section \ref{sec:data}). We finish by highlighting the power of this sample (in addition to existing DEVILS papers) by exploring the impact of local galaxy density on star-formation, when controlled for stellar mass and morphology (Section \ref{sec:density}).                         

\begin{figure*}
\begin{center}
\includegraphics[scale=0.6]{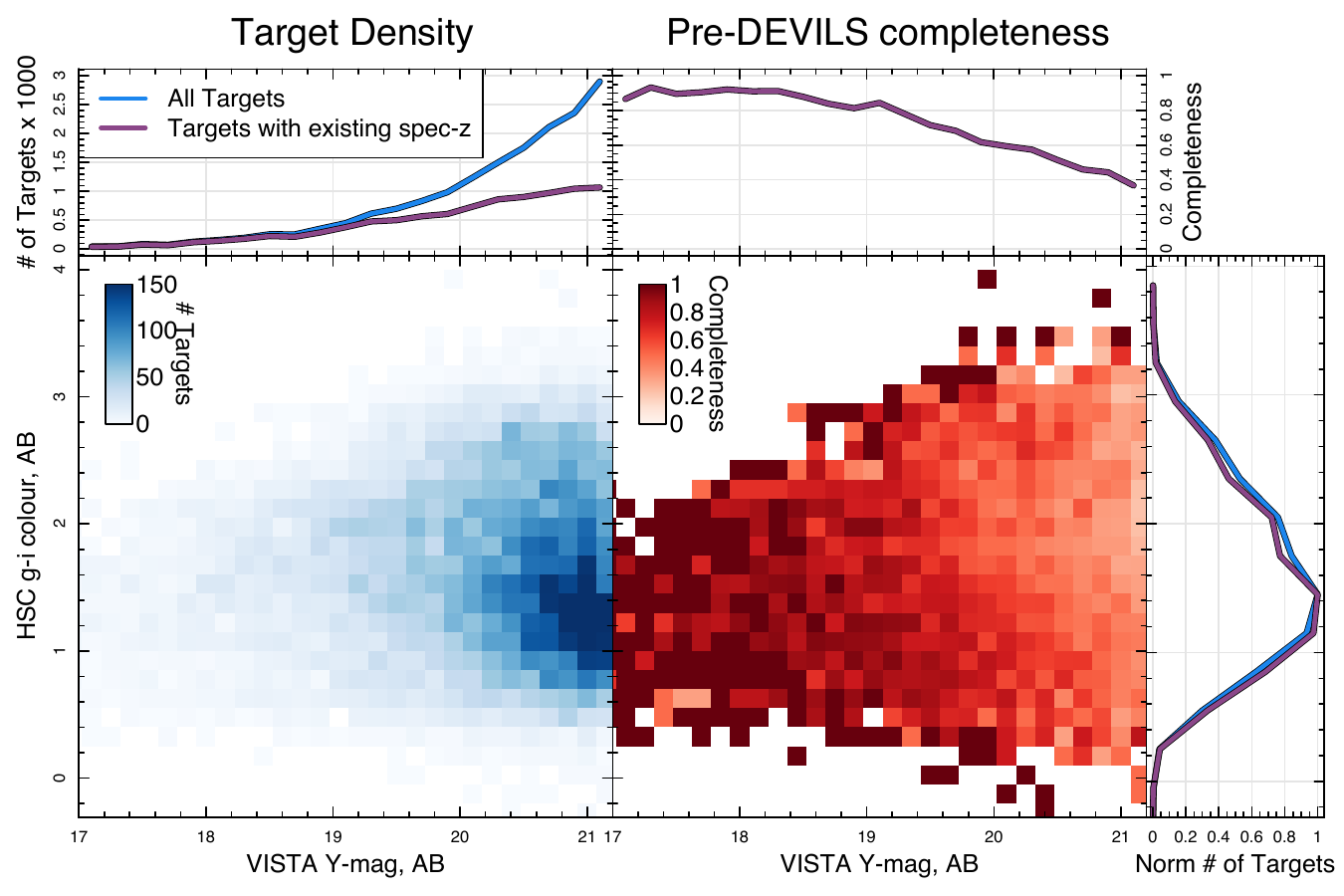}
\caption{Left: The colour-magnitude density of all targets in the D10 region. Right: The colour-magnitude completeness of sources with pre-DEVILS robust spectroscopic redshifts. Targets without a spectroscopic redshift are typically faint, but span a broad range of rest-frame colours. In existing spectroscopic samples, there is a higher fraction of blue galaxies that have a spectroscopic redshift at faint magnitudes - as expected.   }
\label{fig:Sample}
\end{center}
\end{figure*}

\begin{figure*}
\begin{center}
\includegraphics[scale=0.48]{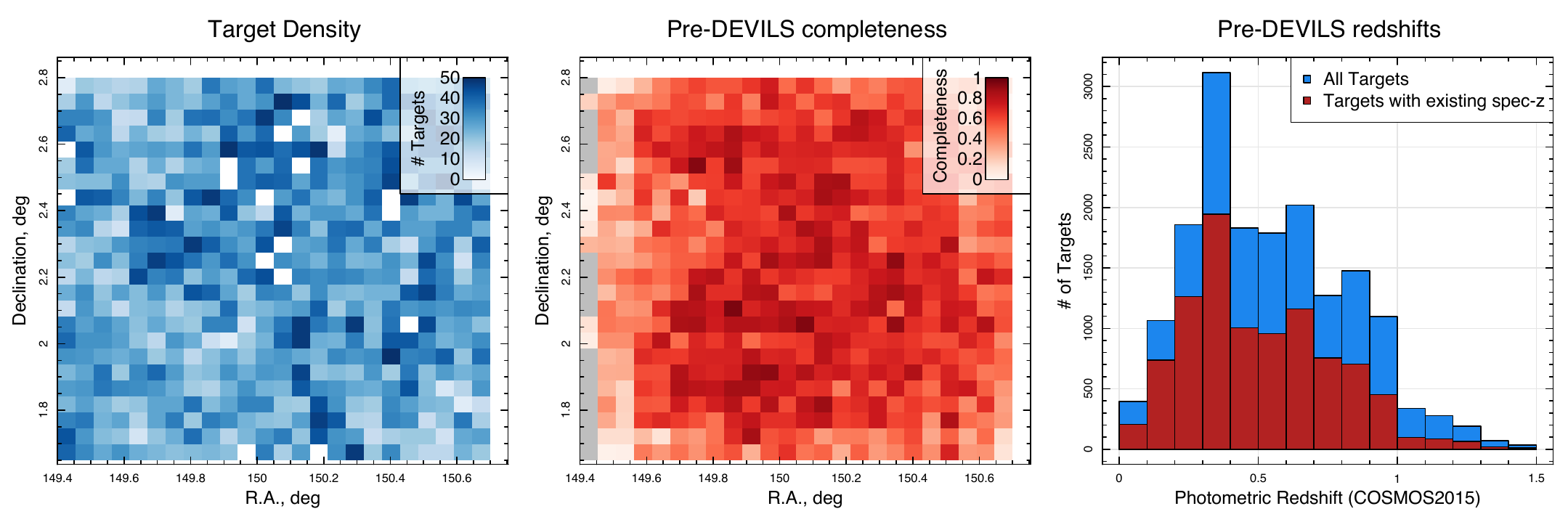}
\caption{Left: The spatial source density of targets in the D10 region. Middle: The spatial completeness of sources with pre-DEVILS robust spectroscopic redshifts - completeness is patchy and higher in the central region of the field. Right: The pre-DEVILS photometric redshift distribution of targets in the D10 region taken from \citet{Laigle16}, all targets are shown in blue, and those with a pre-DEVILS robust spectroscopic redshifts shown in red. Pre-DEVILS completeness to our target sample is $\sim0.5$ at all redshifts. }
\label{fig:Targets}
\end{center}
\end{figure*}

\section{Target Catalogue}
\label{sec:targets}

The DEVILS regions and target catalogues are described extensively in \cite{Davies18}, as such we only briefly summarise the key points here. In all DEVILS regions VISTA Y-band imaging is used for our base target selection. Public imaging in the D10 region is taken from the \textcolor{black}{Ultra-deep Visible and Infrared Survey Telescope for Astronomy (VISTA) survey} \citep[UltraVISTA,][]{McCracken12} and \textcolor{black}{VISTA Deep Extragalactic Observations survey} \citep[VIDEO,][]{Jarvis13}. Source photometry is measured using the \textsc{ProFound} code \citep{Robotham18} which identifies peak flux positions within the image, performs watershed deblending to identify source segments and then iteratively grows (dilates) these segments to measure both total and colour-optimised photometry. These measurements and associated flags are all provided in the DEVILS DR1 data release within the \texttt{devils\_dr1.D10ProFoundPhotometry} catalogue. The input imaging bands used to derive the DEVILS photometry, are also provided via \texttt{Data Central} (\href{https://datacentral.org.au}{\texttt{datacentral.org.au}}).     

For the flags, we use the \textsc{ProFound} NIR photometric measurements to perform star-galaxy separations based on NIR colour and source size. Sources classified as galaxies are assigned \texttt{starFlag}=0, stars assigned \texttt{starFlag}=1 and ambiguous objects \texttt{starFlag=2}.  Next, we use the \textsc{ProFound} output to produce a star and artefact mask. We retain all \textsc{ProFound} sources in our catalogues, but sources in unmasked regions are assigned \texttt{mask}=0, while sources in masked regions have \texttt{mask}$>$0 (with the number indicating the reason the sources was masked and details of these flags are provided in the catalogue). Finally, using \textsc{ProFound} outputs for source colours and sizes we apply an additional artefact flag for a small number of objects that may have been missed in the masking process.  These are assigned flags of \texttt{artefactFlag}=0 (not an artefact) and \texttt{artefactFlag}=1 (is likely an artefact). As such, to build a sample of galaxies with robust photometric measurements from the DEVILS DR1 catalogues, a user would apply the filter:  \texttt{mask}=0 \&  \texttt{starFlag}=0 \& \texttt{artefactFlag}=0. The imaging data, photometric measurements and star-galaxy separation, masking and artefact flagging are described extensively in \cite{Davies18} and \cite{Davies21}. As such, we refer the reader to these works for further information.  

To build the DEVILS target catalogue for spectroscopic observations, we take all sources selected as galaxies in unmasked regions of our field with total magnitudes of Y$_{AB}<21.2$mag. First, the left panel of Figure \ref{fig:Sample} shows the Y magnitude vs observed-frame $g$-$i$ colour for all of these targets, while in the right panel we show the same but coloured by pre-DEVILS redshift completeness. Targets without an existing spectroscopic redshift are typically faint, but span a broad range of rest-frame colours. In existing spectroscopic samples, there is a higher fraction of blue galaxies have a spectroscopic redshift as we move to fainter magnitudes. This is expected as blue galaxies typically have nebular emission lines, for which redshifts are more easily obtained. 

Next, considering the spatial distribution of targets, the left panel of Figure \ref{fig:Targets} shows the target density across the D10 region, where masked regions are displayed as white spaces. In the right panel of Figure \ref{fig:Targets} we show the pre-DEVILS photometric redshift distribution of our targets taken from the COSMOS2015 catalogue \citep{Laigle16}.  This results in 17,209 possible un-masked, extragalactic targets in the D10 region.   

In a final step we removed objects with an existing high-confidence, spectroscopic redshift (at the time of DEVILS observations) to maximise the scientific return from our observations. These sources are described extensively in section 5.1.1 of \cite{Davies18}, but we note here that we remove sources with high-confidence redshifts from VVDS \citep{LeFevre14}, zCOSMOS \citep{Lilly07}, VUDS \citep{LeFevre13} and hCOSMOS \citep{Damjanov18}. Redshifts from all other programs are included in our target catalogue. We find that of our full target catalogue 7,980 sources have an existing high confidence spectroscopic redshift, leaving 9,229 targets with unknown or un-secure redshifts in the field. We then also re-observe a random subsample of zCOSMOS (627) and hCOMOS (303) covering a range of redshifts and magnitudes with which to compare our DEVILS-AAT redshifts. These targets are included in our input catalogue, resulting in a total target sample of 10,159 sources. In our master redshift compilation (see Section \ref{sec:redshiftcomp}), we include all of the existing redshifts to form our highly-spectroscopically complete sample.

\section{Spectroscopic Observations}
\label{sec:obs}

DEVILS spectroscopic observations were undertaken at the 3.9m Anglo-Australian telescope at the Siding Spring Observatory in New South Wales with the AAOmega fibre-fed spectrograph \citep[][]{Saunders04, Sharp06} in conjunction with the Two-degree Field \citep[2dF,][]{Lewis02} positioner. The 2dF positioner has been at the forefront of large galaxy redshift surveys in the local Universe (such as 2dFGRS and GAMA) and has also been used to great success in targeting large numbers of sources to higher redshift \citep[$e.g.$ 2dFLenS and WiggleZ,][]{Blake16, Drinkwater18} and more recently to faint magnitudes \citep[OzDES,][]{Childress17}. As such, it is an ideal facility to perform our deep intermediate redshift survey.  2dF allows the simultaneous observation of $\sim400$ targets with the AAOmega spectrograph, with 2$^{\prime\prime}$ diameter fibres. The 2dF top-end consists of an atmospheric dispersion compensator (ADC) and robot gantry that positions fibres to 0.3$^{\prime\prime}$ accuracy on sky.     

AAOmega observes in two spectral channels (blue and red), both equipped with a 2k$\times$4k E2V CCD detector and an AAO2 CCD controller. We observed with the 5700\AA\ dichroic allowing for simultaneous coverage from 3750\AA\ to 8850\AA. For the blue CCD we use the 580V grating with central wavelength of 4820\AA\ providing a $\sim1.03$\AA/pix dispersion, while for the red CCD we use the 385R grating with central wavelength of 7250\AA\ providing a $\sim1.56$\AA/pix dispersion. This results in a spectral resolution that varies from R$\sim$1000 (blue) to R$\sim$1600 (red). This spectral resolution and wavelength range was selected to enable detection of at least the [OII] (3727\AA) emission line and 4000\AA\ break over our full target redshift range and provide sufficient velocity resolution with which to identify close pairs \citep[$<$50\,km\,sec$^{-1}$, see][]{Robotham14, Fuenblackba25}. 

Spectroscopic observations for DEVILS were taken over $\sim100$ nights between December 2017 and January 2021, covering 25 separate  observing runs. Observations were taken both in person at Siding Spring and remotely from a number of locations.  We note here that the survey was originally planned to complete in 2020, but was significantly delayed due to a number of instrument/facility failures at the AAT and COVID-19 closures.  Details of the DEVILS observational setup and measurements are also described in Section 5 of \cite{Davies18}, so we only briefly summarise key points here: \\

\noindent $\bullet$ \textbf{Calibrations sources:} Sky fibres are selected using blank sky positions taken from the \textsc{ProFound} segmentation maps, and each observation 25 blank sky positions were targeted. Flux standards were selected from SDSS using fibermag\_r $>$ 16.9 and 16.6 $<$ psfmag\_r $<$ 18.4, and classified as either \textsc{SPECTROPHOTO\_STD} or \textsc{REDDEN\_STD}. We observe three spectroscopic standards in each observation.  Finally, for guide stars we select all sources at 13.7$<$R1Mag$<$14.4 from the USNO-B guide-star catalogue and exclude sources with proper motions of >15\,mas\,yr$^{-1}$, using 7-8 in each pointing. \\
  
\noindent $\bullet$ \textbf{Tiling and Fibre Assignment:} Priorities are assigned to targets based on their Y-band magnitude, with fainter sources having higher priority (to allow sources which will require longer integration times to be preferentially observed). Targets are assigned to fibres using the greedy tiling algorithm outlined in \cite{Robotham10} and used in GAMA, which adds additional weights to priorities based on close on-sky clustering to allow complex regions with high levels of potential fibre collisions to be  preferentially targeted. \\

\noindent $\bullet$ \textbf{Nightly Observations:} Fibre flat observations were taken with the Quartz\_75\_A, 75W lamp and arc observations with the FeAr\_1, FeAr\_2, CuAr\_1, CuAr\_2, CuHe\_1, CuNe\_1 lamps. Data are typically observed in a 6\,sec flat, 45\,sec arc, 2$\times$1800\,sec sequence. $\sim30$ dark frames are taken each run and 10 bias frames are observed each day. \\
  
\noindent $\bullet$ \textbf{Redshift feedback:} As discussed in \cite{Davies18}, within the DEVILS survey we use a `redshift feedback' methodology to allow increased survey efficiency when simply aiming to maximise redshift completeness in a given sample. Briefly, in this mode we do not use fixed total integration times based on prior information regarding each source, but instead observe all targets for short $\sim$1h integrations. Following each new observation, we stack all data from a given source, and attempt to measure the source redshift. If a secure redshift is obtained we flag the source as complete and do not re-observe. If no secure redshift is obtained, we prioritise the source for re-observation the following night. In this manner sources are only observed for the minimum (rounded up to the nearest hour) exposure time required to obtain a secure redshift. Note that this process is also proposed for a number of the 4m Multi-Object Spectrograph Telescope \citep[4MOST,][]{deJong14} consortium surveys.  For further details and motivation of this approach, see Section 6.2 of \cite{Davies18} and Section \ref{sec:redshfiting}.

\begin{figure*}
\begin{center}
\includegraphics[scale=0.625]{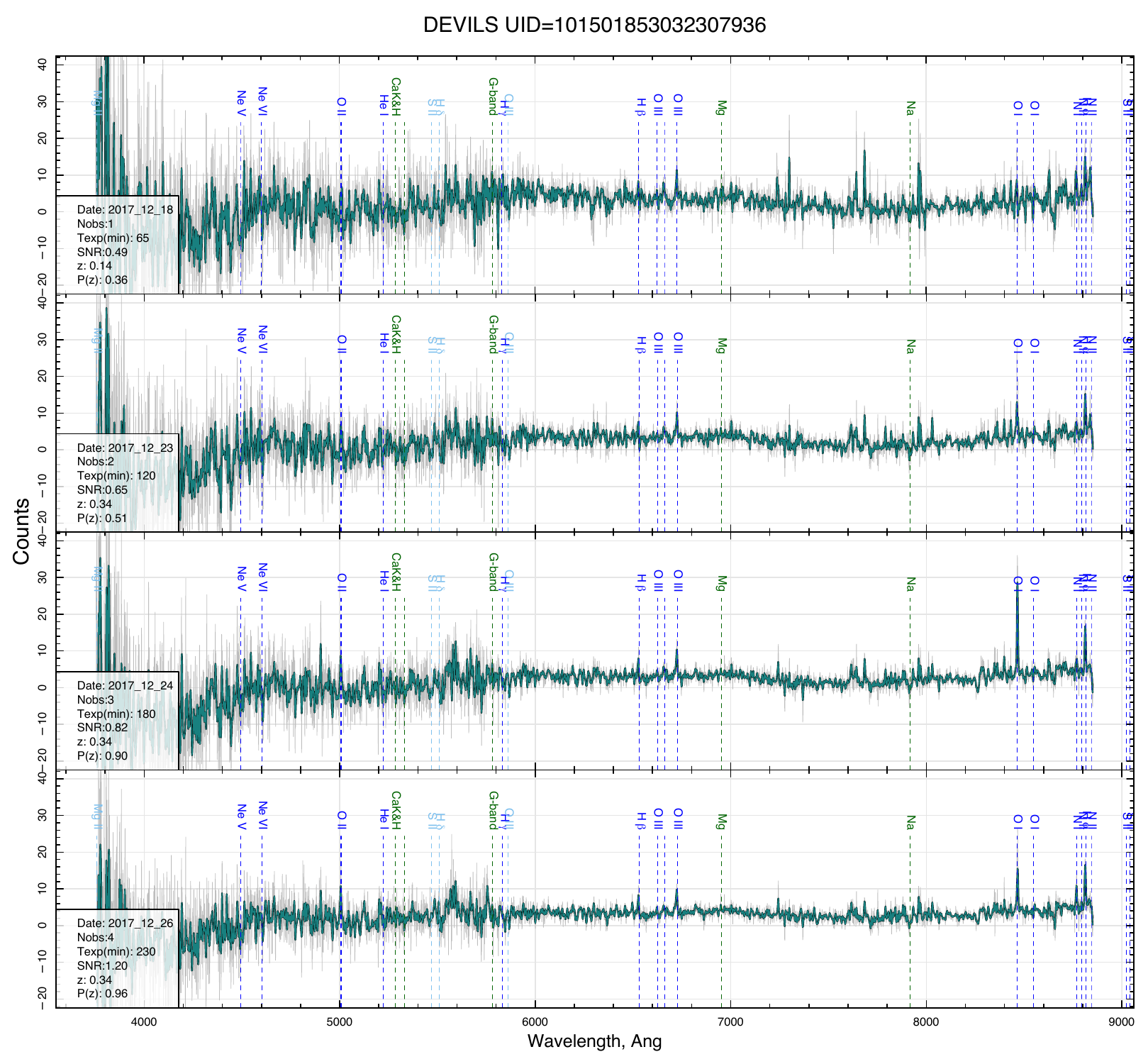}
\caption{Examples of 1D DEVILS spectra for source UID=101501853032307936 indicating the DEVILS inverse variance-weighted stacking procedure. Faint grey lines show the raw AAOmega spectrum, solid green lines show a Hanning smoothed version of this spectrum. Key emission (blue), absorption (green) and AGN (light blue) lines at the final source redshift are shown as vertical dashed lines. Panels from top to bottom show successive stacks of data building up exposure time and increasing signal-to-noise.  The last observation date, number of observations, total exposure time, SNR \textcolor{black}{(measured as a median SNR/pixel between 6500-7000\AA)}, best-fit redshift and redshift probability, P($z$), are given in the legend. In the top panel the source is too low signal-to-noise for the correct redshift to be obtained. In the second/third panels the correct redshift is found, but the P($z$) is too low for this to be secure. In the bottom row, the correct redshift is found at a high probability and the source stops being observed.    }
\label{fig:stack}
\end{center}
\end{figure*}

\section{Nightly Data Reduction and Analysis}
 
Data reduction is primarily undertaken using the DEVILS observing preparation, data reduction and analysis pipeline (Tool for Analysis and Redshifting, TAZ). Given the redshift feedback mode described above, observations during the DEVILS survey must be reduced, combined with existing data, redshifts measured and tiling catalogues updated on a nightly basis. To undertake this, following nightly observations, all data and calibration files are synced to the DEVILS data processing machines. This triggers the automatic pipeline, which then runs the following broad steps.   

\subsection{Raw data reduction}

The raw data were reduced using the AAO's 2DFDR software \citep{Croom04} developed continuously since the advent of the 2dF and optimised for AAOmega. The software is described extensively in a number of AAO documents, so we do not cover it in detail here. Briefly, it performs automated spectral trace (tramline) detection, sky subtraction, wavelength calibration, stacking and splicing. As part of the automated 2DFDR reduction process, the blue and red spectra are flux calibrated to a white dwarf spectrum but with an arbitrary normalisation. 

\subsection{Data Combination}

Following the raw data reduction, 1D target spectra from an individual night are extracted from the reduced data and stored in the DEVILS database. The database is then searched for all existing 1D spectra for a given target. All spectra for a specific target are then stacked in heliocentric-corrected wavelength space using an inverse variance weighting scheme. All individual 1D spectra are retained and the combined total stacked spectrum flagged as the current `best' observations for a given source. All meta data on provenance, total exposure times, etc are retained with the stacked spectrum and included in the DEVILS targeting catalogues. An example of this process is shown for a single source in Figure \ref{fig:stack}. Moving from top to bottom we show successive stacks of source UID=101501853032307936 consisting of 4 $\sim$1h exposures over a period of 8 days in December 2017. We find that the signal-to-noise ratio (measured as a median SNR/pixel between 6500-7000\AA) increases by $\sim1/\sqrt{T_{exp}}$ with successive exposures, as expected.

\subsection{Rough Flux calibration}

Following this, we perform our own additional arm-specific flux calibration on the stacked spectra using the DEVILS broad-band photometric data. For this we use the HSC-g (blue arm) and HSC-i (red arm) photometric filters to measure the spectral flux in each band. We then re-calibrate the AAOmega spectra to have the same spectro-photometric flux as the photometry catalogues and then re-combine spectral arms into a full spliced spectrum. We note that this process is not highly accurate, however flux calibrated spectra are not required for any of the core DEVILS science goals.

\subsection{Redshift Measurements}
\label{sec:redshfiting}

With reduced, calibrated and stacked spectra in hand, we now undertake automatic redshifting on all `best' stacked spectra for each newly-observed target. This is performed using an adapted version of the \textsc{AutoZ} redshifting code \citep{Baldry14}. Full details of this process and its effectiveness in obtaining redshifts from AAOmega spectra using an almost identical set-up to that used in DEVILS are described in the aforementioned work.  The \textsc{AutoZ} code not only provides galaxy redshifts but also a probability that the redshift is correct, P($z$) - see \cite{Baldry14} for details. We note here that in this analysis we retain the top 5 redshift solutions and probabilities for each source. 

Within the DEVILS observations we initially assign a P($z$)>0.95 as a secure redshift. During the nightly redshifting process, objects that obtain a redshift with P($z$)>0.95 are flagged as having a secure redshift and their targeting probability is set to P0 (do not target). Objects which have been observed, but do not have a secure redshift are set to P7 (highest priority). This ensures that an object which has started to be observed, is highly likely to be re-observed in subsequent observations, such that observation time is not wasted. We also enforce the maximum exposure time, $T_ {max}$, for all sources. This ensures that objects do not continue to receive further exposures if they will likely not obtain a redshift with a 4m-class facility. We set $T_ {max}=$15h. Objects which reach a total exposure time of $>$15h but a P($z$)<0.95 are also set to P0. The redshift feedback process is also displayed in the successive spectra in Figure \ref{fig:stack}. Here we see that after the first exposure the P($z$) value is low and the redshift is incorrect compared to the final solution. After the second exposure the correct redshift \textcolor{black}{was} obtained \textcolor{black}{but} the P($z$) is low, so the object is re-observed. After the third exposure the same redshift is obtained, with a higher P($z$), but not above our P($z$)>0.95, defined for a secure redshift. After the fourth exposure P($z$)>0.95, the redshift is deemed secure and we stop observing this source for the remainder of the survey.                                    

\subsection{Tiling catalogue updates}

Once redshifting is complete and new targeting probabilities are assigned, the final stage of the pipeline is to produce new 2dF target fibre assignments and observing catalogues following the process outlined in Section \ref{sec:obs}. These are then returned to the observers at the telescope ready for the next night's observations.    

\begin{figure*}
\begin{center}
\includegraphics[scale=0.85]{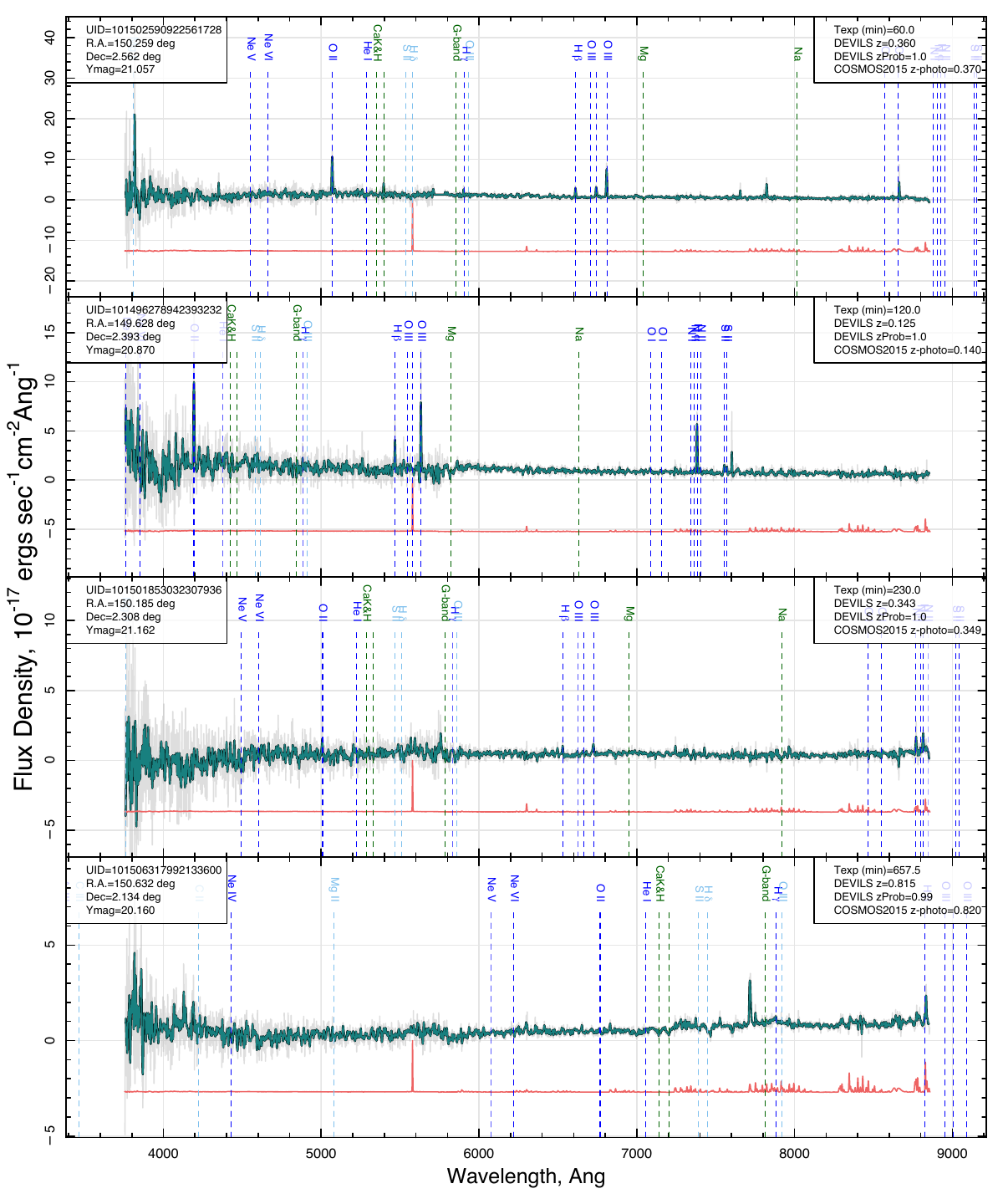}
\caption{Examples of 1D DEVILS spectra from our final reduced sample. Here we show random examples for faint sources at a range of redshift and requiring a range of exposure times to obtain a secure redshift (given in the top left legend). Faint grey lines show the raw AAOmega spectrum, solid green lines show a Hanning smoothed version of this spectrum, red lines show the sky spectrum offset for clarity. Key emission (blue), absorption (green) and AGN (light blue) lines at the source's redshift are shown as vertical dashed lines. }
\label{fig:spec}
\end{center}
\end{figure*}

The process of running the pipeline typically takes a few hours, such that nightly datasets can be reduced and updated target catalogues returned to the observatory in time for the following night's observations. We also note that following the completion of the survey a full reduction of the entire sample was undertaken with slightly modified settings to improve the data reduction. The final DR1 contains data produced by this final re-reduction/analysis which aims to maximise the redshift recovery rate. In Figure \ref{fig:spec} we show a number of random examples of the final-reduced DEVILS spectra, with redshift assignments overlaid.  

\begin{figure}
\begin{center}
\includegraphics[scale=0.55]{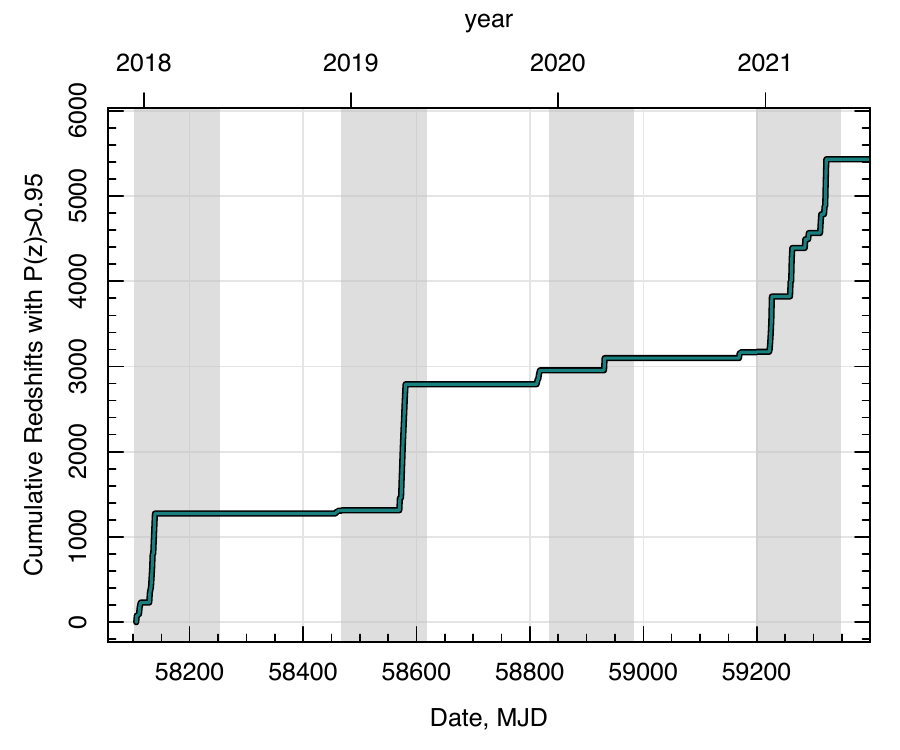}
\caption{Secure, P($z$)$>$0.95, redshift accumulation rate in the D10 as a function of time. Grey shaded region shows the period when the D10 field was observable from the AAT for $>$4h/night. The 2020 observing block was significantly affected by COVID shutdowns and instrument failures.}
\label{fig:progress}
\end{center}
\end{figure}

\begin{figure}
\begin{center}
\includegraphics[scale=0.57]{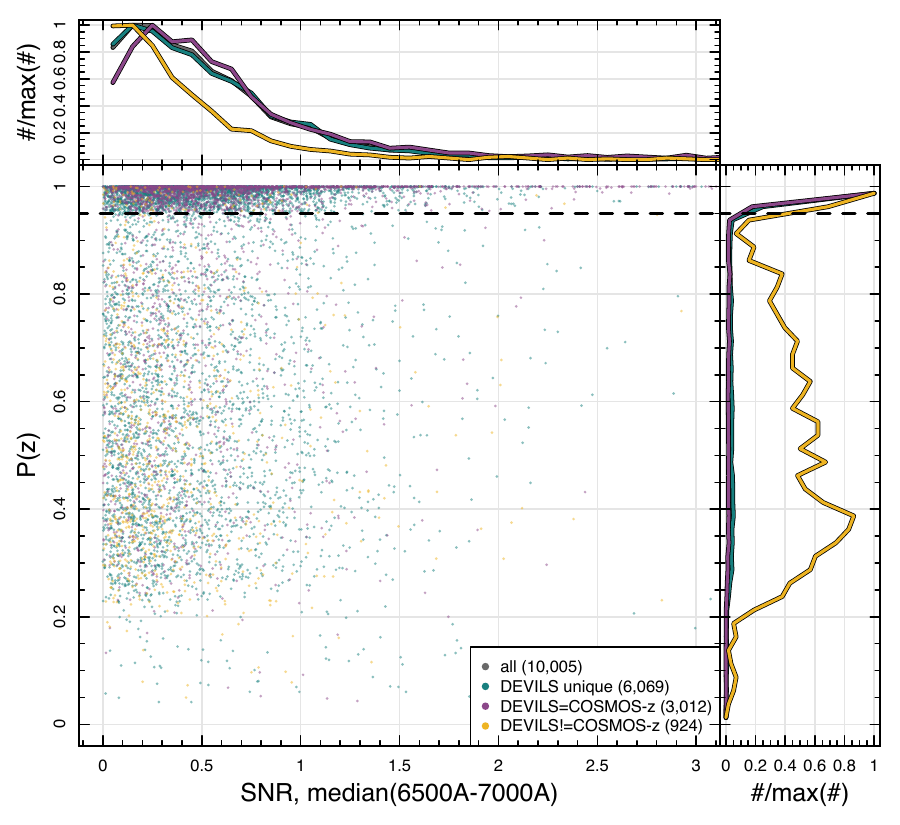}
\caption{Redshift probability measurement, P($z$), as a function of signal-to-noise ratio. Here we also compare DEVILS measurements with those taken from the COSMOS spectroscopic redshift compilation. Green points/lines show sources with unique DEVILS AAT redshifts, purple points/lines show sources which have the same redshift as in the COSMOS spectroscopic redshift compilation, while gold points/lines show sources where our DEVILS-AAT redshifts differ from the COSMOS spectroscopic redshift compilation. The majority of these sit at low P($z$) values. The dashed horizontal line shows P($z$)$>$0.95.    }
\label{fig:pzcomp}
\end{center}
\end{figure}

\section{Sample Diagnostics}
\label{sec:sample}

In the following section we detail a number diagnostics of the final DEVILS DR1 dataset. We note here that our final DEVILS D10 sample contains unique observations of 10,159 sources covering a range of exposure times, over a total of 40,375 fibre-hours of AAT observing time, with a median of 4h exposure time per source. From this sample we initially obtained 5,442 redshifts with a P($z$)$>$0.95. In Figure \ref{fig:progress} we show the redshift accumulation rate as a function of time, highlighting the observability epochs of the D10 region in grey. Redshifts accumulate rapidly in the observable windows. The exception to this is the 2020 observing block, which was impacted by COVID closures and instrument failures. In Figure \ref{fig:pzcomp} we then show the resultant distribution of P($z$) values as a function of signal-to-noise (measured as a median signal-to-noise per pixel between 6500-7000\AA). All objects remain at a relatively low signal-to-noise as we stop observing sources once they reach a P($z$)$>$0.95. As such, sources in this figure that have P($z$)$<$0.95 are those that we did not obtain a secure redshift over the duration of the survey. In figure \ref{fig:pzcomp} we then also match our DEVILS sources to those with secure redshifts from the COSMOS spectroscopic redshift catalogue of \cite{Khostovan25}. There are 3,936 sources from the COSMOS catalogue that were observed in DEVILS (without filtering on P($z$)). We identify sources which have the consistent redshift based off the offset between redshift measurements as: 

\begin{equation}
\frac{|(z_{\mathrm{COSMOS}}-z_{\mathrm{DEVILS}})|}{(1+z_{\mathrm{DEVILS}})}<0.15     
\end{equation}

\noindent We find that 3,012 sources have consistent redshift between DEVILS and COSMOS, and of the remaining 924 sources, $\sim$90\% (826) have a P($z$)$<$0.95. $i.e.$ for a comparison to the COSMOS catalogue only 98 sources have a P($z$)$>$0.95 and a redshift which differs between the two samples. We visually inspect all of these sources and compare to the COSMOS2015 photometric redshifts of \cite{Laigle16}, and find that the reasons for these discrepancies are varied. We find source where the DEVILS redshift appears correct and aligns with the photometric redshift (but differs from the COSMOS spectroscopic redshift), objects where stars are misidentified as galaxies and vice versa, objects where poor sky subtraction has led to line misidentification and objects which show clear AGN-like spectra (which are not accounted for in our redshift measurements with \textsc{AutoZ}). Given the small fraction of objects, we do not aim to manually fix these in our initial DEVILS redshifts. However, in Section \ref{sec:redshiftcomp} when we build our master redshift catalogue, we weight our final redshifts based on a combination of available redshifts for each target, and note that the majority of these cases are updated in this approach - leading to greater consistency between our final redshifts and the COSMOS spectroscopic catalogue.

\begin{figure*}
\begin{center}
\includegraphics[scale=0.48]{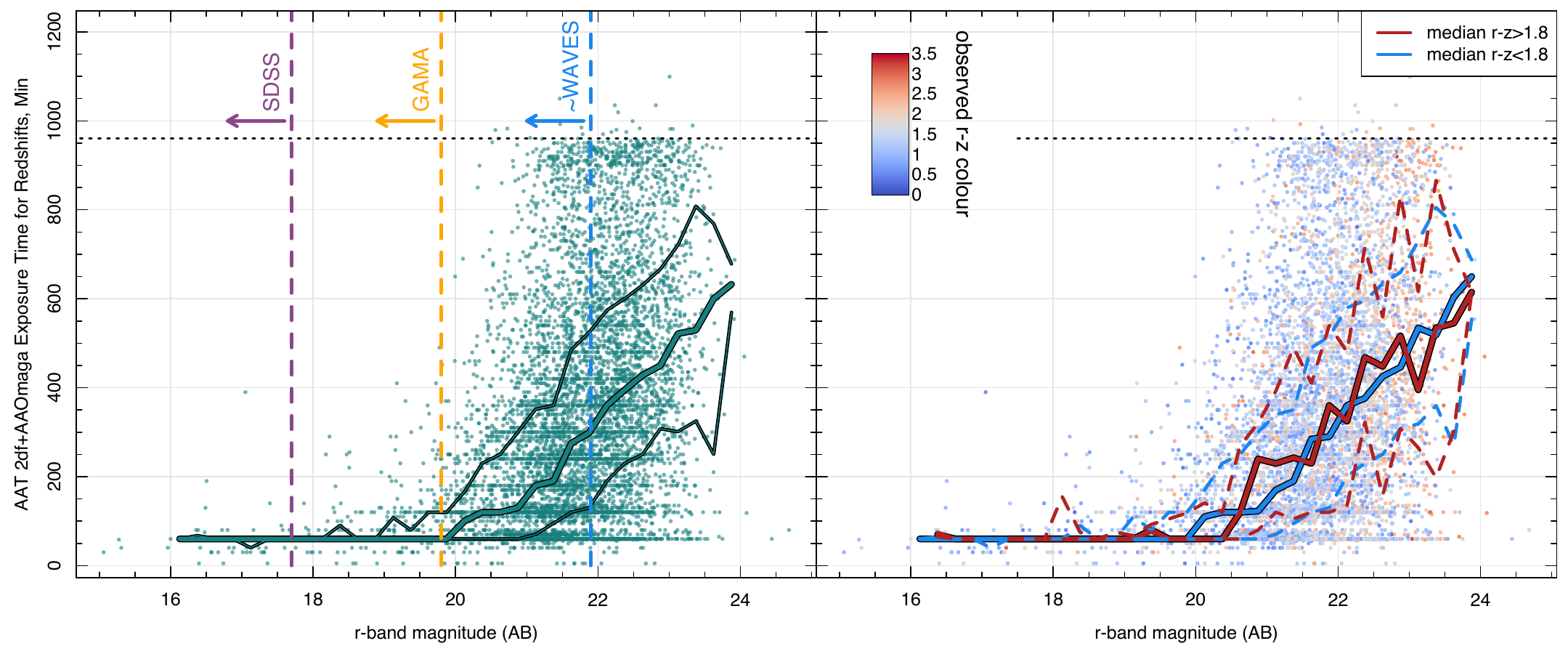}
\caption{Exposure times required to obtain a secure spectroscopic redshift from our DEVILS AAT+AAOmega data as a function of r-band magnitude. The left panel shows all sources, with the green lines displaying the running median and interquartile range. We also over-plot the spectroscopic target limits SDSS (purple), GAMA (orange) and upcoming WAVES (blue). The right panel shows the same but colour-coded by observed-frame $r-Z$ colour. These running median's show very little difference between red and blue populations, suggesting at these faint magnitudes observed-frame colour is not a strong predictor of the exposure time required to obtain a redshift. A redshift-feedback methodology (as described in the text) overcomes all of these issues, by having no fixed exposure time and also not relying on a priori information to determine an exposure time.              }
\label{fig:Texp}
\end{center}
\end{figure*}

\begin{figure*}
\begin{center}
\includegraphics[scale=0.7]{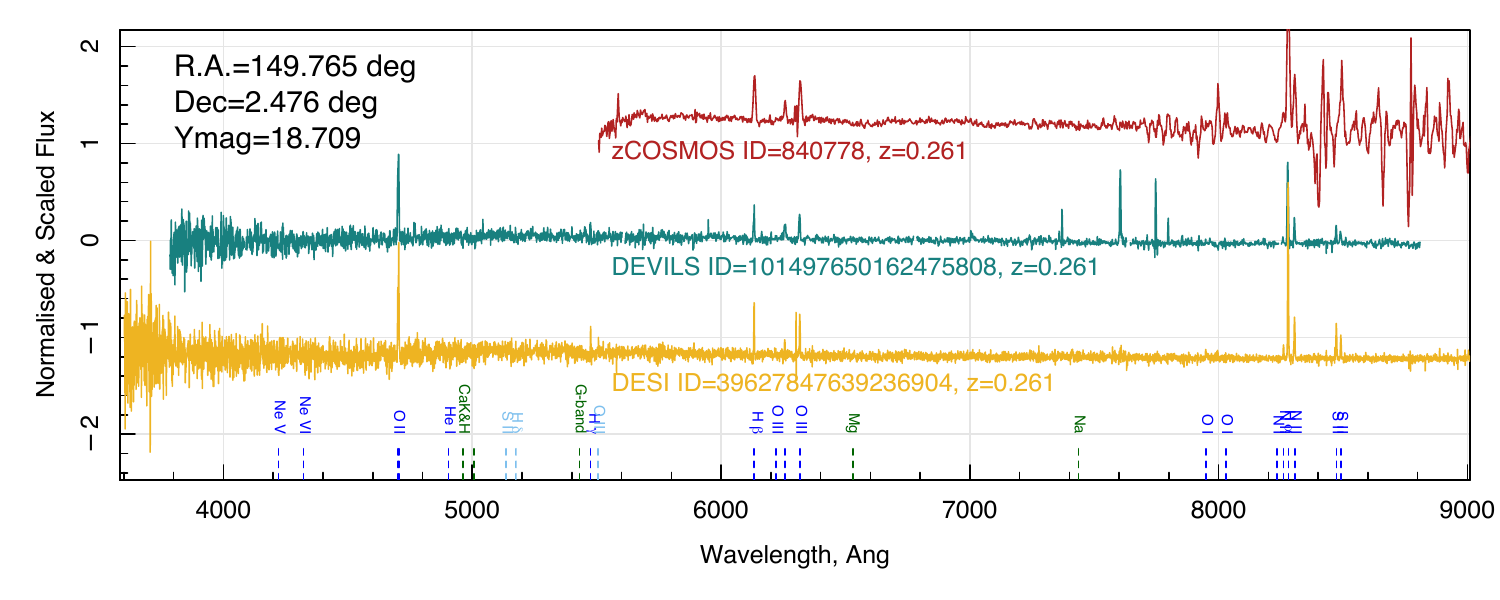}
\caption{Example DEVILS AAT spectrum (green) for source UID=101497650162475808 for which both zCOSMOS (red) and DESI (gold) spectra are also available. Spectra are scaled to the same noise level and normalised to offset flux values for clarity. All surveys obtain the same redshift assignment for this source. }
\label{fig:SpecComp}
\end{center}
\end{figure*}

\begin{figure*}
\begin{center}
\includegraphics[scale=0.53]{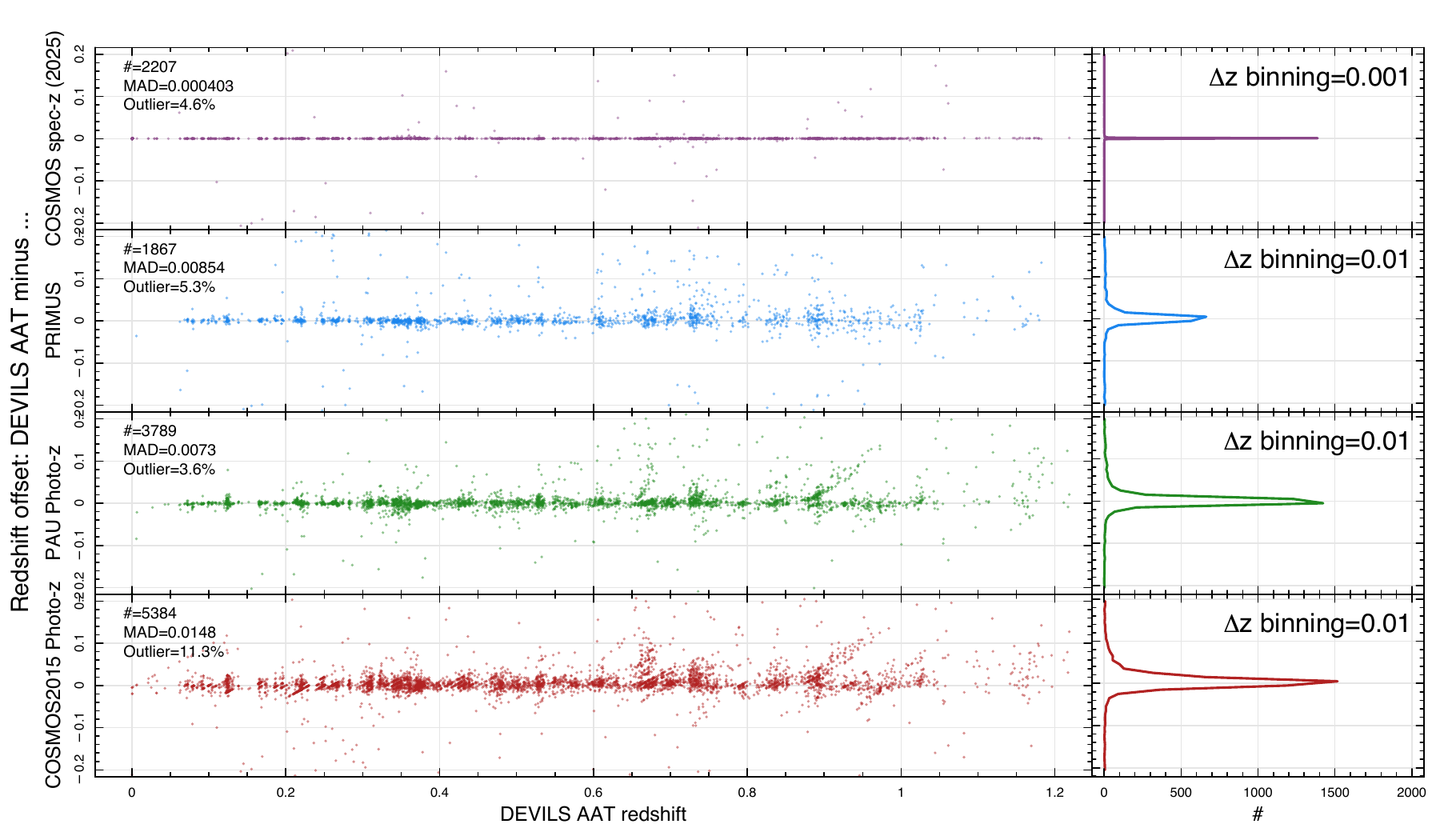}
\caption{Comparison between our DEVILS-AAT spectroscopic redshifts and other redshift measurements for the same sources. The left panels show $\Delta z$ of DEVILS-AAT - X as a function of the DEVILS-AAT redshifts. The number of sources in the comparison, mean absolute deviation (MAD) and outlier fraction are given in the top left. Outlier fraction is defined as $|(z_{X}-z_{\mathrm{DEVILS}})/(1+z_{DEVILS})|>0.15$ .The right panels show a histogram of $\Delta z$ values. Note, binning for $\Delta z$ histograms is finer for the spectroscopic comparison in the top row to show the tighter relation - binning is noted as $\Delta z$Offset. The top row shows a comparison with the recent COSMOS spectroscopic complication of \citet{Khostovan25} in purple, next row shows a comparison to PRIMUS grism redshifts in blue, next row shows a comparison to PAU photometric redshifts in green, and the bottom row shows a comparison to COSMOS2015 photometric redshifts.}
\label{fig:RedComp}
\end{center}
\end{figure*}

\begin{figure*}
\begin{center}
\includegraphics[scale=0.53]{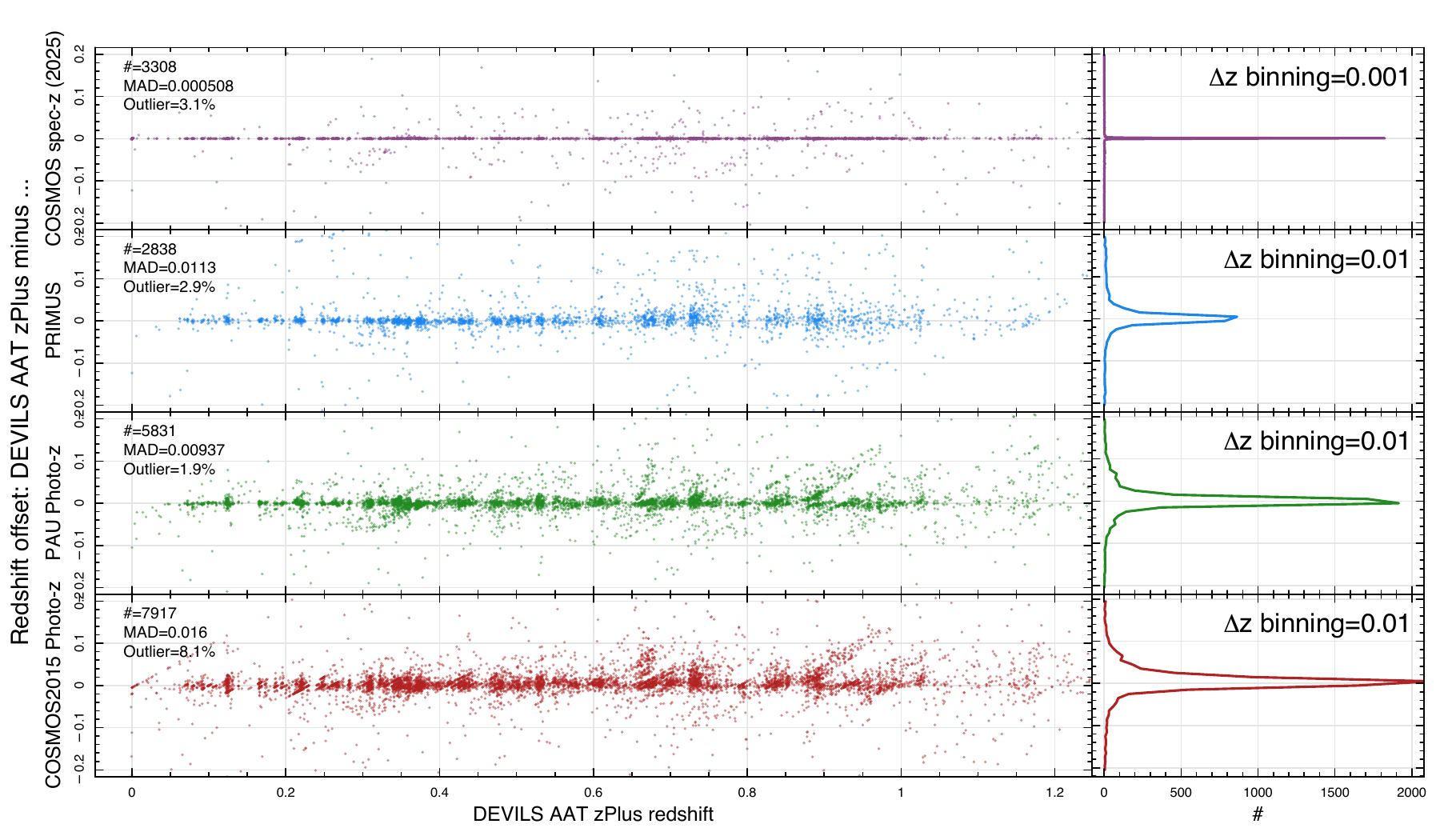}
\caption{The same as Figure \ref{fig:RedComp} but now showing comparisons to our \texttt{DEVILS\_zPlus} redshifts (DEVILS-AAT redshifts weighted by other survey redshifts in the COMSOS region).}
\label{fig:RedComp2}
\end{center}
\end{figure*}

\subsection{ExposureTimes for Faint Sources}

Given the nature of DEVILS, which aims to provide a deep but high-completeness spectroscopic sample of galaxies at intermediate redshifts, it is interesting to explore the observational difficulties with obtaining redshifts at these magnitudes/epochs. This not only describes one of the core diagnostics for the resultant DEVILS sample, but also provides key information required for the planning and execution of future deep redshift surveys, such as 4MOST-WAVESdeep which will target similar populations to DEVILS. As described in Section \ref{sec:redshfiting}, one of the biggest complications with this process is our inability to define the required exposure time to obtain a redshift for faint sources. To highlight this, Figure \ref{fig:Texp} shows the final total exposure times required to obtain a secure redshift, as a function of r-band magnitude. In the left panel we highlight the running median and interquartile range of exposure times for a secure redshifts. We also over-plot the $\sim$r-band limits of the SDSS, GAMA and WAVES spectroscopic samples. Clearly we see here that for surveys such as SDSS and GAMA, little is gained from going beyond fixed short exposure times. However, as we move to fainter galaxies, the exposure time required to obtain a redshift can vary highly, such that using fixed exposure times is no longer practical. As such, exposure times are problematic to estimate a priori and using a redshift-feedback methodology significantly improves survey efficiency.  A potential solution to this is to use colour information to better predict required exposure times. However, as shown, in the right panel of Figure \ref{fig:Texp}, these exposure times are not significantly different for red and blue populations (in the observed frame), such that this information largely does not help in a priori predicting exposure times for faint sources - further motivating the redshift feedback methodology.

\begin{figure}
\begin{center}
\includegraphics[scale=0.6]{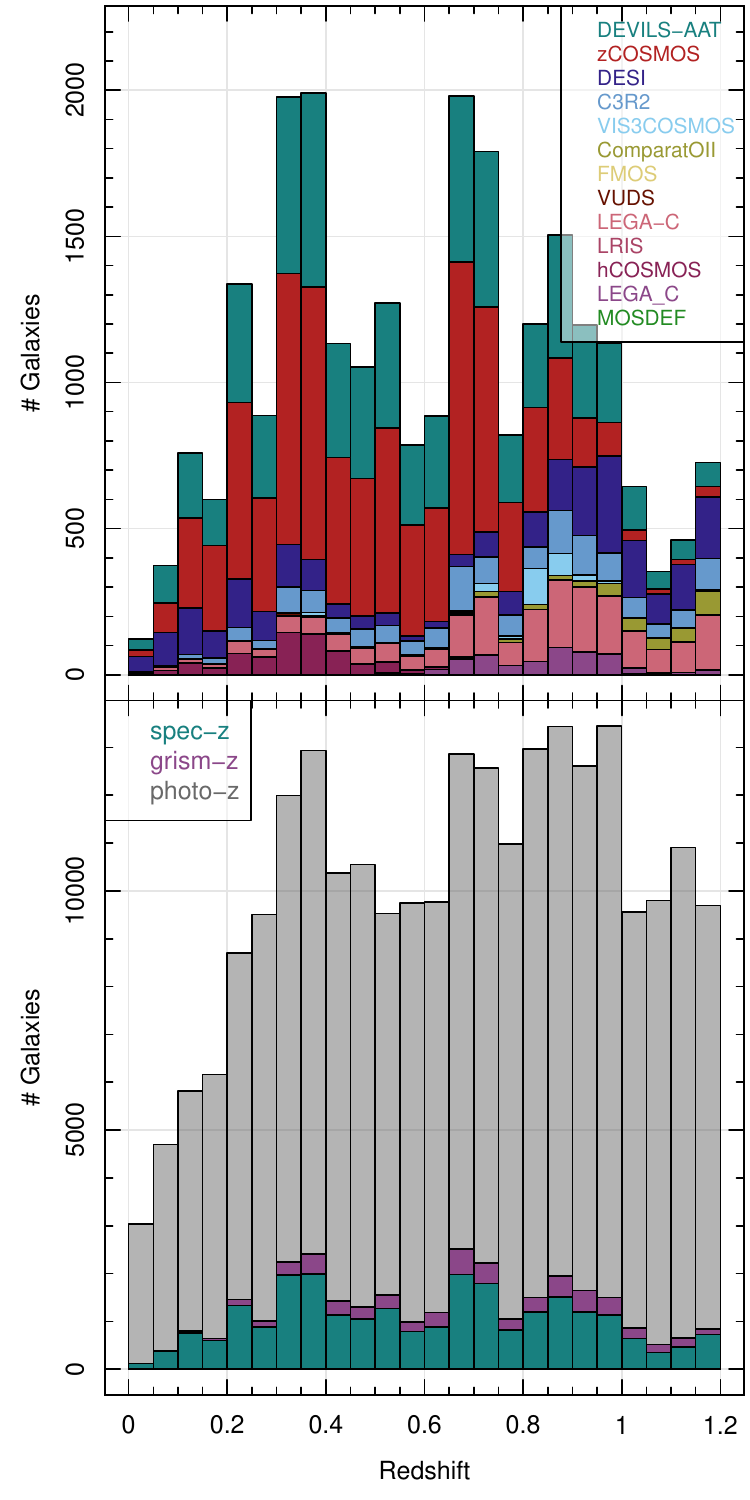}
\caption{Redshift samples included in the \texttt{devils\_dr1.D10MasterRedshifts} catalogue. Top panel shows the breakdown of \texttt{zBest} values for various spectroscopic samples over various campaigns (given as \texttt{zBestSource} in the catalogue), bottom panel shows a comparison of spectroscopic, grism and photometric redshifts (given as \texttt{zBestType} in the catalogue) going into the final \texttt{zBest} value. \textcolor{black}{Both panels show the full distribution of sources over this redshift range, with no magnitude limit imposed.}  }
\label{fig:redshifts}
\end{center}
\end{figure}

\begin{table*}
\caption{Surveys used to compile redshifts in our \texttt{devils\_dr1.D10MasterRedshifts} catalogue. $^{1}$ Number of sources with any redshift measurement in the D10 region.  $^{2}$ filter used to define a robust redshift for this survey. $^{3}$ Number of sources used as \texttt{zBest} in the final \texttt{devils\_dr1.D10MasterRedshifts} catalogue. Objects are assigned \texttt{zBest} in order of priority from top to bottom in the table with the filter column applied. $^{4}$ Median VISTA Y-band AB magnitude of  \texttt{zBest} objects from this sample. $^{5}$ Notes on sample containing the method used to measure redshifts and the probability measure given in the \texttt{devils\_dr1.D10MasterRedshifts} catalogue. Redshifting approaches: \textsc{AutoZ}=Automatic redshift code \citep{Baldry14}, \textsc{EZ}=Easy Redshift \citep{Garilli10},  cross-corr= cross-correlation between templates and galaxy spectra, \textsc{SpecPro}= Interactive tool for visual classification \citep{Masters11}, Visual = visual inspection of galaxy redshifts, \textsc{EAZY}=photometric redshift fitting code \citep{Brammer08}, \textsc{Le Phare}=photometric redshift fitting code \citep{Arnouts11}. Probability types:  P($z$)=correct redshift probability, flag=numeric flag assigned to redshift quality (can be different for different surveys, but DR1 documentation describes flag meanings for each survey), correlation=strength of the cross-correlation peak between the best-fit template and galaxy spectrum.}
\begin{center} 
\begin{scriptsize}
\begin{tabular}{l c l c c l c c l l}
Name  &  Type & Facility & \#$^{1}$  & $z_{med}$ & filter$^{2}$ &  \# zBest$^{3}$ & Y$_{med}$$^{4}$ & Ref. & Note$^{5}$ \\
\hline
DEVILS & Spec & AAT-AAOmega & 10,005 & 0.47 & DEVILS\_probPlus$>$0.95 OR & 7,946 & 20.46 & This work & \textsc{AutoZ} redshifts, prob=P($z$) \\
 &  &  &  &  &  DEVILS\_probPlus$>$0.90 if &  &  &  &  \\
  &  &  &  &  &  zBestType!=`spec' &  &  &  &  \\
zCOSMOS & Spec & VLT-VIMOS & 18,957 & 0.56 & 3$\le$zCOSMOS\_prob$<$5 OR  & 10,940 & 20.76 & \cite{Lilly07} & \textsc{EZ} redshifts, prob=flag \\
 &  &  &  &  &  13$\le$zCOSMOS\_prob$<$15 OR &  &  &  &  \\
 &  &  &  &  &  23$\le$zCOSMOS\_prob$<$25 &  &  &  &  \\
DESI & Spec & Mayall-DESI & 10,206 & 0.73 & DESI\_prob$\ge$9 \& ZWARN==0 & 6,428 & 21.71 & \cite{DESI24} & \textsc{RedRock}, prob=DELTACHI \\
hCOSMOS & Spec & MMT-Hectospec & 4,208 & 0.32 & None & 754 & 19.67 & \cite{Damjanov18} & cross-corr, prob=correlation \\
LEGA\_C & Spec & VLT-VIMOS & 1,880 & 0.76 & None & 679 & 21.44 & \cite{vanderWel16} & cross-corr, prob=correlation \\
VVDS & Spec & VLT-VIMOS & 4 & 0.69 & VUDS\_prob=1.5,2,3,4,$>$10 & 1 & 22.18 & \cite{LeFevre13} & \textsc{EZ}, prob=flag \\
VUDS & Spec & VLT-VIMOS & 156 & 2.51 & 3$\le$VVDS\_prob$\le$5 & 148 & 24.13 & \cite{LeFevre15} & \textsc{EZ}, prob=flag \\
FMOS & Spec & Subaru-FMOS & 1,150 & 1.56 & 3$\le$FMOS\_prob$\le$5 & 288 & 22.64 & \cite{Silverman15} & \textsc{SpecPro}, prob=flag \\
MOSDEF & Spec & Keck-MOSFIRE & 564 & 2.29 & MOSDEF\_prob$\ge$4 & 357 & 24.01 & \cite{Kriek15} & Visual, prob=flag \\
C3R2 & Spec & Keck-Multi & 2,622 & 0.86 & 3$\le$C3R2\_prob$\le$4 & 2,238 & 22.97 & \cite{Masters19} & Visual, prob=flag \\
DEIMOS & Spec & Keck-DEIMOS & 7,440 & 0.89 & DEIMOS\_prob=2 OR 1.5 & 4,739 & 22.41 & \cite{Hasinger18} & \textsc{SpecPro}, prob=flag \\
LRIS & Spec & Keck-LRIS & 433 & 2.50 & 3$\le$LRIS\_prob$\le$4 & 234 & 23.74 & \cite{Lee18} & Visual, prob=flag \\
ComparatOII & Spec & VLT-FORS2 & 1,440 & 1.14 & None & 657 & 22.11 & \cite{Comparat15} & Visual, prob=flag \\
 VIS3COSMOS & Spec & VLT-VIMOS & 653 & 0.84 & None & 354 & 22.11 & \cite{Paulino-Afonso18} & \textsc{SpecPro}, prob=flag \\
PRIMUS & Grism & Baade-IMACS & 27,466 &  0.57 & None & 5,590 & 22.06 &\cite{Cool13} & Grism+photometry cross-corr \\
3DHST & Grism & HST-G141 & 3,797 &  0.94 & None & 1,463 & 23.19 &\cite{Momcheva16} & Grism+photometry \textsc{EAZY} \\
PAU & Photo & WHT-PAUCam & 37,474 &  0.64 & None & 13,885 & 22.03 &\cite{Alarcon21} & 40 Narrow-band SED-fits \\
COSMOS2015 & Photo & Multi & 507,395 &  1.15 & None & 451,276 & 24.71 &\cite{Laigle16} & \textsc{Le Phare}  \\
MIGHTEEtmp & Photo & Multi & 494,179 &  1.13 & None & 221,980 & 26.09 &\cite{Adams20} & MIGHTEE team \textsc{Le Phare}  \\

\end{tabular}
\end{scriptsize}
\end{center}
\label{tab:surveys}
\end{table*}

\begin{figure*}
\begin{center}
\includegraphics[scale=0.48]{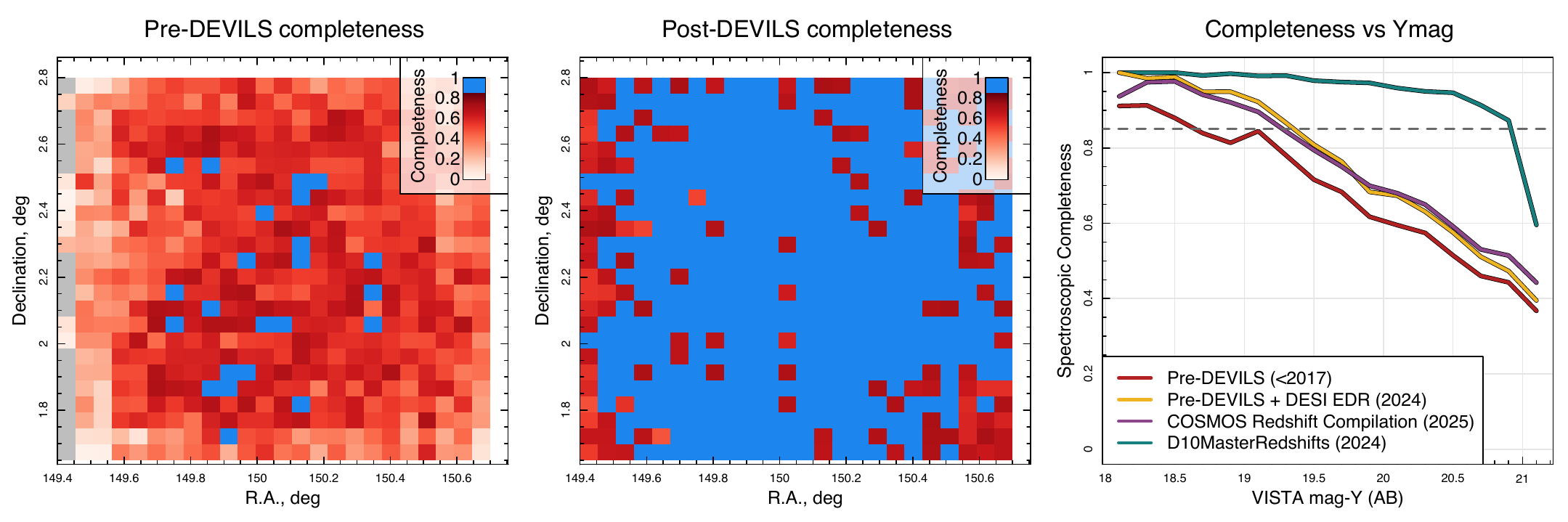}
\caption{Left: The spatial completeness of DEVILS targets with pre-DEVILS robust spectroscopic redshifts. Middle: The spatial completeness of DEVILS targets with post-DEVILS robust spectroscopic redshifts. Grid cells with completeness $>$ 85\% are shown in blue. Right: The spectroscopic completeness as a function of Ymag for a pre-DEVILS sample,   pre-DEVILS sample + DESI EDR redshifts, the recent COSMOS Spectroscopic Redshift Compilation Catalog and full DEVILS sample taken from the \texttt{devils\_dr1.D10MasterRedshifts} catalogue. The dashed horizontal line shows a completeness of 85\%, which is required for high completeness science ($e.g.$ group finding). The DEVILS sample extends to roughly 1.5 magnitudes fainter than other samples at this completeness.   }
\label{fig:FinalSpatialComp}
\end{center}
\end{figure*}

\subsection{Comparison to existing redshifts}

Next we further compare our DEVILS-AAT redshifts to existing known redshifts in the DEVILS-D10/COSMOS region. There are a large number of redshift campaigns in COSMOS. These will be used to build our master redshift catalogue and will be described extensively in Section \ref{sec:redshiftcomp}. Here we simply show a comparison to a number of key programs in order to validate the redshift measurements obtained from our DEVILS AAT spectra. 

First, for an initial visual comparison we identify sources that have a common observation across multiple surveys where 1D spectra are publicly available. In our DR1 data release we ultimately provide matched and standardised DEVILS-AAT, zCOSMOS, and DESI EDR 1D spectra for our D10 targets. As such, for initial visual comparison we identify 171 sources which have an observation, and secure redshifts, in all of DEVILS-AAT, zCOSMOS and DESI.  This number is small, as we deliberately do not target the majority of zCOSMOS objects with our DEVILS-AAT observations - but provides a direct comparison point for the three surveys. We visually inspect these spectra and find 97.5\% agreement between DESI EDR and DEVILS-AAT redshifts and 95.2\% agreement between zCOSMOS and DEVILS-AAT  - and also note that the small number of sources which disagree on redshifts are different between DESI EDR and zCOSMOS (96.8\%). An example of this visual comparison, selected where all redshifts agree, is shown in Figure \ref{fig:SpecComp} for a single source.   

Next, we take DEVILS galaxies with  P($z$)$>$0.95 and perform a 2\,arcsec positional match to the COSMOS spectroscopic redshift compilation of \citet{Khostovan25}. We note that this sample also contains redshifts from the DESI EDR and zCOSMOS as noted above. In the top panel of Figure \ref{fig:RedComp} we show a comparison of redshift measurements for this matched sample, reiterating the agreement discussed perviously. Next in Figure \ref{fig:RedComp} we show comparisons to robust (Q=3-4) grism redshift from the PRIsm MUlti-object Survey \citep[PRIMUS,][]{Cool13}, highly-precise narrow-band photometric redshifts from the Physics of the Accelerated Universe Survey using PAUCam \cite[PAUS,][]{Alarcon21} and COSMOS2015 photometric from \cite{Laigle16}. These redshifts are matched to DEVILS-AAT sources using the methodology outlined in Section \ref{sec:redshiftcomp} (using the DEVILS segmentation maps).  In the top right legend in Figure \ref{fig:RedComp} we note the number of sources matched between each sample, the mean absolute deviation (MAD) of the redshift offsets and outlier fraction, with outliers defined as sources with $|(z_{X}-z_{\mathrm{DEVILS}})/(1+z_{DEVILS})|>0.15$ (where $X$ is the survey being compared to). All samples show good agreement with our DEVILS-AAT redshifts, with low outlier fraction and low MAD (see figure for details).  As such, overall we find good agreement between our new DEVILS-AAT redshifts and those for other surveys in the D10 region. We see no systematic biases in our redshift assignments, which would bias any scientific analyses with this sample.

\section{Compilation of D10 Redshift Catalogue}
\label{sec:redshiftcomp}

As noted previously, there are a large number of surveys that have measured redshifts in the D10 region. These range from relatively low precision/accuracy broad-band photometric redshifts, to high precision/accuracy spectroscopic redshifts. In order to build science-ready \texttt{devils\_dr1.D10MasterRedshifts} catalogue, we combine all available redshifts for sources in the region. In this section we describe the origin of redshifts used in the \texttt{devils\_dr1.D10MasterRedshifts} catalogue, how they are matched to sources in the DEVILS photometry catalogue (\texttt{devils\_dr1.D10ProFoundPhotometry}), how we derive `improved' DEVILS-AAT redshifts and probabilities using existing redshift information and finally how we assign a `best' redshift to each source in the \texttt{devils\_dr1.D10ProFoundPhotometry} catalogue. 

First, in Table \ref{tab:surveys} we outline the various surveys used in our redshift compilation. Note this is an updated version of the compilation  presented in \cite{Thorne21}. In the \texttt{devils\_dr1.D10MasterRedshifts} catalogue we provide all individual redshift measures that are associated with a given source. The number of sources with redshifts provided by a particular sample and the median redshift is given in  Table \ref{tab:surveys}.  In order to match redshifts for the same source, we utilise the \textsc{ProFound} segmentation maps derived in our photometric analyses discussed in detail in \cite{Davies21}. Briefly, these maps associate imaging pixels to a particular source in a deep NIR stacked image using UltraVISTA imaging. To match sources, we first identify all sources from a given redshift catalogue from Table \ref{tab:surveys} which lie within the segment of a particular DEVILS photometric source. If multiple sources from the same catalogue fall within a single DEVILS photometric source, we take the closest object to the flux-weighted centre of the segment. The matched source redshift values are then attached to this DEVILS source, and added to the \texttt{devils\_dr1.D10MasterRedshifts} catalogue.

\subsection{Weighting based on redshift agreement}

To this point, we have assigned the DEVILS-AAT redshifts in isolation. However, the additional redshift information for other surveys in the COSMOS region can help better constrain the correct galaxy redshifts from our AAT spectra. This is particularly true when including information from low precision measurements from $e.g.$ photometric redshifts. For example, if a low P($z$) redshift solution from our \textsc{AutoZ} analysis is consistent with the photometric redshift (or grism or existing spectroscopic redshift), this adds additional weight to that particular solution. As such, we can use the existing redshifts in the COSMOS region to further improve both our redshift assignment and probabilities. We do not apply this process in our initial analysis, as we wish to retain our direct \textsc{AutoZ} redshifts measured only from of DEVILS-AAT spectra, but add additional columns to the final \texttt{devils\_dr1.D10MasterRedshifts} catalogue for these new redshifts and probabilities (hereafter, \texttt{DEVILS\_zPlus} and \texttt{DEVILS\_probPlus}). 

To undertake this process, we return to our original \textsc{AutoZ} outputs containing the top 5 redshift solutions for each source. We then compare each of these solutions to the matched redshifts from all of the other surveys described above. First, we do not include the MIGHTEEtmp photometric redshifts in this analysis as the photometric redshift have low precision and are largely used to fill the faint sample where no COSMOS2015 photometric exists. For the COSMOS2015 photometric sample, we use the lower and upper $1\sigma$ bound to the photometric redshift P($z$), provided in the COSMOS2015 catalogue, and check if any of our DEVILS \textsc{AutoZ} redshifts fall within this range. This is then deemed as a match. For all other surveys we use the metric for redshift agreement as:

\begin{equation*}
| ln(1+z_{\mathrm{DEVILS}})-ln(1+z_{X}) | < lim_{\mathrm{type}}
\end{equation*}

where $z_{X}$ is the redshift from the survey in question, and $lim_{\mathrm{type}}$ is a variable limit depending on the redshift type (spec, grism, photo) where:  $lim_{\mathrm{spec}}$=0.005, $lim_{\mathrm{grism}}$=0.025, and $lim_{\mathrm{photo}}$=0.05. We initially define the new \texttt{DEVILS\_probPlus} variable equal to the raw \textsc{AutoZ} P($z$) values for that solution. We then apply the following additions to this value: \\

\noindent $\bullet$ If the DEVILS \textsc{AutoZ} redshift aligns with any existing photometric redshift,  \texttt{DEVILS\_probPlus} = \texttt{DEVILS\_probPlus}+1 \\

\noindent $\bullet$ If the DEVILS \textsc{AutoZ} redshift aligns with any existing grism redshift,  \texttt{DEVILS\_probPlus} = \texttt{DEVILS\_probPlus}+2 \\  

\noindent $\bullet$ If the DEVILS \textsc{AutoZ} redshift aligns with any existing spectroscopic redshift,  \texttt{DEVILS\_probPlus} = \texttt{DEVILS\_probPlus}+4 \\

In this process, any source with a \texttt{DEVILS\_probPlus}$>$1.0 delineates that multiple redshift sources agree, and the decimal values indicate the original \textsc{AutoZ} P($z$). All of these are also additive, such that a source with a DEVILS \textsc{AutoZ} redshift that aligns with a photometric, grism and spectroscopic redshift will have a value of \texttt{DEVILS\_probPlus}$=$7.xx. We then determine the new best DEVILS \textsc{AutoZ} redshift based on the highest \texttt{DEVILS\_probPlus} value and assign this as \texttt{DEVILS\_zPlus}.     

We find that through this process, 2,539 sources are added as having a robust redshift. Of these, 1,432 DEVILS sources change their initial redshift assignment ($i.e.$ \texttt{DEVILS\_z}!=\texttt{DEVILS\_zPlus}), the rest simply have their initial redshift up-weighted in probability. Of the sources that change redshift, only 217 were initially assigned a P($z$)>0.95 - such that we are largely just adding to the sample from low confidence redshifts which now have an up-weighted probability based of existing redshift information. Of the sources where the redshift is changed, 845 have a redshift consistent with an existing photometric redshift, 488 are consistent with an existing grism redshift and 372 are consistent with an existing spectroscopic redshift. In Figure \ref{fig:RedComp2} we show the same as Figure \ref{fig:RedComp}, but now using our new \texttt{DEVILS\_zPlus} redshifts. We find the number of robust redshift matches between samples goes up considerably, and outlier fractions are reduced for all comparisons.

Once sources are matched we then aim to identify the `best' redshift available for a particular photometric source (hereafter \texttt{zBest}). To do this will apply an order of priority from bottom to top in Table \ref{tab:surveys}, with the filters applied. For example, sources are initially assigned a MIGHTEEtmp photometric redshift, then if a COSMOS2015 redshift exists, \texttt{zBest} is updated, next if a PAU redshift exists, \texttt{zBest} is updated, and so on. Note that for this process we now use the \texttt{DEVILS\_zPlus} and \texttt{DEVILS\_probPlus} values over the initial DEVILS redshifts.  Quality filters are applied to the spectroscopic catalogues as to ensure a robust redshift is used. These filters are provided in Table \ref{tab:surveys} and notes on the *\_prob value in these filters are given in the final column ($i.e.$ zCOSMOS provides a quality flag, while DESI provides a $\Delta chi^{2}$ value). In our final catalogue we provide all of these values. Once the filters are applied and \texttt{zBest} obtained, the number of sources from each sample used in this \texttt{zBest} are given in the table, along with the median Y-mag for the sources used. In Figure \ref{fig:redshifts} we show the resultant distribution of \texttt{zBest} values. The bottom panels show the breakdown between spectroscopic, grism and photometric redshifts, while the top panel shows the breakdown from individual spectroscopic samples.

\subsection{Completeness}

Using the \texttt{zBest} values obtained above, it is interesting to consider how our final spectroscopic redshift sample compares to existing samples. In Figure \ref{fig:FinalSpatialComp} we show the pre-DEVILS spatial spectroscopic redshift completeness to Ymag$<21.2$  DEVILS targets (left) and the final spatial spectroscopic redshift completeness using the \texttt{zBest} values defined above. In the right panel of Figure \ref{fig:FinalSpatialComp} we show this completeness as a function of Ymag. Here we show the DEVILS \texttt{devils\_dr1.D10MasterRedshifts} completeness in green, the pre-DEVILS completeness in red, the pre-DEVILS + DESI EDR completeness in gold and the completeness from the COSMOS spectroscopic redshift compilation of \citet{Khostovan25} in purple. DEVILS obtains a spectroscopic redshift completeness of $>85\%$ to $\sim$1.5 magnitudes fainter than other samples - achieving our goal.

\begin{figure}
\begin{center}
\includegraphics[scale=0.6]{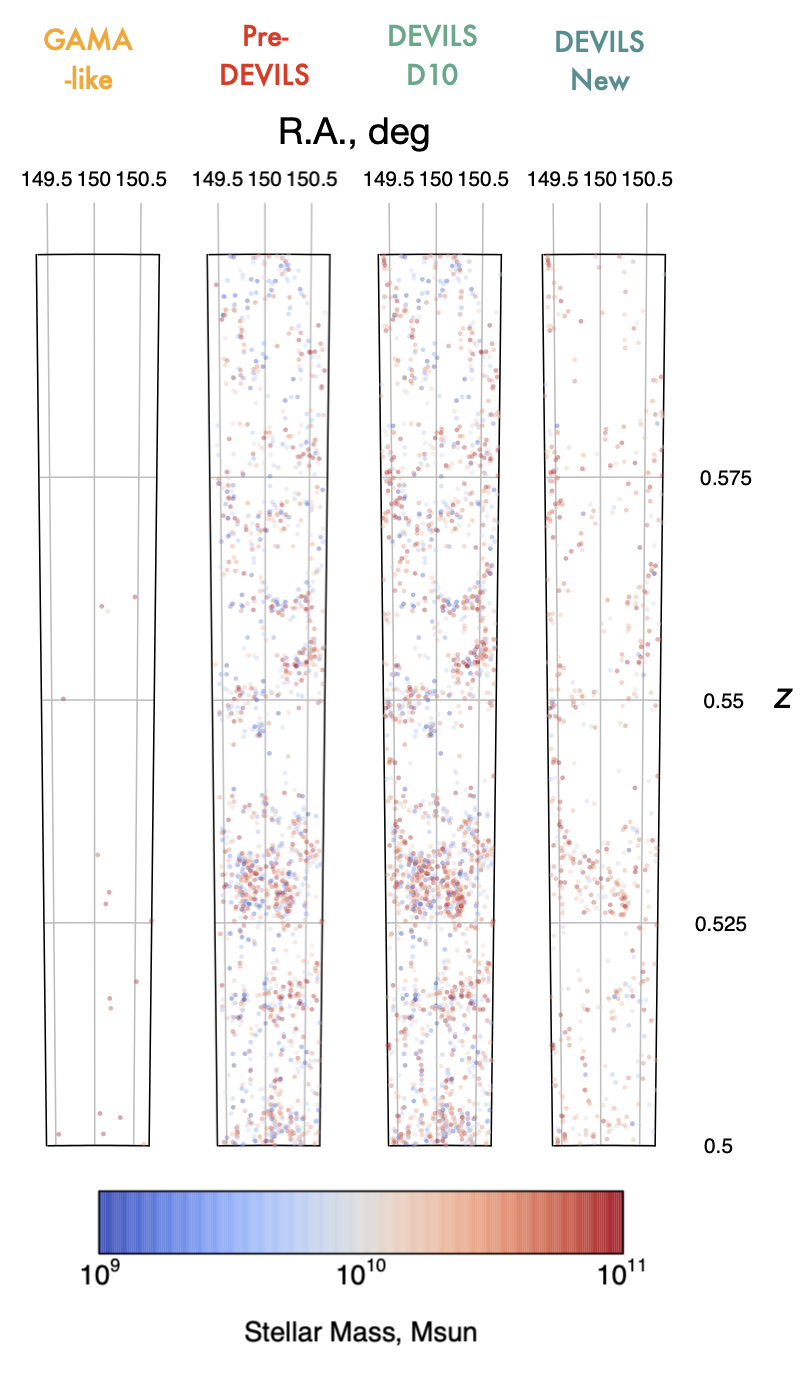}
\caption{Light-cones showing large-scale-structure in the D10 region at 0.5$<$$z$$<$0.6. Points are coloured by log$_{10}$(stellar mass). Left shows the sample which would be obtained using GAMA-like spectroscopic limits ($.i.e.$ comparable completeness sample at a brighter magnitude limit), where no large-scale-structure could be defined at these epochs. Middle-left shows the pre-DEVILS sample, where large-scale-structure is visible for the most massive structures, but a significant fraction of the population is missing. Middle-right shows the DEVILS D10 sample, where we are robustly tracing the large-scale-structure, including filaments, and identifying many more galaxies associated with each over-density (which will allow more robust dark matter halo measurements). Right shows the galaxies added by the DEVILS observations. }
\label{fig:Lightcones}
\end{center}
\end{figure}

To highlight the benefit of this high completeness, in Figure \ref{fig:Lightcones} we then show the light-cone distribution of galaxies in the D10 region at 0.5$<$$z$$<$0.6. Here points are coloured by their stellar mass taken from the \texttt{devils\_dr1.D10ProSpectAGN} catalogue described in the following section. On the left, we show the distribution of objects assuming a GAMA-like selection ($i.e.$ high spectral redshift completeness in the local Universe), which can not map any large-scale-structure at these redshifts. In the middle-left we show the pre-DEVILS sample, where the large-scale-structure can be seen, but individual structures contain a smaller number of objects and faint filaments are hard to discern. On the middle-right we show the resultant sample from the \texttt{devils\_dr1.D10MasterRedshifts} catalogue where individual structures contain a larger number of objects (allowing more robust measurements of halo mass, etc) and faint filaments are visible out to $z$$=$0.6. On the right we show new objects added by the DEVILS observations, which trace the same large-scale-structure. Interestingly, DEVILS predominantly adds higher stellar mass galaxies, which are typically passive and have therefore been missed by existing surveys.       

\begin{figure}
\begin{center}
\includegraphics[scale=0.5]{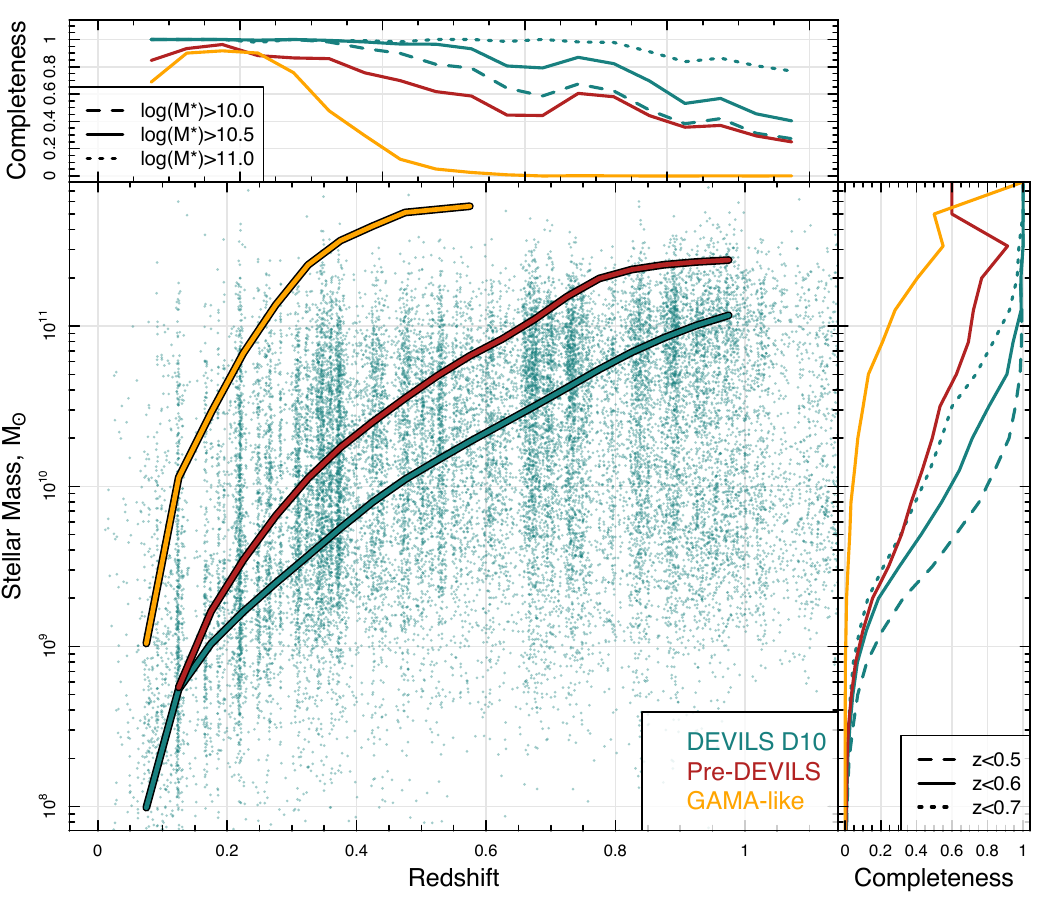}
\caption{The resultant stellar mass - redshift distribution from combining the \texttt{devils\_dr1.D10MasterRedshifts}  catalogue and \texttt{devils\_dr1.D10ProSpectAGN} catalogue. Top and side panels show the spectroscopic completeness as a function of redshift and stellar mass respectively. For DEVILS, these are shown for a range of stellar mass and redshift limits (given the respective legends). We show the samples obtained from a GAMA-like ($r<19.8$) selection limit (orange), the COSMOS sample prior to DEVILS spectroscopic observations (red) and the final DEVILS D10 sample (green). To estimate the stellar mass - redshift completeness limits, we use the full photometric sample of objects and compare to those with spectroscopic redshifts. Lines in the main panel, are shown where the sample drops below 85\% completeness. These lines can be used to define a stellar-mass complete sample at a given epoch. As such, in comparison to previous samples, the DEVILS D10 sample allows stellar mass complete samples to be extended by $\sim0.5\,dex$ in stellar mass at all epochs out to $z\sim1$.}
\label{fig:Mz}
\end{center}
\end{figure}

Finally, we then explore the stellar-mass completeness of our final \texttt{devils\_dr1.D10MasterRedshifts} catalogue in combination with the stellar masses taken from the \texttt{devils\_dr1.D10ProSpectAGN} catalogue (see Section \ref{sec:SEDs}).  Figure \ref{fig:Mz} show the redshift-stellar mass distribution of our final sample. We then also show the stellar mass completeness limits by calculating the point at which the spectroscopic sample drops below 85\% completeness in comparison to the full  photometric redshift sample. These lines are shown for a GAMA-like ($r<19.8$) selection limit (orange), the COSMOS sample prior to DEVILS spectroscopic observations (red) and the final \texttt{devils\_dr1.D10MasterRedshifts} sample (green). These lines can then be used to define a science-ready stellar-mass complete sample at a given epoch. In comparison to previous samples, the \texttt{devils\_dr1.D10MasterRedshifts} sample allows stellar mass complete samples to be extended by $\sim0.5\,dex$ in stellar mass at all epochs out to $z\sim1$.

\section{DR1 Data Products}
\label{sec:data}

In the following section we detail all of the DEVILS DR1 data products which are made publicly available via \texttt{Data Central}. These have been discussed in other key works, as such we do not go into detail regarding the derivation of these data products, we only present the data provided and notes how best to use said data. 

\subsection{Catalogues}

Within the DEVILS DR1 we release 8 catalogues of measured and derived galaxy properties, discussed below. A unique identifier (\texttt{UID}) in applied to each photometric source. This \texttt{UID} is generated as \texttt{UID}=field number+round(\texttt{RAmax}*1000000) + | round(\texttt{DECmax*}1000000) |. For example, a source in the D10 region, at \texttt{RAmax}=149.717842907065 and \texttt{DECmax}=1.94849275929888 has a resultant UID=101497178431948496. Many of the DEVILS catalogue tables also contains a standard set of starting columns to allow easy selection of science-quality extra-galactic targets. These columns are outlined in Table \ref{tab:baseColumns}. All tables described below can be accessed through the \texttt{Data Central} archive using a number of tools (Structured Query Language, Application Programming Interface, schema browser, etc) along with all meta data and documentation associated with the catalogue parameters.  

\begin{table*}
\caption{Base DR1 catalogue parameters used across multiple DR1 catalogues.}
\begin{center}
\begin{tabular}{l l l c l }
Name  &  Description & UCD & Unit & Type \\
\hline
UID & Unique identifier based on DEVILS filed, RA \& DEC & meta.id & None & integer \\ 
RAcen & Right ascension position of the \textsc{ProFound} segment centre in the VISTA Y-band image & pos.eq.ra & deg & double \\
DECcen & Declination position of the  segment centre in the VISTA Y-band image & pos.eq.ra & deg & double \\
RAmax & Right ascension position of the segment brightest pixel in the VISTA Y-band image & pos.eq.ra & deg & double \\
DECmax & Declination position of the segment brightest pixel in the VISTA Y-band image & pos.eq.ra & deg & double \\
mask & Flag to identify if the segment falls in a masked region. 0=not masked, >0=masked & meta.id & none & integer\\
artefactFlag & Flag to identify if the segment is an artefact (see \cite{Davies21}). 0=Not artefact, >0=artefact & meta.id & none & integer \\
starFlag & Star/galaxy photometric classification (see \cite{Davies21}). 0=galaxy, 1=star & meta.id & None & integer \\
mag\_Y & Total magnitude measured in VISTA Y band segment  & phot.mag;em.IR.Y & AB mag & double \\ 

\end{tabular}
\end{center}
\label{tab:baseColumns}
\end{table*}

\begin{figure*}
\begin{center}
\includegraphics[scale=0.195]{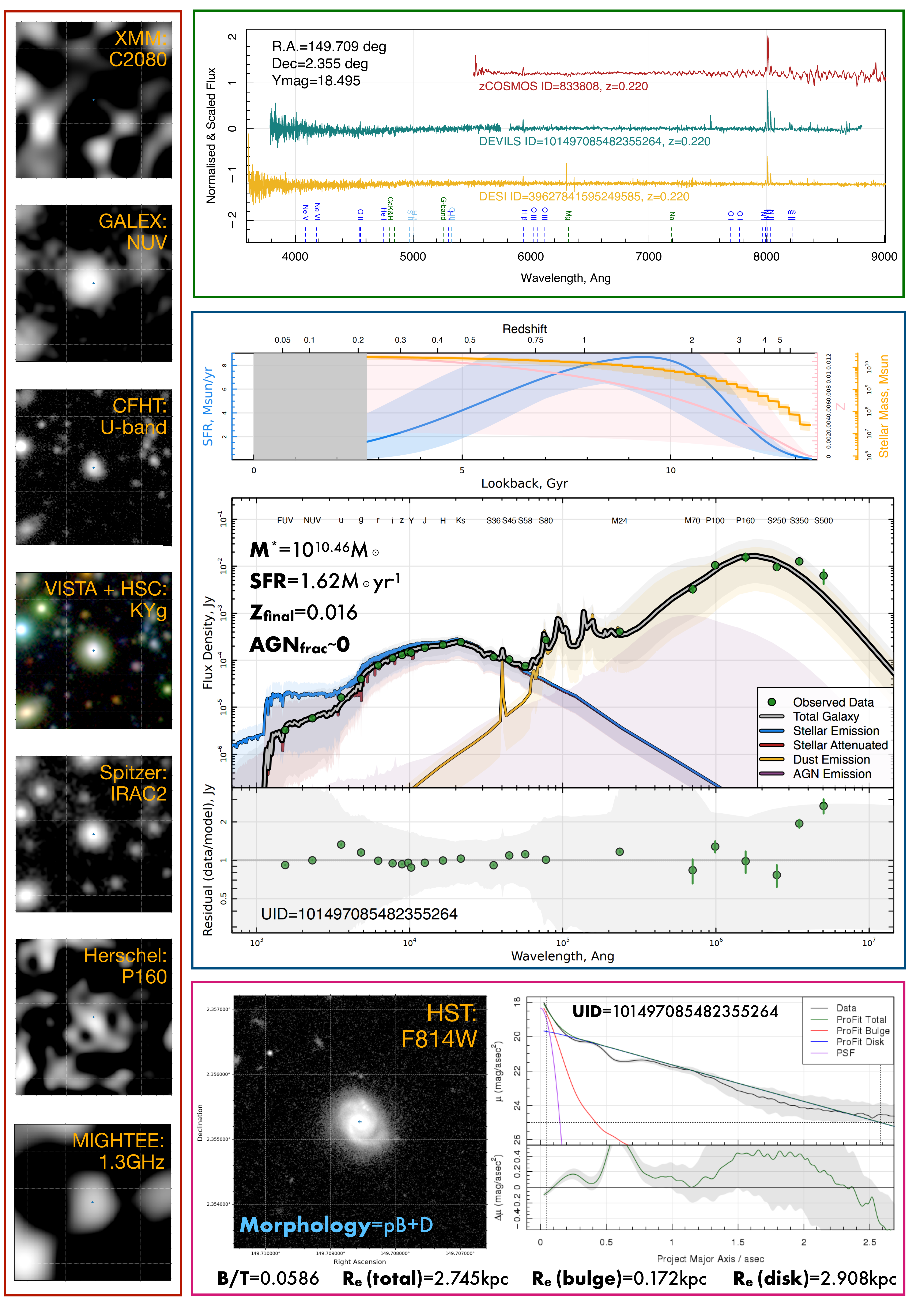}
\vspace{-2mm}
\caption{DEVILS DR1 data available for  UID=101497085482355264. Left column displays selected imaging produced using the \texttt{Data Central} cutout tool. Top: spectroscopic observations from zCOSMOS, DEVILS-AAT and DESI-EDR. Middle: Output from the SED fitting presented in \citet{Thorne22}.  In the bottom panel photometric data (including those from the left column) are shown in green. Photometry is fit with \textsc{ProSpect} for a stellar (blue line), dust (gold line) and AGN (purple line - in this case essentially zero) model. The stellar model is dust-attenuated to give the red line. Overall combined SED fit to the data is shown as the grey line, with residual shown in the bottom panel. Polygons show the upper and lower bounds of the MCMC chains produced in the fit. Resultant best-fit parameters for stellar mass, SFR, metallicity and AGN fraction are given. The top panel shows the best fit SFH (blue), stellar mass growth (orange) and metallicity evolution (pink). The x-axis here shows look-back time from today, such that the right-most edge of the grey-shaded region is the galaxy's observation epoch.  Bottom: HST morphological and structural parameters. The left panel shows the HST image for this galaxy. In the morphological analysis of \citet{Hashemizadeh21}, this galaxy is visually classified as a pseudo bulge + disk (pB+D). In the right panel we show the \textsc{ProFit} structural decomposition of this galaxy from \citet{Cook25}, showing the radial surface brightness profile of bulge and disk components as a function of angular major axis, below the plot we show key structural parameters of the light-wighted bulge-to-total ratio (B/T), physical total/disk/bulge effective radius (R$_{e}$) in kpc.}
\label{fig:summary}
\end{center}
\end{figure*}

\subsubsection{Photometric Measurements: \texttt{devils\_dr1.D10ProFoundPhotometry}}

Photometric measurements and associated flags etc for the D10 sample are described extensively in \cite{Davies21}. Briefly, this work takes photometric data across 28 bands from the UV to FIR and measures source photometry using \textsc{ProFound}. Initial base segments are defined using a deep Y+J+H+Ks stack from the UltraVISTA images. These segments are then used to measure photometry in UV through MIR imaging. In the FIR, PSF-based photometry is measured at the positions of 24$\mu m$-detected sources. Catalogues are then combined to provide multi-wavelength photometry for all sources detected in the NIR stack. This forms the base catalogue rows on which all other catalogues are generated. Following this, an object mask is defined excluding the regions around bright stars etc, and potential artefacts are flagged. Finally, sources are classified as either stars or galaxies based on their NIR colours, sizes and magnitudes. In the current data release we provide the full catalogues of these photometric measurements and flags as \texttt{devils\_dr1.D10ProFoundPhotometry}. Columns in this catalogue (which are not already described in Table \ref{tab:baseColumns}), are outlined in Table \ref{tab:D10ProFoundPhotometry}.            

\subsubsection{Redshifts: \texttt{devils\_dr1.D10MasterRedshifts}}
\label{redshifts}

Redshift measurements detailed extensively in this paper are provided in catalogue \texttt{devils\_dr1.D10MasterRedshifts}. This contains all redshift measurements discussed in Section \ref{sec:redshiftcomp}. For each survey we include, from the original catalogue: the team-internal ID, R.A., Dec, redshift, flag or metric associated with redshift quality, and the redshift type ($i.e.$ spec, grism, photo). In addition to the survey-specific redshift measurements,  we then also include \texttt{zBest}, the survey source of the \texttt{zBest} (\texttt{zBestSource}) and the type of \texttt{zBest} (\texttt{zBestType}), based on the process outlined in Section \ref{sec:redshiftcomp}. Columns in this catalogue (which are not already described in Table \ref{tab:baseColumns}), are outlined in Table \ref{tab:D10MasterRedshifts2}.

\subsubsection{SED Fitting: \texttt{devils\_dr1.D10ProSpectNoAGN} \& \texttt{devils\_dr1.D10ProSpectAGN}}
\label{sec:SEDs}

SED fitting in the DEVILS D10 region is described extensively in \cite{Thorne21} and \cite{Thorne22}. Briefly, \cite{Thorne21} use the \textsc{ProSpect} \citep{Robotham20} SED fitting code to estimate galaxy properties such as stellar mass, SFR and star-formation history (SFH). In \cite{Thorne22}, this processes is updated to include an Active Galactic Nuclei (AGN) model, which allows for the identification of sources hosting bright AGN and improvements to the other derived properties for AGN host galaxies. The \textsc{ProSpect} analysis of \cite{Thorne21, Thorne22} uses a parametric skewed-log-Normal truncated SFH for all galaxies, which assumes that all SFHs are smoothly evolving with look-back time. Note that between the publication of the Thorne et al works and the compilation of the redshift catalogues, 0.3\% of sources changed their redshift assignment. The majority of these come from the DESI EDR, which was not publicly available at the time of the Thorne et al work. For this data release, we have re-run these objects though \textsc{ProSpect} using an identical set up to the Thorne et al work and updated the catalogue parameters. In our DR1 data release we provide both catalogues containing SED fit parameters for fits with and without a possible AGN model (\texttt{devils\_dr1.D10ProSpectNoAGN} and \texttt{devils\_dr1.D10ProSpectAGN} respectively). Columns in this catalogue (which are not already described in Table \ref{tab:baseColumns}), are outlined in Tables \ref{tab:D10ProSpectNoAGN} and \ref{tab:D10ProSpectAGN}.

\subsubsection{Visual Morphology: \texttt{devils\_dr1.D10VisualMorphology}}

Visual morphological classifications are outlined in \cite{Hashemizadeh21} who use HST-ACS imaging from the COSMOS survey \citep{Scoville07} to visually classify $\sim$36,000 galaxies in the D10 region. Galaxies are selected for visual classification at log$_{10}$(M$_{\star}$/M$_{\odot}$)$>$9.5 and $z$$<$1.0. Outside of this range of stellar mass even HST resolution imaging does not have adequate spatial resolution to visually constrain galaxy morphology, and above this redshift the traditional picture of galaxy morphological classes begins to break down as more and more galaxies become clumpy and asymmetric (both caused by true structural evolution and observational effects from observations probing bluer rest-frame emission). Galaxies within the sample limits are then classified as either elliptical, pure disk, disk + diffuse/pseudo bulge (pDB), disk + classical bulge (cBD),  compact (too visually small to classify) and hard (asymmetric, merging, clumpy, extremely compact, and low-S/N systems) based on their F814W HST imaging. Galaxies were independently assessed by 4 classifiers, and objects with three or more votes in one category were adopted. Where classifiers disagreed, classes were debated until a consensus was obtained. Details of potential biases and systematics in this classification are described in \cite{Hashemizadeh21}. Columns in this catalogue (which are not already described in Table \ref{tab:baseColumns}), are outlined in Table \ref{tab:D10VisualMorphology}.

\subsubsection{Structural Fitting: \texttt{devils\_dr1.D10ProFitStructure}}

Galaxy structural parameters in the D10 region are outlined in \cite{Cook25}. Briefly, this work used the \textsc{ProFit} \citep{Robotham17} and \textsc{ProFuse} \citep{Robotham22} packages to decompose the F814W HST imaging for DEVILS galaxies into bulge and disk components and provide various S\'{e}rsic fitting parameters for each source such as S\'{e}rsic indices, sizes, bulge-to-total ratios and classifications based on the likelihood of the system being single- or two-component. Note that this analysis is once again only performed on M$_{\star}>$10$^{9.5}$M$_{\odot}$ and $z$$<$1.0 galaxies. Columns in this catalogue (which are not already described in Table \ref{tab:baseColumns}), are outlined in Table \ref{tab:D10ProFitStructure}.

\subsubsection{Group Environments: \texttt{devils\_dr1.D10GroupGals} \& \texttt{devils\_dr1.D10Groups}}

Finally, we also included catalogues detailing the group environmental metrics outlined in \cite{Bravo25}. Briefly, this work uses a similar approach to that outlined in \cite{Robotham11} based on a bespoke friends-of-friends grouping algorithm.  \cite{Bravo25} optimises friends-of-friends linking lengths for DEVILS specifically and tests the success and failures of group recovery using updated DEVILS mock galaxy light cones from the \textsc{shark} semi-analytic model \citep{Lagos24}. First, galaxies are linked to common halos and assigned a group ID and then group halo masses are measured from the galaxy kinematics, corrected to remove biases when compared to the mock galaxy light cones. Columns in this catalogue (which are not already described in Table \ref{tab:baseColumns}), are outlined in Table \ref{tab:D10GroupGals} for the properties of group galaxies ($i.e.$ group assignments, central/satellite classifications, group-centric locations) and Table \ref{tab:D10Groups} for the associated group properties ($i.e.$ sizes, velocity dispersions, halo masses, etc).

\subsection{Imaging}

In addition to the catalogue photometric measurements for D10 galaxies, we also release all UV-FIR imaging from which this photometry is derived. Details of the majority of these imaging datasets are described extensively in Table 1 of \cite{Davies21}, and as such we do not recover them again here. However, we note that all imaging is standardised to the same D10 field size and pixel matched to the 0.15\,arcsec/pixel VISTA-VIRCam UltraVISTA data. In addition to the imaging described in \cite{Davies21} in our DR1 data release we also include XMM-Newton x-ray imaging for both C0520 and C2080 bands taken from the XXM-COSMOS survey program presented in \cite{Cappelluti09}, the HST imaging taken from the COSMOS survey \citep{Scoville07} mosaic images (this is not pixel-matched to UltraVISTA to retain the HST resolution imaging) and the 1.3\,GHz radio continuum maps taken as part of the MeerKAT International Gigahertz Tiered Extragalactic Explorations \citep[MIGHTEE,][]{Jarvis16} and detailed in \cite{Heywood22}. Full characteristics and details of this imaging are provided with the DEVILS DR1 release documentation on \texttt{Data Central}. This imaging can be accessed via various tools on the \texttt{Data Central} archive such as a multi-band and RGB cutout service, cone search and single object viewer.

\subsection{Spectra}

Finally we also provide flux and wavelength calibrated 1D spectra associated with DEVILS sources in the D10 region. These consist of the AAT+AAOmega spectra discussed in this paper, and publicly-available zCOSMOS \citep{Lilly07} and DESI EDR \citep{DESI24} spectra where available. Spectra are tagged to a specific DEVILS UID, standardised to the same wavelength scale and converted to the same format with the same header keywords. Details of this format  and associated meta data are made available via \texttt{Data Central}. The spectra can also be dynamically visualised through the \texttt{Data Central} single object viewer and/or downloaded directly from the \texttt{Data Central} archive.         

\begin{figure*}
\begin{center}
\includegraphics[scale=0.32]{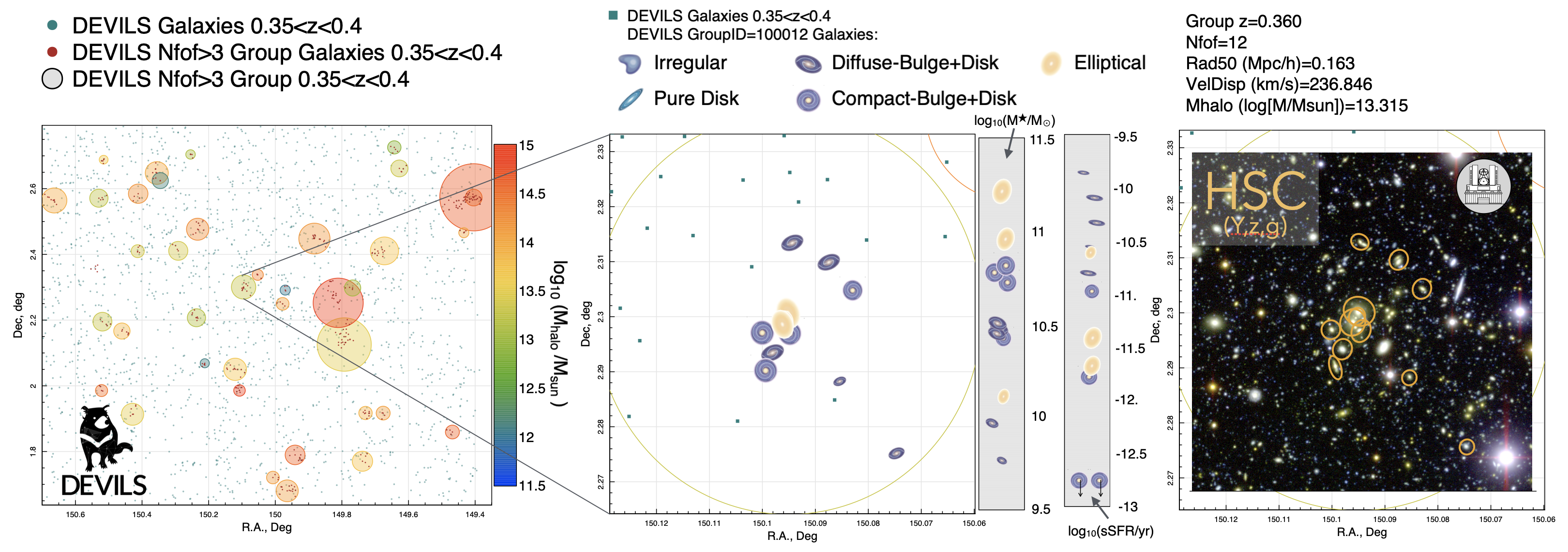}
\caption{Example of group diagnostics available in DEVILS DR1. Left shows the distribution of isolated galaxies (green points), group galaxies (red points) and dark matter haloes (coloured circles) at 0.35$<$$z$$<$0.4. Halos are coloured by halo mass, and their circle size represents the group physical extent. Middle shows a zoom in on group 100012. Group galaxies are now shown based on their morphological type, and grey bars show the stellar mass and SFR distribution of the group galaxies. Right shows the Subaru HSC Yzg image of the group, with group galaxies shown in orange. Key group measurements are given above the image. }
\label{fig:groups}
\end{center}
\end{figure*}

\vspace{2mm}

To highlight the breadth of data available for this sample, in Figure \ref{fig:summary} we show a summary overview of the data products available in the DEVILS DR1 data release for a single source (UID=101497085482355264). This comprises of examples of multi-wavelength imaging (left), 1D spectra from DEVILS-AAT, zCOSMOS and DESI-EDR (top), SED fits and associated parameters (middle), and morphological and structural characteristics (bottom). In Figure \ref{fig:groups}, we then show an example of group diagnostics for a group in the D10 region (GroupID=100012). Data of this quality and scope exists for a large fraction of sources in the DEVILS DR1 data release.

 \begin{figure*}
\begin{center}
\includegraphics[scale=0.6]{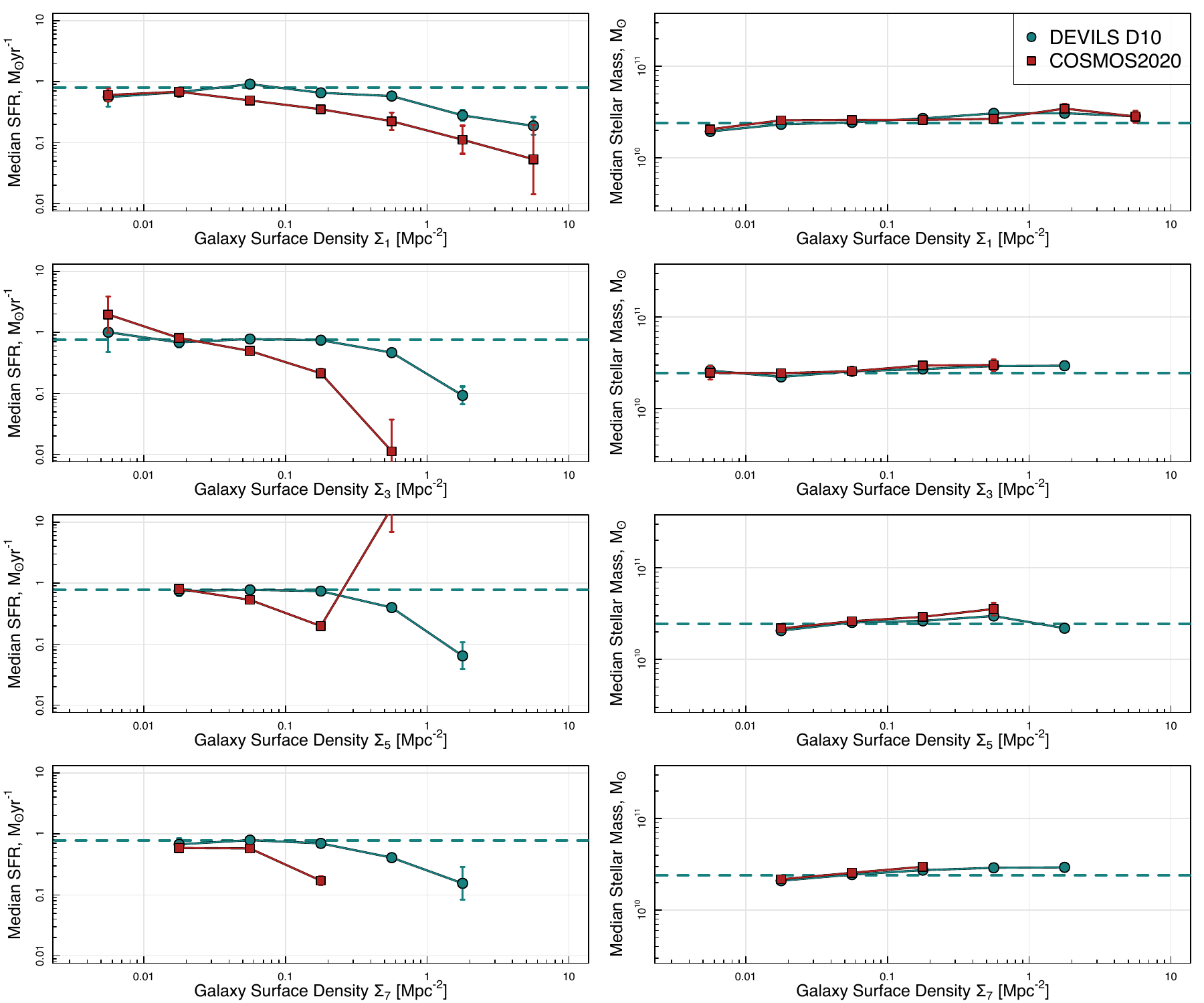}
\vspace{-2mm}
\caption{The median SFR (left column) and stellar mass (right column) of 10$<$log$_{10}$(M$_{\star}$/M$_{\odot}$)$<$11 as a function of nearest-neighbour density, $\Sigma_{N}$, for DEVILS galaxies at 0.2$<$$z$$<$0.5 (black) and COSMOS2020 galaxies with the same selection (red - see text for details). The dashed horizontal lines show the median value for all galaxies with $\Sigma_{N}<0.1$ ($i.e.$ isolated systems) in DEVILS. Error bars are calculated as the standard error on the median ($\sigma/\sqrt{\#}$, where $\#$ is the number of galaxies in the bin). Rows show nearest-neighbour density calculated for different $N^{th}$ nearest neighbours. Only bins containing $\geq$10 galaxies are shown. In DEVILS strong trends are observed of SFR decreasing with local density and weak trends of stellar mass increasing with local density. While these trends are also found in the COSMOS2020 catalogue, particularly at low $\Sigma_{N}$, photometric redshift samples have larger errors and show different level of suppression to the full spectroscopic DEVILS sample. }
\label{fig:Density}
\end{center}
\end{figure*}

\section{The star-formation - density relation at intermediate redshift}
\label{sec:density}

In the final parts of this paper, we highlight the power of the extensive data set available in the DEVILS DR1 data release by exploring the impact of local galaxy density on the star-formation in galaxies, split by both stellar mass and morphology. We choose this science case as it utilises the DR1 photometric catalogues (\texttt{devils\_dr1.D10ProFoundPhotometry}), redshift catalogue (\texttt{devils\_dr1.D10MasterRedshifts}), SED fitting catalogue (\texttt{devils\_dr1.D10ProSpectAGN}) and morphology catalogue (\texttt{devils\_dr1.D10VisualMorphology}). These results are simply designed to highlight the power of combining data from the DEVILS DR1 data release to produce new and interesting results, to supplement the existing body of work already published using the DEVILS DR1 data.       

Briefly,  galaxies can be broadly classified into two main categories: blue, gas-rich, star-forming systems, and red, gas-poor, quiescent systems with little or no ongoing star formation \citep[e.g.,][]{Blanton03, Kauffmann03a, Kauffmann04, Baldry04, Balogh04, Brinchmann04, Taylor15, Davies19b}. Throughout the history of the Universe, the relative abundance of galaxies in these two categories has evolved significantly. All galaxies are believed to begin as low-mass, blue, star-forming systems, gradually transitioning into quiescent systems over time \citep[e.g.,][]{Bell04, Faber07, Martin07}. This transformation is ongoing, with an increasing fraction of galaxies becoming quiescent as the Universe ages. Eventually, star formation is expected to cease entirely, leaving all galaxies in a passive state. One of the primary drivers of galaxy quenching is local environment. In dense environments—such as galaxy clusters, groups, or even close galaxy pairs \citep[see][]{Patton11, Robotham14, Davies16a}—the supply of cold gas needed for ongoing star formation can be heated, disrupted or removed, ultimately leading to quenching \citep[e.g.,][]{Peng10, Peng12, McGee14, Haines15, Schaefer17, Bluck21a, Bluck21b, Cortese21, Cleland21}. A variety of physical mechanisms can contribute to this process, including ram-pressure stripping and tidal interactions \citep{Gunn72, Moore99, Poggianti17, Brown17, Barsanti18}, strangulation (or starvation) \citep{Larson1980, Moore99, Peng15, Nichols11}, and galaxy harassment \citep{Moore96}. 

Many studies have found a strong correlation between suppression of star-formation and over-dense environments, witnessing environmental quenching in action \cite[$e.g.$][]{Wetzel13, Barsanti18, Oxland24, Davies19b, Davies25c}. However, most of these studies are limited to the local Universe, and as such open questions remain as to the ubiquity of environmental quenching at earlier times. Some works have studied the impact of environment on galaxies at earlier times. However, these works focus on using sparse spectroscopic redshifts and/or photometric redshift-dominated samples \citep[$e.g.$][]{Fossati17, Peng10} or target the most over-dense and extreme environments \citep[$e.g.$][]{Butcher84, Muzzin14, Haines15, Foltz18, vanderBurg20, Turner21, Mao22, Siudek22, Figueira24}.   In addition, it is also well-known that stellar mass, star formation and galaxy morphological types all correlate with environment \citep[][]{Bamford09, Alpaslan15,vanderWel10, Pfeffer23, Davies25a, Davies25b, Davies25d}. As such, to disentangle the impact of environment on the star-forming properties of galaxies alone, we must also control for the varying distribution of both stellar mass and morphology as a function of environment.  

  \begin{figure*}
\begin{center}
\includegraphics[scale=0.6]{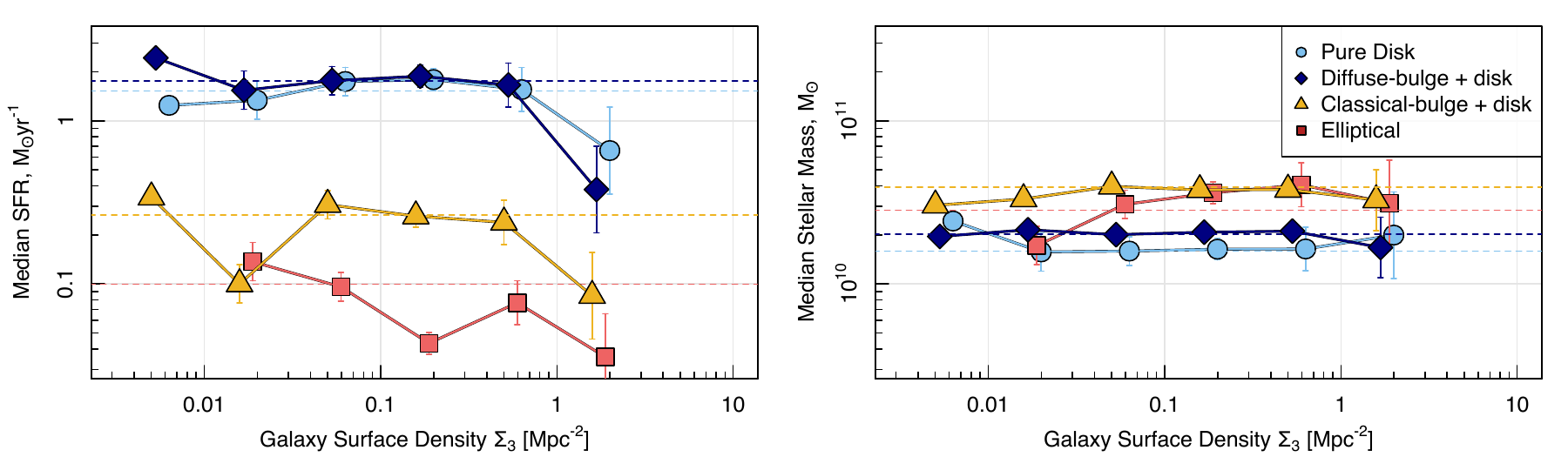}
\vspace{-2mm}
\caption{The same as Figure \ref{fig:Density}, but now split into individual morphological classes. We see suppression of star-formation in high density environments for all morphological classes. }
\label{fig:DensityMorph}
\end{center}
\end{figure*}

Here we utilise the DEVILS DR1 sample described in this paper to explore the impact of over-dense environments on star-formation at intermediate redshift (0.2$<$$z$$<$0.5) and in more typical environments. We then split this sample as a function of stellar mass and morphology, to determine if star-formation suppression in over-dense environments is a ubiquitous feature of the galaxy population, or limited to a specific stellar mass or morphological type. 

First, we take all DEVILS DR1 galaxies at intermediate redshift (0.2$<$$z$$<$0.5) and 10$<$log$_{10}$(M$_{\star}$/M$_{\odot}$)$<$11 ($i.e.$ above the 80\% spectroscopic completeness limits at all epochs in Figure \ref{fig:Mz}) and for a relatively narrow range of stellar masses. This allows us to study the suppression of star-formation in a controlled and largely stellar-mass complete sample.  Next, to define the local galaxy density for this sample, we take each galaxy in turn and only consider sources within a  radial velocity separation of $\pm1000$\,km\,s$^{-1}$ \citep[$e.g.$ as in][]{Baldry06}. We then calculate the projected local galaxy density as: 

\begin{equation}
\Sigma_{N}=\frac{N}{\pi d_{N}^{2}}, \mathrm{Mpc}^{-2}
\end{equation}

\noindent where $d_{N}$ is the projected distance to the Nth nearest neighbour. In our analysis we calculate this for the 1st ($i.e$ separation to closest neighbour), 3rd,  5th and 7th nearest neighbour ($\Sigma_{1}$, $\Sigma_{3}$, $\Sigma_{5}$, $\Sigma_{7}$). We initially explore this using a range of $\Sigma_{N}$ values, to show that the choice of $N$ does not significantly impact our results.   

We then bin the sample in $\Delta$log$_{10}$($\Sigma_{N}$)=0.5 bins, and measure the median SFR and stellar mass of galaxies in each bin. In Figure \ref{fig:Density} as black points/lines we show the median SFR as a function of $\Sigma_{N}$ in the left column and median stellar mass as a function of $\Sigma_{N}$ in the right column. Rows show the different values of $N$ used to calculate $\Sigma_{N}$. Error bars are calculated as the standard error on the median and only bins containing $>$10 galaxies are plotted. The black dashed horizontal line shows the median values for all isolated galaxies (defined as  $\Sigma_{N}$$<$0.1). We see that for all metrics of $\Sigma_{N}$, there is a decline in median SFR with increasing galaxy density. Median SFRs are relatively constant at $\sim0.8$M$_{\odot}$\,yr$^{-1}$ until the local galaxy density reaches above $\sim0.2$\,Mpc$^{-2}$, where we then see a decline of $\sim$1\,dex in median SFRs to the most over-dense regions. In the right column, we see that there is a weak trend of the median stellar mass increasing slight as we move to higher density environment. However, for the fixed stellar mass sample used here (10$<$log$_{10}$(M$_{\star}$/M$_{\odot}$)$<$11), this change is small, and not as significant as the change seen in the median SFR.  First this indicates that high-density environments are having an impact on star-formation in the overall galaxy population at 0.2$<$$z$$<$0.5, and that this suppression can be observed irrespective of the choice of $\Sigma_{N}$. This is consistent with the interpretation that over-dense environments lead to the suppression of star-formation via processes such as stripping, tidal interactions, starvation, etc.          

To highlight the power of our highly-complete spectroscopic sample in DEVILS when exploring the environmental impact on galaxy evolution, we repeat the process above using the publicly available COSMOS2020 catalogue in the COSMOS region \cite{Weaver22}. First we take all sources that are flagged as having \textcolor{black}{good} photometry as \texttt{FLAG\_COMBINED=0}. The COSMOS2020 catalogue provides \textsc{LePhare} best-fit photometric redshifts, stellar masses and SFRs \citep{Ilbert06}. We then repeat the process outlined in the previous paragraph selecting the same stellar mass and redshift range, and determining local galaxy densities in an identical manner. We note that we cannot easily supplement this with the spectroscopic redshift catalogue in COSMOS as stellar masses and SFRs are calculated assuming the photometric redshift is correct. In Figure \ref{fig:Density} we then also show median SFR and stellar mass as a function of local density for the COSMOS2020 sample as red points/lines. While we do see  evidence of SFR suppression in over-dense environments for $\Sigma_{1}$ and $\Sigma_{3}$ local density metrics, these have larger errors than seen in the DEVILS D10 sample, and are only very weakly observed at higher values of $\Sigma_{N}$. We also find that the observed strength of the impact of local over-density on star-formation in galaxies differs significantly between these samples, where the photometric redshift samples of the COSMOS2020 catalogue over-estimate the decline in median SFR at higher over-densities in comparison to DEVILS. This could be due to differences in the SED fitting approaches. However, the fact that the weak trends between stellar mass and local environmental density are very similar between the two samples, and the median SFRs at low density environments are very similar, potentially suggests that some of this difference is driven by the use of photometric redshifts over spectroscopic sample to define local density. \textcolor{black}{We also note that using the metric for local galaxy density defined above may bias any results derived from photometric redshifts alone. Errors on photometric redshifts are typically $>$1000\,km\,s$^{-1}$, and thus excluding galaxies above $\pm1000$\,km\,s$^{-1}$ in our local galaxy density calculation, may miss true galaxy nearest neighbours with large photometric redshift errors. This will typically lead to local galaxy densities being underestimated. However, if we relax this constraint to $\pm5000$\,km\,s$^{-1}$ for the COSMOS2020 sample, the results do not significantly change. In combination,} this indicates that less-biased and highly complete spectroscopic samples are essential in determining the environmental impact on galaxy evolution.                

Following this, we can now also utilise the \texttt{devils\_dr1.D10VisualMorphology} catalogue to split this sample based on morphological type (which cannot be undertaken for the COSMOS2020 sample). In Figure \ref{fig:DensityMorph} we display the median SFR and stellar mass as a function of $\Sigma_{3}$, now split by individual visual morphological types. We note that other $\Sigma_{N}$ values show similar trends. Here we see that the decline in median SFR in higher density environments remains for all morphological types. This suggests that the suppression in star-formation in over-dense environments is not limited to a specific galaxy type, or is driven by the varying contributions of different morphological types as a function of environment \citep[$e.g.$][]{Davies25d}.  

However, we do see that potentially different morphological classes have different trends with galaxy surface density.  Elliptical and disk + classical bulge morphological types shows a gradual decline in median SFR with increasing over-density while pure disk and  disk + diffuse/pseudo bulge galaxies only show a decline in median SFR in the most over-dense bin.  We do also see a weak trend of stellar masses increasing towards over-dense environments for the elliptical population, which could potentially drive some of this result (as higher stellar mass galaxies have lower SFRs). However, it does appear that the suppression of star-formation in high density environments is not limited to a specific morphological type and is seen in all classes.   

\begin{figure*}
\begin{center}
\includegraphics[scale=0.6]{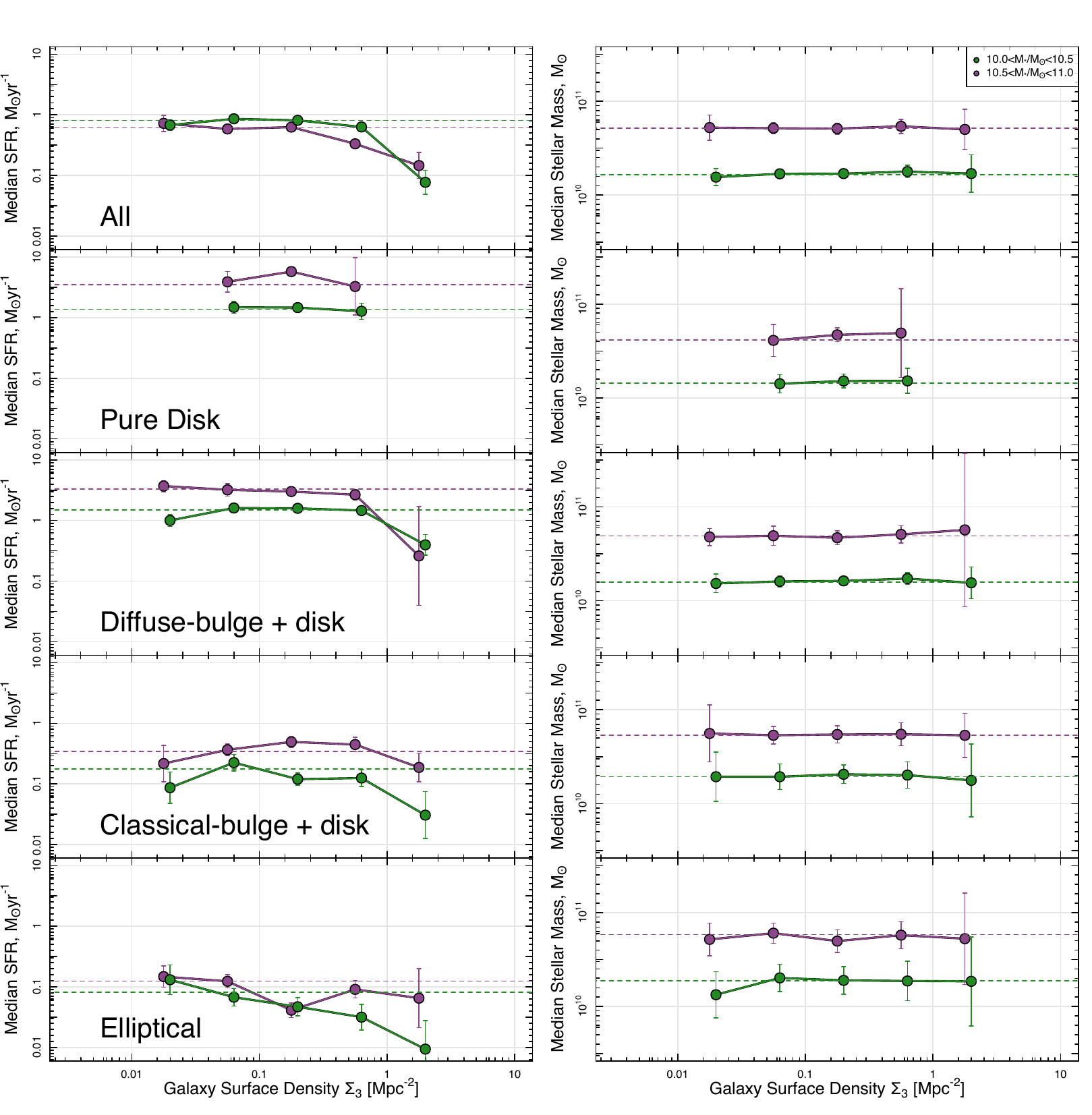}
\vspace{-2mm}
\caption{Similar to Figure \ref{fig:Density}, but now split into two different stellar mass ranges and showing all morphological types separately. Given the smaller number of galaxies in this sub-selection, we relax our criteria to show all bins containing $\geq$5 galaxies. We find typically higher level of suppression in lower stellar mass galaxies, even when controlled for morphological type.  }
\label{fig:DensityMass}
\end{center}
\end{figure*}

Finally, to explore how changes in the stellar mass distribution as a function of local over-density may impact these trends, in Figure \ref{fig:DensityMass}, we then split our sample further into two stellar mass ranges, and show panels split by morphological type. Here we see that in both stellar mass ranges the full sample of all morphologies (top row) displays a suppression of star-formation. We now also see no trend in stellar mass as a function of local galaxy density - indicating this is a true environmental effect on star-formation. We also find that this suppression of star-formation in over-dense environments is also observed when split by both morphological type and stellar mass, $i.e.$ there is a decline in median star formation in higher density environments at all stellar mass and morphologies. However, for all morphological classes, we do find that lower stellar mass galaxies appear more impacted by their environment - showing a greater degree of suppression. 

Overall, these result indicate that the suppression of star-formation (quenching) in over-dense environments is a ubiquitous feature of the Universe at 0.2$<$$z$$<$0.5) and 10$<$log$_{10}$(M$_{\star}$/M$_{\odot}$)$<$11, occurring at a range of stellar masses and in all morphological types. We also find that lower stellar mass galaxies show more suppression than higher stellar mass galaxies for all morphological types. This is likely expected as the processes which lead to quenching in satellite galaxies (ram-pressure stripping, tidal interactions, etc) are likely dependant on satellite stellar mass, where in a fixed mass halo, lower stellar mass galaxies are more likely to impacted by these processes.

\section{Conclusions}

In this work we present the first data release of the Deep Extragalactic VIsible Legacy Survey, covering all data products in the D10 (COSMOS) region. We discuss the spectroscopic DEVILS observations at the AAT and completion of the main redshift catalogues used in DEVILS science papers - providing 4,859 new unique spectroscopic redshifts for faint sources in the COSMOS region. We then outline details of all data products which are made available including catalogues of measure/derived properties, imaging and 1D spectra. All data is made available through \texttt{Data Central} (\url{datacentral.org.au}). Finally we highlight a scientific application of this extensive data set, exploring the suppression of star-formation in over-dense environments at intermediate redshift (0.2$<$$z$$<$0.5) when split by both stellar mass and visual morphology.    

\section*{Acknowledgements}

LJMD acknowledges support from the Australian Research Councils Future Fellowship and Discovery Project schemes (FT200100055 and DP250104611). ASGR, and SB acknowledge support from the Australian Research Council's Future Fellowship scheme (FT200100375). MB is funded by McMaster University through the William and Caroline Herschel Fellowship. M.S. acknowledges support by the State Research Agency of the Spanish Ministry of Science and Innovation under the grants 'Galaxy Evolution with Artificial Intelligence' (PGC2018-100852-A-I00) and 'BASALT' (PID2021-126838NB-I00) and the Polish National Agency for Academic Exchange (Bekker grant BPN/BEK/2021/1/00298/DEC/1). This work was partially supported by the European Union's Horizon 2020 Research and Innovation program under the Maria Sklodowska-Curie grant agreement (No. 754510).  Parts of this research were conducted by the Australian Research Council Centre of Excellence for All Sky Astrophysics in 3 Dimensions (ASTRO 3D), through project number CE170100013. DEVILS is an Australian project based around a spectroscopic campaign using the Anglo-Australian Telescope. DEVILS is part funded via Discovery Programs by the Australian Research Council and the participating institutions. The DEVILS website is \url{devils.research.org.au}. This paper includes data that has been provided by the AAO \texttt{Data Central} Science Platform (\url{datacentral.org.au}).

\section{Data Availability}

All data products used in this paper are made available through the Data Central (\url{datacentral.org.au}) archive.

\appendix

\section{DR1 table descriptions} 

In this appendix we detail the columns provided in the DR1 data release catalogues.

\begin{table*} 
\caption{\textcolor{black}{Additional c}olumns in the \texttt{devils\_dr1.D10ProFoundPhotometry} catalogue (which are not in Table \ref{tab:baseColumns}). \textbf{Note:} all photometry measurements are provided for \textit{total} flux measurements (no suffix) and colour-optimised flux measurements (suffix *\_c) - see \citet{Davies21} for details.   } 
\begin{center}\begin{scriptsize}
\begin{tabular}{l l l c l }
Name  &  Description & UCD & Unit & Type \\ 
\hline
NewSegID & Segment ID derived from segmentation maps - unique over full DEVILS region & meta.id &  & integer \\ 
field & DEVILS Field identifier (D02/D03/D10) & meta.id &  & character \\ 
R50\_Y & The segment radius contain 50\% of the VISTA Y-band flux & phys.angSize;em.IR.Y & arcsec & double \\ 
N100\_Y & Number of segment pixels in the VISTA Y-band & phys.area &  & integer \\ 
subFrame & ProFound processing subFrame within the field that the segment is taken from & meta.id &  & integer \\ 
segIDFrame & Segment ID within subFrame (non-unique across DEVILS region) & meta.id &  & integer \\ 
xcenFrame & x position of the centre of the segment within the subFrame in the VISTA Y-band image & instr.pixel & pixel & double \\ 
ycenFrame & y position of the centre of the segment within the subFrame in the VISTA Y-band image & instr.pixel & pixel & double \\ 
xmaxFrame & x position of the brightest pixel of the segment within the subFrame in the VISTA Y-band image & instr.pixel & pixel & double \\ 
ymaxFrame & y position of the brightest pixel of the segment within the subFrame in the VISTA Y-band image & instr.pixel & pixel & double \\ 
EBV & Galactic Extinction correction coefficient applied to photometry - derived from Planck 2018 maps & phys.excitParam &  & double \\ 
flux\_FUV & Total flux measured in GALEX FUV band & phot.flux.density;em.UV.100-200nm & Jy & double \\ 
flux\_err\_FUV & Error on total flux measured in GALEX FUV band & stat.error;phot.flux.density;em.UV.100-200nm & Jy & double \\ 
flux\_NUV & Total flux measured in GALEX NUV band & phot.flux.density;em.UV.200-300nm & Jy & double \\ 
flux\_err\_NUV & Error on total flux measured in GALEX NUV band & stat.error;phot.flux.density;em.UV.200-300nm & Jy & double \\ 
flux\_u & Total flux measured in CFHT-Megaprime u band & phot.flux.density;em.opt.u & Jy & double \\ 
flux\_err\_u & Error on total flux measured in CFHT-Megaprime u band & stat.error;phot.flux.density;em.opt.u & Jy & double \\ 
flux\_g & Total flux measured in Subaru-HSC g band  & phot.flux.density;em.opt.g & Jy & double \\ 
flux\_err\_g & Error on total flux measured in Subaru-HSC g band & stat.error;phot.flux.density;em.opt.g & Jy & double \\ 
flux\_r & Total flux measured in Subaru-HSC r band & phot.flux.density;em.opt.r & Jy & double \\ 
flux\_err\_r & Error on total flux measured in Subaru-HSC r band  & stat.error;phot.flux.density;em.opt.r & Jy & double \\ 
flux\_i & Total flux measured in Subaru-HSC i band & phot.flux.density;em.opt.i & Jy & double \\ 
flux\_err\_i & Error on total flux measured in Subaru-HSC i band & stat.error;phot.flux.density;em.opt.i & Jy & double \\ 
flux\_z & Total flux measured in Subaru-HSC z band & phot.flux.density;em.opt.z & Jy & double \\ 
flux\_err\_z & Error on total flux measured in Subaru-HSC z band & stat.error;phot.flux.density;em.opt.z & Jy & double \\ 
flux\_HY & Total flux measured in Subaru-HSC Y band & phot.flux.density;em.IR.Y & Jy & double \\ 
flux\_err\_HY & Error on total flux measured in Subaru-HSC Y band & stat.error;phot.flux.density;em.IR.Y & Jy & double \\ 
flux\_Y & Total flux measured in VISTA-VIRCAM Y band & phot.flux.density;em.IR.Y & Jy & double \\ 
flux\_err\_Y & Error on total flux measured in VISTA-VIRCAM Y band & stat.error;phot.flux.density;em.IR.Y & Jy & double \\ 
flux\_J & Total flux measured in VISTA-VIRCAM J band & phot.flux.density;em.IR.J & Jy & double \\ 
flux\_err\_J & Error on total flux measured in VISTA-VIRCAM J band & stat.error;phot.flux.density;em.IR.J & Jy & double \\ 
flux\_H & Total flux measured in VISTA-VIRCAM H band & phot.flux.density;em.IR.H & Jy & double \\ 
flux\_err\_H & Error on total flux measured in VISTA-VIRCAM H band & stat.error;phot.flux.density;em.IR.H & Jy & double \\ 
flux\_Ks & Total flux measured in VISTA-VIRCAM Ks band & phot.flux.density;em.IR.Ks & Jy & double \\ 
flux\_err\_Ks & Error on total flux measured in VISTA-VIRCAM Ks band & stat.error;phot.flux.density;em.IR.Ks & Jy & double \\ 
flux\_S36 & Total flux measured in Spitzer-IRAC 3.6micron band & phot.flux.density;em.IR.3-4um & Jy & double \\ 
flux\_err\_S36 & Error on total flux measured in Spitzer-IRAC 3.6micron band & stat.error;phot.flux.density;em.IR.3-4um & Jy & double \\ 
flux\_S45 & Total flux measured in Spitzer-IRAC 4.5micron band & phot.flux.density;em.IR.4-8um & Jy & double \\ 
flux\_err\_S45 & Error on total flux measured in Spitzer-IRAC 4.5micron band & stat.error;phot.flux.density;em.IR.4-8um & Jy & double \\ 
flux\_S58 & Total flux measured in Spitzer-IRAC 5.8micron band & phot.flux.density;em.IR.4-8um & Jy & double \\ 
flux\_err\_S58 & Error on total flux measured in Spitzer-IRAC 5.8micron band & stat.error;phot.flux.density;em.IR.4-8um & Jy & double \\ 
flux\_S80 & Total flux measured in Spitzer-IRAC 8.0micron band & phot.flux.density;em.IR.4-8um & Jy & double \\ 
flux\_err\_S80 & Error on total flux measured in Spitzer-IRAC 8.0micron band & stat.error;phot.flux.density;em.IR.4-8um & Jy & double \\ 
flux\_M24 & Total flux measured in Spitzer-MIPS 24micron band & phot.flux.density;em.IR.15-30um & Jy & double \\ 
flux\_err\_M24 & Error on total flux measured in Spitzer-MIPS 24micron band & stat.error;phot.flux.density;em.IR.15-30um & Jy & double \\ 
flux\_M70 & Total flux measured in Spitzer-MIPS 70micron band & phot.flux.density;em.IR.60-100um & Jy & double \\ 
flux\_err\_M70 & Error on total flux measured in Spitzer-MIPS 70micron band & stat.error;phot.flux.density;em.IR.60-100um & Jy & double \\ 
flux\_P100 & Total flux measured in Herschel-PACS 100micron band & phot.flux.density;em.IR.FIR & Jy & double \\ 
flux\_err\_P100 & Error on total flux measured in Herschel-PACS 100micron band & stat.error;phot.flux.density;em.IR.FIR & Jy & double \\ 
flux\_P160 & Total flux measured in Herschel-PACS 160micron band & phot.flux.density;em.IR.FIR & Jy & double \\ 
flux\_err\_P160 & Error on total flux measured in Herschel-PACS 160micron band & stat.error;phot.flux.density;em.IR.FIR & Jy & double \\ 
flux\_S250 & Total flux measured in Herschel-SPIRE 250micron band & phot.flux.density;em.IR.FIR & Jy & double \\ 
flux\_err\_S250 & Error on total flux measured in Herschel-SPIRE 250micron band & stat.error;phot.flux.density;em.IR.FIR & Jy & double \\ 
flux\_S350 & Total flux measured in Herschel-SPIRE 350micron band & phot.flux.density;em.IR.FIR & Jy & double \\ 
flux\_err\_S350 & Error on total flux measured in Herschel-SPIRE 350micron band & stat.error;phot.flux.density;em.IR.FIR & Jy & double \\ 
flux\_S500 & Total flux measured in Herschel-SPIRE 500micron band & phot.flux.density;em.IR.FIR & Jy & double \\ 
flux\_err\_S500 & Error on total flux measured in Herschel-SPIRE 500micron band & stat.error;phot.flux.density;em.IR.FIR & Jy & double \\ 
\end{tabular} 
\end{scriptsize} 
\end{center} 
\label{tab:D10ProFoundPhotometry} 
\end{table*}

\begin{table*} 
\caption{\textcolor{black}{Additional c}olumns in the \texttt{devils\_dr1.D10MasterRedshifts} catalogue (which are not in Table \ref{tab:baseColumns}). Continued in Table \ref{tab:D10MasterRedshifts2} } 
\begin{center} 
\begin{scriptsize} 
\begin{tabular}{l l l c l } 
Name  &  Description & UCD & Unit & Type \\ 
\hline 
DEVILS\_ID & DEVILS spectroscopic ID taken from DEVILS-internal DMCat (DMCat2022-01-20) & meta.id &  & character \\ 
DEVILS\_RA & DEVILS spectroscopic RA taken from DEVILS-internal DMCat (DMCat2022-01-20) & pos.eq.ra & deg & double \\ 
DEVILS\_DEC & DEVILS spectroscopic DEC taken from DEVILS-internal DMCat (DMCat2022-01-20) & pos.eq.dec & deg & double \\ 
DEVILS\_z & DEVILS current best spectroscopic redshift taken from DEVILS-internal DMCat (DMCat2022-01-20) & src.redshift &  & double \\ 
DEVILS\_prob & DEVILS current best spectroscopic redshift probability (0-1) taken from DMCat (DMCat2022-01-20) & stat.probability;src.redshift &  & double \\ 
DEVILS\_zPlus & DEVILS spectroscopic redshifts weighted by matches existing redshift information & src.redshift &  & double \\ 
DEVILS\_probPlus & DEVILS weighted probability. DEVILS\_prob +1 (if consistent with photo-z), +2 (if consistent with grism redshift), + 4 (if  & stat.probability;src.redshift &  & double \\ 
& consistent with spec redshift). Numbers are adaptive, such that 7.xxx means redshift is consistent with existing photo-z, grism-z and spec-z &  &  &  \\ 
DEVILS\_type & DEVILS redshift type  & meta.id &  & character \\ 
zCOSMOS\_ID & zCOSMOS spectroscopic ID taken from zCOSMOS Bright DR3 catalogue (zcosbrightspec20k\_dr3\_catalog\_1595990782.csv - \cite{Lilly07}) & meta.id &  & character \\ 
zCOSMOS\_RA & zCOSMOS spectroscopic RA taken from zCOSMOS Bright DR3 catalogue (zcosbrightspec20k\_dr3\_catalog\_1595990782.csv) & pos.eq.ra & deg & double \\ 
zCOSMOS\_DEC & zCOSMOS spectroscopic DEC taken from zCOSMOS Bright DR3 catalogue (zcosbrightspec20k\_dr3\_catalog\_1595990782.csv) & pos.eq.dec & deg & double \\ 
zCOSMOS\_z & zCOSMOS spectroscopic redshift taken from zCOSMOS Bright DR3 catalogue (zcosbrightspec20k\_dr3\_catalog\_1595990782.csv) & src.redshift &  & double \\ 
zCOSMOS\_prob & zCOSMOS spectroscopic confidence taken from zCOSMOS Bright DR3 catalogue (zcosbrightspec20k\_dr3\_catalog\_1595990782.csv). & meta.id &  & double \\
&  4=very secure, 3=secure, 2=probable, 1=insecure, 9=single line, 18=single line BL AGN.... &  &  &  \\ 
zCOSMOS\_type & zCOSMOS redshift type & meta.id &  & character \\ 
DESI\_ID & DESI spectroscopic ID taken from DESI EDR catalogue (zall-pix-edr-vac.fits - \cite{DESI24}) & meta.id &  & character \\ 
DESI\_RA & DESI spectroscopic RA taken from DESI EDR catalogue (zall-pix-edr-vac.fits) & pos.eq.ra & deg & double \\ 
DESI\_DEC & DESI spectroscopic DEC taken from DESI EDR catalogue (zall-pix-edr-vac.fits) & pos.eq.dec & deg & double \\ 
DESI\_z & DESI spectroscopic redshift taken from DESI EDR catalogue (zall-pix-edr-vac.fits) & src.redshift &  & double \\ 
DESI\_prob & DESI spectroscopic confidence taken from DELTACHI2 in the DESI EDR catalogue (zall-pix-edr-vac.fits).Values >9 are classed as secure. & meta.id &  & double \\ 
DESI\_type & DESI redshift type & meta.id &  & character \\ 
hCOSMOS\_ID & hCOSMOS spectroscopic ID generated for hCOSMOS catalogue (a sequence using row order from hcosmosNew.cat.txt - \cite{Damjanov18}) & meta.id &  & character \\ 
hCOSMOS\_RA & hCOSMOS spectroscopic RA taken from hCOSMOS catalogue (hcosmosNew.cat.txt) & pos.eq.ra & deg & double \\ 
hCOSMOS\_DEC & hCOSMOS spectroscopic DEC taken from hCOSMOS catalogue (hcosmosNew.cat.txt) & pos.eq.dec & deg & double \\ 
hCOSMOS\_z & hCOSMOS spectroscopic redshift taken from hCOSMOS catalogue (hcosmosNew.cat.txt) & src.redshift &  & double \\ 
hCOSMOS\_prob & hCOSMOS spectroscopic error taken from hCOSMOS catalogue (hcosmosNew.cat.txt). Actual $\Delta z$. & stat.error;src.redshift &  & double \\ 
hCOSMOS\_type & hCOSMOS redshift type & meta.id &  & character \\ 
DEIMOS\_ID & DEIMOS spectroscopic ID taken from DEIMOS catalogue COSMOS\_DEIMOS.csv - \cite{Hasinger18}) & meta.id &  & character \\ 
DEIMOS\_RA & DEIMOS spectroscopic RA taken from DEIMOS catalogue (COSMOS\_DEIMOS.csv) & pos.eq.ra & deg & double \\ 
DEIMOS\_DEC & DEIMOS spectroscopic DEC taken from DEIMOS catalogue (COSMOS\_DEIMOS.csv) & pos.eq.dec & deg & double \\ 
DEIMOS\_z & DEIMOS spectroscopic redshift taken from DEIMOS catalogue (COSMOS\_DEIMOS.csv) & src.redshift &  & double \\ 
DEIMOS\_prob & DEIMOS spectroscopic quality flag taken from DEIMOS catalogue (COSMOS\_DEIMOS.csv).  & meta.id &  & double \\ 
&  2=Reliable, 1.5=uncertain but agrees with photo-z, 1=uncertain&  &  &  \\ 
DEIMOS\_type & DEIMOS redshift type & meta.id &  & character \\ 
ComparatOII\_ID & ComparatOII spectroscopic ID generated from ComparatOII catalogue (Comparat\_catalog.dat - \cite{Comparat15}) & meta.id &  & character \\ 
ComparatOII\_RA & ComparatOII spectroscopic RA taken from ComparatOII catalogue (Comparat\_catalog.dat) & pos.eq.ra & deg & double \\ 
ComparatOII\_DEC & ComparatOII spectroscopic DEC taken from ComparatOII catalogue (Comparat\_catalog.dat) & pos.eq.dec & deg & double \\ 
ComparatOII\_z & ComparatOII OII-derived spectroscopic Redshift taken from ComparatOII catalogue (Comparat\_catalog.dat) & src.redshift &  & double \\ 
ComparatOII\_prob & **NOTE** ComparatOII catalogue does not contain quality flags, added here to match data structure of other surveys  & meta.id &  & double \\ 
ComparatOII\_type & ComparatOII redshift type & meta.id &  & character \\ 
FMOS\_ID & FMOS spectroscopic ID taken from FMOS catalogue FMOS\_COSMOS\_v1.0.fits - \cite{Silverman15}) & meta.id &  & character \\ 
FMOS\_RA & FMOS spectroscopic RA taken from FMOS catalogue (FMOS\_COSMOS\_v1.0.fit) & pos.eq.ra & deg & double \\ 
FMOS\_DEC & FMOS spectroscopic DEC taken from FMOS catalogue (FMOS\_COSMOS\_v1.0.fit) & pos.eq.dec & deg & double \\ 
FMOS\_z & FMOS spectroscopic redshift taken from FMOS catalogue (FMOS\_COSMOS\_v1.0.fit) & src.redshift &  & double \\ 
FMOS\_prob & FMOS spectroscopic quality flag taken from FMOS catalogue (FMOS\_COSMOS\_v1.0.fit). & meta.id &  & double \\ 
&  4=Very secure, 3=secure, 2=probable, 1=not secure, 0=nothing &  &  &  \\ 
FMOS\_type & FMOS redshift type & meta.id &  & character \\ 
MOSDEF\_ID & MOSDEF spectroscopic ID (ID\_V4) taken from MOSDEF catalogue mosdef\_zcat.final.fits - \cite{Kriek15}) & meta.id &  & character \\ 
MOSDEF\_RA & MOSDEF spectroscopic RA taken from MOSDEF catalogue (mosdef\_zcat.final.fits) & pos.eq.ra & deg & double \\ 
MOSDEF\_DEC & MOSDEF spectroscopic DEC taken from MOSDEF catalogue (mosdef\_zcat.final.fits) & pos.eq.dec & deg & double \\ 
MOSDEF\_z & MOSDEF spectroscopic redshift taken from MOSDEF catalogue (mosdef\_zcat.final.fits) & src.redshift &  & double \\ 
MOSDEF\_prob & MOSDEF spectroscopic quality flag taken from MOSDEF catalogue (mosdef\_zcat.final.fits) & meta.id &  & double \\ 
MOSDEF\_type & MOSDEF redshift type & meta.id &  & character \\ 
C3R2\_ID & C3R2 spectroscopic ID taken from C3R2 catalogue c3r2\_DR1+DR2\_2019april11.txt - \cite{Masters19}) & meta.id &  & character \\ 
C3R2\_RA & C3R2 spectroscopic RA taken from C3R2 catalogue (c3r2\_DR1+DR2\_2019april11.txt) & pos.eq.ra & deg & double \\ 
C3R2\_DEC & C3R2 spectroscopic DEC taken from C3R2 catalogue (c3r2\_DR1+DR2\_2019april11.txt) & pos.eq.dec & deg & double \\ 
C3R2\_z & C3R2 spectroscopic redshift taken from C3R2 catalogue (c3r2\_DR1+DR2\_2019april11.txt) & src.redshift &  & double \\ 
C3R2\_prob & C3R2 spectroscopic quality flag taken from C3R2 catalogue (c3r2\_DR1+DR2\_2019april11.txt). & meta.id &  & double \\ 
&  4=certain, 3.5=very high confidence, 3=high confidence  &  &  &  \\ 
C3R2\_type & C3R2 redshift type & meta.id &  & character \\ 
LRIS\_ID & LRIS spectroscopic ID taken from LRIS catalogue cl2017\_valueadded\_release\_20171009.txt - \cite{Lee18}) & meta.id &  & character \\ 
LRIS\_RA & LRIS spectroscopic RA taken from LRIS catalogue (cl2017\_valueadded\_release\_20171009.txt) & pos.eq.ra & deg & double \\ 
LRIS\_DEC & LRIS spectroscopic DEC taken from LRIS catalogue (cl2017\_valueadded\_release\_20171009.txt) & pos.eq.dec & deg & double \\ 
LRIS\_z & LRIS spectroscopic redshift taken from LRIS catalogue (cl2017\_valueadded\_release\_20171009.txt) & src.redshift &  & double \\ 
LRIS\_prob & LRIS spectroscopic quality flag taken from LRIS catalogue (cl2017\_valueadded\_release\_20171009.txt).  & meta.id &  & double \\ 
&  4=high confidence, 3=reasonable confidence, 2=low-confidence, 1=guess &  &  &  \\ 
LRIS\_type & LRIS redshift type & meta.id &  & character \\ 
VIS3COSMOS\_ID & VIS3COSMOS spectroscopic ID taken from VIS3COSMOS catalogue VIS3COSMOS\_tableb1.dat - \cite{Paulino-Afonso18}) & meta.id &  & character \\ 
VIS3COSMOS\_RA & VIS3COSMOS spectroscopic RA taken from VIS3COSMOS catalogue (VIS3COSMOS\_tableb1.dat) & pos.eq.ra & deg & double \\ 
VIS3COSMOS\_DEC & VIS3COSMOS spectroscopic DEC taken from VIS3COSMOS catalogue (VIS3COSMOS\_tableb1.dat) & pos.eq.dec & deg & double \\ 
VIS3COSMOS\_z & VIS3COSMOS spectroscopic redshift taken from VIS3COSMOS catalogue (VIS3COSMOS\_tableb1.dat) & src.redshift &  & double \\ 
VIS3COSMOS\_prob & VIS3COSMOS catalogue (VIS3COSMOS\_tableb1.dat) has no quality flag & meta.id &  & double \\ 
VIS3COSMOS\_type & VIS3COSMOS redshift type & meta.id &  & character \\ 
LEGA\_C\_ID & LEGA\_C spectroscopic ID taken from LEGA\_C catalogue ADP.2018-06-13T13:17:01.234.fits - \cite{vanderWel16}) & meta.id &  & character \\ 
LEGA\_C\_RA & LEGA\_C spectroscopic RA taken from LEGA\_C catalogue (ADP.2018-06-13T13:17:01.234.fits) & pos.eq.ra & deg & double \\ 
LEGA\_C\_DEC & LEGA\_C spectroscopic DEC taken from LEGA\_C catalogue (ADP.2018-06-13T13:17:01.234.fits) & pos.eq.dec & deg & double \\ 
LEGA\_C\_z & LEGA\_C spectroscopic redshift taken from LEGA\_C catalogue (ADP.2018-06-13T13:17:01.234.fits) & src.redshift &  & double \\ 
LEGA\_C\_prob & LEGA\_C catalogue redshift quality flag (ADP.2018-06-13T13:17:01.234.fits). 0=good redshift, 1=bad redshift & meta.id &  & double \\ 
LEGA\_C\_type & LEGA\_C redshift type & meta.id &  & character \\ 
\end{tabular} 
\end{scriptsize} 
\end{center} 
\label{tab:D10MasterRedshifts} 
\end{table*}

\begin{table*} 
\caption{Table \ref{tab:D10MasterRedshifts} continued... \textcolor{black}{Additional c}olumns in the \texttt{devils\_dr1.D10MasterRedshifts} catalogue (which are not in Table \ref{tab:baseColumns}). } 
\begin{center} 
\begin{scriptsize} 
\begin{tabular}{l l l c l } 
Name  &  Description & UCD & Unit & Type \\ 
\hline 
VVDS\_ID & VVDS spectroscopic ID taken from VVDS catalogue VVDS\_COSMOS.csv - \cite{LeFevre13}) & meta.id &  & character \\ 
VVDS\_RA & VVDS spectroscopic RA taken from VVDS catalogue (VVDS\_COSMOS.csv) & pos.eq.ra & deg & double \\ 
VVDS\_DEC & VVDS spectroscopic DEC taken from VVDS catalogue (VVDS\_COSMOS.csv) & pos.eq.dec & deg & double \\ 
VVDS\_z & VVDS spectroscopic redshift taken from VVDS catalogue (VVDS\_COSMOS.csv) & src.redshift &  & double \\ 
VVDS\_prob & VVDS spectroscopic quality flag taken from VVDS catalogue (VVDS\_COSMOS.csv). & meta.id &  & double \\ 
&  4=high confidence, 3=reasonable confidence, 2=low-confidence, 1=guess  &  &  &  \\ 
VVDS\_type & VVDS redshift type & meta.id &  & character \\ 
VUDS\_ID & VUDS spectroscopic ID taken from OzDES master catalogue OzDES\_GRC\_2017\_08\_30.csv -  \cite{LeFevre15}) & meta.id &  & character \\ 
VUDS\_RA & VUDS spectroscopic RA taken from VUDS catalogue (OzDES\_GRC\_2017\_08\_30.csv) & pos.eq.ra & deg & double \\ 
VUDS\_DEC & VUDS spectroscopic DEC taken from VUDS catalogue (OzDES\_GRC\_2017\_08\_30.csv) & pos.eq.dec & deg & double \\ 
VUDS\_z & VUDS spectroscopic redshift taken from VUDS catalogue (OzDES\_GRC\_2017\_08\_30.csv) & src.redshift &  & double \\ 
VUDS\_prob & VUDS spectroscopic quality flag taken from VUDS catalogue (OzDES\_GRC\_2017\_08\_30.csv). reliable = 1.5,2,3,4,9 & meta.id &  & double \\ 
VUDS\_type & VUDS redshift type & meta.id &  & character \\ 
PRIMUS\_ID & PRIMUS grism ID taken from PRIMUS catalogue PRIMUS\_2013\_zcat\_v1.fits - \cite{Cool13}) & meta.id &  & character \\ 
PRIMUS\_RA & PRIMUS grism RA taken from PRIMUS catalogue (PRIMUS\_2013\_zcat\_v1.fits) & pos.eq.ra & deg & double \\ 
PRIMUS\_DEC & PRIMUS grism DEC taken from PRIMUS catalogue (PRIMUS\_2013\_zcat\_v1.fits) & pos.eq.dec & deg & double \\ 
PRIMUS\_z & PRIMUS grism redshift taken from PRIMUS catalogue (PRIMUS\_2013\_zcat\_v1.fits) & src.redshift &  & double \\ 
PRIMUS\_prob & PRIMUS grism quality flag taken from PRIMUS catalogue (PRIMUS\_2013\_zcat\_v1.fits).  & meta.id &  & double \\ 
&  44=high quality (dz/(1+z)~0.003), 3=good quality dz/(1+z)~0.015), 2=probable dz/(1+z)~0.03, -1=star  &  &  &  \\ 
PRIMUS\_type & PRIMUS redshift type & meta.id &  & character \\ 
HST3D\_ID & HST3D grism ID taken from HST3D catalogue cosmos\_3dhst.v4.1.5.zbest.fits - \cite{Momcheva16} - only use\_zgrism=1 & meta.id &  & character \\ 
HST3D\_RA & HST3D grism RA taken from HST3D catalogue (cosmos\_3dhst.v4.1.5.zbest.fits) & pos.eq.ra & deg & double \\ 
HST3D\_DEC & HST3D grism DEC taken from HST3D catalogue (cosmos\_3dhst.v4.1.5.zbest.fits) & pos.eq.dec & deg & double \\ 
HST3D\_z & HST3D grism redshift taken from HST3D catalogue (cosmos\_3dhst.v4.1.5.zbest.fits) & src.redshift &  & double \\ 
HST3D\_l68 & HST3D grism 1sigma lower confidence boundary taken from HST3D catalogue (cosmos\_3dhst.v4.1.5.zbest.fits). & stat.error;src.redshift &  & double \\ 
HST3D\_u68 & HST3D grism 1sigma upper confidence boundary taken from HST3D catalogue (cosmos\_3dhst.v4.1.5.zbest.fits). & stat.error;src.redshift &  & double \\ 
HST3D\_type & HST3D redshift type & meta.id &  & character \\ 
PAU\_ID & PAU PhotoZ ID taken from PAU catalogue pau\_cosmos\_photoz\_v0\_4\_c.fits - \cite{Alarcon21}. & meta.id &  & character \\ 
PAU\_RA & PAU PhotoZ RA taken from PAU catalogue (pau\_cosmos\_photoz\_v0\_4\_c.fits) & pos.eq.ra & deg & double \\ 
PAU\_DEC & PAU PhotoZ DEC taken from PAU catalogue (pau\_cosmos\_photoz\_v0\_4\_c.fits) & pos.eq.dec & deg & double \\ 
PAU\_z & PAU PhotoZ redshift taken from PAU catalogue (pau\_cosmos\_photoz\_v0\_4\_c.fits) & src.redshift &  & double \\ 
PAU\_l68 & **NOTE** PAU PhotoZ catalogue has no error bound. Added here to match structure of other surveys & stat.error;src.redshift &  & double \\ 
PAU\_u68 & **NOTE** PAU PhotoZ catalogue has no error bound. Added here to match structure of other surveys & stat.error;src.redshift &  & double \\ 
PAU\_type & PAU redshift type & meta.id &  & character \\ 
COSMOS2015\_ID & COSMOS2015 PhotoZ ID (ID2013) taken from COSMOS2015 catalogue COSMOS2015\_Laigle+\_v1.1.fits - \cite{Laigle16} & meta.id &  & character \\ 
COSMOS2015\_RA & COSMOS2015 PhotoZ RA taken from COSMOS2015 catalogue (COSMOS2015\_Laigle+\_v1.1.fits) & pos.eq.ra & deg & double \\ 
COSMOS2015\_DEC & COSMOS2015 PhotoZ DEC taken from COSMOS2015 catalogue (COSMOS2015\_Laigle+\_v1.1.fits) & pos.eq.dec & deg & double \\ 
COSMOS2015\_z & COSMOS2015 PhotoZ (ZPDF) redshift taken from COSMOS2015 catalogue (COSMOS2015\_Laigle+\_v1.1.fits) & src.redshift &  & double \\ 
COSMOS2015\_l68 & COSMOS2015 PhotoZ 1sigma lower confidence boundary taken from COSMOS2015 catalogue (COSMOS2015\_Laigle+\_v1.1.fits). & stat.error;src.redshift &  & double \\ 
COSMOS2015\_u68 & COSMOS2015 PhotoZ 1sigma upper confidence boundary taken from COSMOS2015 catalogue (COSMOS2015\_Laigle+\_v1.1.fits). & stat.error;src.redshift &  & double \\ 
COSMOS2015\_type & COSMOS2015 redshift type & meta.id &  & character \\ 
MIGHTEEtmp\_ID & MIGHTEEtmp PhotoZ ID taken from MIGHTEEtmp catalogue COS\_Master\_hatfield\_GPz\_V1.fits - \cite{Adams20}  & meta.id &  & character \\ 
MIGHTEEtmp\_RA & MIGHTEEtmp PhotoZ RA taken from MIGHTEEtmp catalogue (COS\_Master\_hatfield\_GPz\_V1.fits) & pos.eq.ra & deg & double \\ 
MIGHTEEtmp\_DEC & MIGHTEEtmp PhotoZ DEC taken from MIGHTEEtmp catalogue (COS\_Master\_hatfield\_GPz\_V1.fits) & pos.eq.dec & deg & double \\ 
MIGHTEEtmp\_z & MIGHTEEtmp PhotoZ (Z\_GPz\_all) redshift taken from MIGHTEEtmp catalogue (COS\_Master\_hatfield\_GPz\_V1.fits) & src.redshift &  & double \\ 
MIGHTEEtmp\_l68 & **NOTE** No error provided for MIGHTEEtmp PhotoZ. Added here to match structure of other surveys (COS\_Master\_hatfield\_GPz\_V1.fits). & stat.error;src.redshift &  & double \\ 
MIGHTEEtmp\_u68 & **NOTE** No error provided for MIGHTEEtmp PhotoZ. Added here to match structure of other surveys(COS\_Master\_hatfield\_GPz\_V1.fits). & stat.error;src.redshift &  & double \\ 
MIGHTEEtmp\_type & MIGHTEEtmp redshift type & meta.id &  & character \\ 
zBest & Best Redshift for source from all available redshifts (see text for details) & src.redshift &  & double \\ 
zBestSource & Survey from which best redshift comes, identifier in column names & meta.id &  & character \\ 
zBestType & Type of redshift: spec, grism, photoz & meta.id &  & character \\ 
\end{tabular} 
\end{scriptsize} 
\end{center} 
\label{tab:D10MasterRedshifts2} 
\end{table*} 

\begin{table*} 
\caption{\textcolor{black}{Additional c}olumns in the \texttt{devils\_dr1.D10ProSpectNoAGN} catalogue (which are not in Table \ref{tab:baseColumns}). \textbf{Note:} for all parameters in this table the catalogue also provides the $1\sigma$ upper and lower bounds of the MCMC chains with suffix *\_UB and *\_LB respectively.  }  
\begin{center} 
\begin{scriptsize} 
\begin{tabular}{l l l c l } 
Name  &  Description & UCD & Unit & Type \\ 
\hline 
z & Redshift used in \textsc{ProSpect} SED fitting (can differ from \texttt{devils\_dr1.D10MasterRedshifts}) & src.redshift &  & double \\ 
mSFR & \textsc{ProSpect} parameter that controls the maximum star formation rate in the skewed normal - logged & stat.param & log(solMass/yr) & double \\ 
mpeak & \textsc{ProSpect} parameter that controls where the peak of star formation occurs & stat.param & Gyr & double \\ 
mperiod & \textsc{ProSpect} parameter that controls the width of the star formation skewed Normal - logged & stat.param & log(Gyr) & double \\ 
mskew & \textsc{ProSpect} Parameter that controls the skew of the skewed Normal & stat.param &  & double \\ 
tau\_birth & \textsc{ProSpect} parameter to describe the column density of dust in the birth clouds - logged & stat.param &  & double \\ 
tau\_screen & \textsc{ProSpect} parameter to describe the column density of dust in the ISM- logged & stat.param &  & double \\ 
alpha\_SF\_birth & \textsc{ProSpect} parameter to describe the temperature of the dust in the birth clouds & stat.param &  & double \\ 
alpha\_SF\_screen & \textsc{ProSpect} parameter to describe the temperature of the dust in the ISM & stat.param &  & double \\ 
Zfinal & Final gas-phase metallicity, modelled in \textsc{ProSpect} - logged & stat.param;phys.abund.Z &  & double \\ 
LP & log-posterior of best fit model in \textsc{ProSpect} & stat.likelihood &  & double \\ 
StellarMass & Remaining galaxy stellar mass (assuming returned fraction), generally this should be the value to use for stellar masses & phys.mass & solMass & double \\ 
StellarMassFormed & Total stellar mass formed, not remaining ($i.e.$ assuming no returned fraction) & phys.mass & solMass & double \\ 
dustmass.birth & Dust mass in birth clouds & phys.mass & solMass & double \\ 
dustmass.screen & Dust mass in ISM & phys.mass & solMass & double \\ 
dustmass.total & Dust mass total & phys.mass & solMass & double \\ 
dustlum.birth & Dust luminosity in birth clouds & phys.luminosity & solLum & double \\ 
dustlum.screen & Dust luminosity in ISM & phys.luminosity & solLum & double \\ 
dustlum.total & Dust luminosity total & phys.luminosity & solLum & double \\ 
SFRburst & SFR measured over 100Myr baseline, generally this should be the value uses for SFRs & phys.SFR  & solMass/yr & double \\ 
flux.galex\_fuv & best fit model flux in galex\_fuv & phot.flux.density & Jy & double \\ 
flux.galex\_nuv & best fit modelflux in galex\_nuv & phot.flux.density & Jy & double \\ 
flux.cfht\_u & best fit model flux in cfht\_u & phot.flux.density & Jy & double \\ 
flux.hsc\_g & best fit model flux in hsc\_g & phot.flux.density & Jy & double \\ 
flux.hsc\_r & best fit model flux in hsc\_r & phot.flux.density & Jy & double \\ 
flux.hsc\_i & best fit model flux in hsc\_i & phot.flux.density & Jy & double \\ 
flux.hsc\_z & best fit model flux in hsc\_z & phot.flux.density & Jy & double \\ 
flux.vista\_Y & best fit model flux in vista\_Y & phot.flux.density & Jy & double \\ 
flux.vista\_J & best fit model flux in vista\_J & phot.flux.density & Jy & double \\ 
flux.vista\_H & best fit model flux in vista\_H & phot.flux.density & Jy & double \\ 
flux.vista\_Ks & best fit model flux in vista\_Ks & phot.flux.density & Jy & double \\ 
flux.irac\_1 & best fit model flux in irac\_1 & phot.flux.density & Jy & double \\ 
flux.irac\_2 & best fit model flux in irac\_2 & phot.flux.density & Jy & double \\ 
flux.irac\_3 & best fit model flux in irac\_3 & phot.flux.density & Jy & double \\ 
flux.irac\_4 & best fit model flux in irac\_4 & phot.flux.density & Jy & double \\ 
flux.mips\_24 & best fit model flux in mips\_24 & phot.flux.density & Jy & double \\ 
flux.mips\_70 & best fit model flux in mips\_70 & phot.flux.density & Jy & double \\ 
flux.pacs\_100 & best fit model flux in pacs\_100 & phot.flux.density & Jy & double \\ 
flux.pacs\_160 & best fit model flux in pacs\_160 & phot.flux.density & Jy & double \\ 
flux.spire\_250 & best fit model flux in spire\_250 & phot.flux.density & Jy & double \\ 
flux.spire\_350 & best fit model flux in spire\_350 & phot.flux.density & Jy & double \\ 
flux.spire\_500 & best fit model flux in spire\_500 & phot.flux.density & Jy & double \\ 
\end{tabular} 
\end{scriptsize} 
\end{center} 
\label{tab:D10ProSpectNoAGN} 
\end{table*}

\begin{table*} 
\caption{\textcolor{black}{Additional c}dditional columns in the \texttt{devils\_dr1.D10ProSpectAGN} catalogue (which are not in Table \ref{tab:baseColumns} and are not already listed in Table \ref{tab:D10ProSpectNoAGN}). \textbf{Note:} for all parameters in this table the catalogue also provides the $1\sigma$ upper and lower bounds of the MCMC chains with suffix *\_UB and *\_LB respectively.  }  
\begin{center} 
\begin{scriptsize} 
\begin{tabular}{l l l c l } 
Name  &  Description & UCD & Unit & Type \\ 
\hline 
AGNlum & Bolometric luminosity of AGN source  -  logged & phys.luminosity.bol;stat.min & log(erg/s) & double \\ 
AGNan & Angle of observation of AGN source & phys.ang;stat.min & deg & double \\ 
AGNta & Optical depth tau of AGN source  -  logged & phys.absorption.opticalDepth;stat.min &  & double \\ 
LP & Log posterior of best fit model in \textsc{ProSpect} & stat.likelihood;stat.min &  & double \\ 
AGNfrac & fraction of flux contributed by the AGN component between 5-20micron & phys.luminosity.bol;stat.weight &  & double \\ 
Type & AGN type 1 or 2 & meta.code.class &  & integer \\ 
Iteration & Iteration for convergence & meta.number &  & integer \\ 
flux\_3GHz & predicted flux at 3GHz from \textsc{ProSpect} fit & em.radio.3-6GHz;stat.value & Jy & double \\ 
HalfMassAge & The age taken to form half the galaxy's mass based on the best fit SFH & phys.age & Gyr & double \\ 
\end{tabular} 
\end{scriptsize} 
\end{center} 
\label{tab:D10ProSpectAGN} 
\end{table*}

\begin{table*} 
\caption{\textcolor{black}{Additional c}olumns in the \texttt{devils\_dr1.D10ProFitStructure} catalogue (which are not in Table \ref{tab:baseColumns}).} 
\begin{center} 
\begin{scriptsize} 
\begin{tabular}{l l l c l } 
Name  &  Description & UCD & Unit & Type \\ 
\hline 
vis\_morph & Visual morphology class from \texttt{devils\_dr1.D10VisualMorphology} catalogue & src.morph.type &  & character \\ 
best\_model & The preferred model based on the outputs of the model fit: twocomps, puredisk, spheroid, failed \citep[see][]{Cook25} & meta.id &  & character \\ 
class & The preferred model based on the visual inspections and adjusted accordingly based on the best-fitting model parameters:  & meta.id &  & character \\ 
&  twocomps, puredisk, spheroid, failed \citep[see][]{Cook25}  &  &  &  \\ 
magBulge & Final Class: Bulge Profile integrated HST F814W magnitude & phot.mag;em.opt.I & mag & double \\ 
magDisk & Final Class: Disk Profile integrated HST F814W magnitude & phot.mag;em.opt.I & mag & double \\ 
reBulge\_pix & Final Class: Bulge Observed effective radius (R50) in pixels & phys.size.radius;em.opt.I & pixel & double \\ 
reDisk\_pix & Final Class: Disk Observed effective radius (R50) in pixels & phys.size.radius;em.opt.I & pixel & double \\ 
reTotal\_pix & Final Class: Total (disk+bulge) Observed effective radius (R50) in pixels & phys.size.radius;em.opt.I & pixel & double \\ 
reBulge\_kpc & Final Class: Bulge physical effective radius (R50) in kpc & phys.size.radius;em.opt.I & kpc & double \\ 
reDisk\_kpc & Final Class: Disk physical effective radius (R50) in kpc & phys.size.radius;em.opt.I & kpc & double \\ 
reTotal\_kpc & Final Class: Total (disk+bulge) Observed effective radius (R50) in kpc & phys.size.radius;em.opt.I & kpc & double \\ 
nserBulge & Final Class: Bulge Sersic Index & src.morph.type;em.opt.I &  & double \\ 
nserDisk & Final Class: Disk Sersic Index & src.morph.type;em.opt.I &  & double \\ 
axratBulge & Final Class: Bulge Axial Ratio & phys.size.axisRatio;em.opt.I &  & double \\ 
axratDisk & Final Class: Disk Axial Ratio & phys.size.axisRatio;em.opt.I &  & double \\ 
angBulge & Final Class: Bulge Position angle (increasing clockwise 0 deg from North) & pos.posAng;em.opt.I & deg & double \\ 
angDisk & Final Class: Disk Position angle (increasing clockwise 0 deg from North) & pos.posAng;em.opt.I & deg & double \\ 
B2T\_corr & Final Class: Light-weighted Bulge-to-total ratio & src.morph.type &  & double \\ 
xcen\_single & Single component fit: Central x pixel position & pos.centroid & pixel & double \\ 
ycen\_single & Single component fit: Central y pixel position & pos.centroid & pixel & double \\ 
mag\_single & Single component fit: Profile integrated HST F814W magnitude & phot.mag;em.opt.I & mag & double \\ 
re\_single\_kpc & Single component fit: Effective radius (R50) in kpc & phys.size.radius;em.opt.I & kpc & double \\ 
nser\_single & Single component fit: Sersic Index & src.morph.type;em.opt.I &  & double \\ 
ang\_single & Single component fit: Position angle (increasing clockwise 0 deg from North) & pos.posAng;em.opt.I & deg & double \\ 
axrat\_single & Single component fit: Axial Ratio & phys.size.axisRatio;em.opt.I &  & double \\ 
xcen\_err\_single & Single component fit: Central x pixel position Error & stat.error;pos.centroid & pixel & double \\ 
ycen\_err\_single & Single component fit: Central y pixel position Error & stat.error;pos.centroid & pixel & double \\ 
mag\_err\_single & Single component fit: Profile integrated HST F814W magnitude Error & stat.error;phot.mag;em.opt.I & mag & double \\ 
re\_err\_single & Single component fit: Effective radius (R50) in kpc Error & stat.error;phys.size.radius;em.opt.I & kpc & double \\ 
nser\_err\_single & Single component fit: Sersic Index Error & stat.error;src.morph.type;em.opt.I &  & double \\ 
ang\_err\_single & Single component fit: Position angle (increasing clockwise 0 deg from North) Error & stat.error;pos.posAng;em.opt.I & deg & double \\ 
axrat\_err\_single & Single component fit: Axial Ratio Error & stat.error;phys.size.axisRatio;em.opt.I &  & double \\ 
LP\_single & Single component fit: LP for best fit & stat.fit.goodness &  & double \\ 
DIC\_single & Single component fit: deviance information criterion for best fit & stat.fit.goodness &  & double \\ 
time\_single & Single component fit: time to run ProFuse fit & time.duration & sec & double \\ 
xcen\_serexp & Sersic Bulge + Exponential Disk fit: Central x pixel position & pos.centroid & pixel & double \\ 
ycen\_serexp & Sersic Bulge + Exponential Disk fit: Central y pixel position & pos.centroid & pixel & double \\ 
mag1\_serexp & Sersic Bulge + Exponential Disk fit: Comp1 Profile integrated HST F814W magnitude & phot.mag;em.opt.I & mag & double \\ 
mag2\_serexp & Sersic Bulge + Exponential Disk fit: Comp2 Profile integrated HST F814W magnitude & phot.mag;em.opt.I & mag & double \\ 
re1\_serexp\_kpc & Sersic Bulge + Exponential Disk fit: Comp1 Effective radius (R50) in kpc & phys.size.radius;em.opt.I & kpc & double \\ 
re2\_serexp\_kpc & Sersic Bulge + Exponential Disk fit: Comp2 effective radius (R50) in kpc & phys.size.radius;em.opt.I & kpc & double \\ 
nser1\_serexp & Sersic Bulge + Exponential Disk fit: Comp1 Sersic Index & src.morph.type;em.opt.I &  & double \\ 
ang2\_serexp & Sersic Bulge + Exponential Disk fit: Comp2 Position angle (increasing clockwise 0 deg from North) & pos.posAng;em.opt.I & deg & double \\ 
axrat2\_serexp & Sersic Bulge + Exponential Disk fit: Comp2 Axial Ratio & phys.size.axisRatio;em.opt.I &  & double \\ 
xcen\_err\_serexp & Sersic Bulge + Exponential Disk fit: Central x pixel position Error & stat.error;pos.centroid & pixel & double \\ 
ycen\_err\_serexp & Sersic Bulge + Exponential Disk fit: Central y pixel position Error & stat.error;pos.centroid & pixel & double \\ 
mag1\_err\_serexp & Sersic Bulge + Exponential Disk fit: Comp1 Profile integrated HST F814W magnitude Error & stat.error;phot.mag;em.opt.I & mag & double \\ 
mag2\_err\_serexp & Sersic Bulge + Exponential Disk fit: Comp2 Profile integrated HST F814W magnitude Error & stat.error;phot.mag;em.opt.I & mag & double \\ 
re1\_err\_serexp & Sersic Comp1 + Exponential Disk fit: Comp1 Effective radius (R50) in kpc Error & stat.error;phys.size.radius;em.opt.I & kpc & double \\ 
re2\_err\_serexp & Sersic Bulge + Exponential Disk fit: Comp2 effective radius (R50) in kpc Error & stat.error;phys.size.radius;em.opt.I & kpc & double \\ 
nser1\_err\_serexp & Sersic Bulge + Exponential Disk fit: Comp1 Sersic Index Error & stat.error;src.morph.type;em.opt.I &  & double \\ 
ang2\_err\_serexp & Sersic Bulge + Exponential Disk fit: Comp2 Position angle (increasing clockwise 0 deg from North) Error & stat.error;pos.posAng;em.opt.I & deg & double \\ 
axrat2\_err\_serexp & Sersic Bulge + Exponential Disk fit: Comp2 Axial Ratio Error & stat.error;phys.size.axisRatio;em.opt.I &  & double \\ 
LP\_serexp & Sersic Bulge + Exponential Disk fit: LP for best fit & stat.fit.goodness &  & double \\ 
DIC\_serexp & Sersic Bulge + Exponential Disk fit: deviance information criterion for best fit & stat.fit.goodness &  & double \\ 
time\_serexp & Sersic Bulge + Exponential Disk fit: time to run ProFuse fit & src.morph.type & sec & integer \\ 
B2T\_serexp & Sersic Bulge + Exponential Disk fit: Light-weighted Comp1-to-Comp1+2 ratio & time.duration &  & integer \\ 
nexps & Number of HST exposures included in fit & stat.value &  & double \\ 
npixels & Number of HST pixels used in the profile fitting & instr.pixel &  & double \\ 
\end{tabular} 
\end{scriptsize} 
\end{center} 
\label{tab:D10ProFitStructure} 
\end{table*}

\begin{table*} 
\caption{\textcolor{black}{Additional c}olumns in the \texttt{devils\_dr1.D10VisualMorphology} catalogue (which are not in Table \ref{tab:baseColumns}). Note: some columns in this catalogue as replicated from other catalogues, and are superseded by the other base catalogue values. They are retained here as they were used in the selection of the visual morphological sample.  } 
\begin{center} 
\begin{scriptsize} 
\begin{tabular}{l l l c l } 
Name  &  Description & UCD & Unit & Type \\ 
\hline 
ProSpect\_Mstar & \textsc{ProSpect} stellar mass taken from \cite{Thorne21} - used for sample selection & phys.mass.stars & log(solMass) & double \\ 
ProSpect\_Mstar\_LB & \textsc{ProSpect} stellar mass lower bound of final MCMC chain taken from \cite{Thorne21}  - used for sample selection  & phys.mass.stars;stat.min & log(solMass) & double \\ 
ProSpect\_Mstar\_UB & \textsc{ProSpect} stellar mass upper bound of final MCMC chain taken from \cite{Thorne21} - used for sample selection  & phys.mass.stars;stat.max & log(solMass) & double \\ 
ProSpect\_SFR & \textsc{ProSpect} burst SFR taken from \cite{Thorne21} - used for sample selection  & phys.sfr & log(solMass/yr) & double \\ 
ProSpect\_SFR\_LB & \textsc{ProSpect} burst SFR lower bound of final MCMC chain taken from \cite{Thorne21} - used for sample selection  & phys.sfr;stat.min & log(solMass/yr) & double \\ 
ProSpect\_SFR\_UB & \textsc{ProSpect} burst SFR upper bound of final MCMC chain taken from \cite{Thorne21} - used for sample selection  & phys.sfr;stat.max & log(solMass/yr) & double \\ 
DEVILS\_ID & DEVILS spectroscopic ID from \texttt{devils\_dr1.D10MasterRedshifts} catalogue & meta.id &  & character \\ 
Laigle\_logMstar & Stellar mass measurement from the COSMOS2015 catalogue  - used for sample selection & phys.mass.stars & log10(solMass) & double \\ 
G10\_zBest & Best redshift from the \cite{Davies15} G10/COSMOS catalogue & src.redshift &  & double \\ 
SERSIC\_n\_SINGLE & Sersic index from initial \textsc{ProFit} single component Sersic fit to HST F814W band & phys.size.sersic &  & double \\ 
AUTO\_CLASS & Automatic morphology class from Galaxy Zoo and ANN combination & meta.id &  & character \\ 
FIRST\_CLASS & First rough classification & meta.id &  & character \\ 
CHECKER & Second pass name of visual checker & meta.id &  & character \\ 
SPD\_CLASS & Simon Driver visual classification & meta.id &  & character \\ 
SB\_CLASS & Sabine Bellstedt visual classification & meta.id &  & character \\ 
LJMD\_CLASS & Luke Davies visual classification & meta.id &  & character \\ 
FINAL\_CLASS & Final visual classification based on combination of all classifiers & meta.id &  & character \\ 
HSTreg & Is galaxy within the useable HST-COSMOS region? 1=yes, 0=no & meta.id &  & integer \\ 
sampSel & Is galaxy within the DEVILS morphology sample selection  1=yes, 0=no & meta.id &  & integer \\ 
& (ProSpect\_Mstar>10$^{9.5}$ \& zBest<1.0 \& RAcen>=149.5)? &  &  &  \\ 
\end{tabular} 
\end{scriptsize} 
\end{center} 
\label{tab:D10VisualMorphology} 
\end{table*}

\begin{table*} 
\caption{\textcolor{black}{Additional c}olumns in the \texttt{devils\_dr1.D10GroupGals} catalogue (which are not in Table \ref{tab:baseColumns}).} 
\begin{center} 
\begin{scriptsize} 
\begin{tabular}{l l l c l } 
Name  &  Description & UCD & Unit & Type \\ 
\hline 
apmag\_Y & Apparent Y-band magnitude (AB). & phot.mag; em.IR.Y & mag & double \\ 
DM\_737 & Distance modulus assuming a 737 cosmology ($h$$_0$ = 0.7, $\Omega$$_m$ = 0.3, and $\Omega$$_{\Lambda}$ = 0.7). & phot.mag.distMod & mag & double \\ 
SigErr & Velocity uncertainty on redshift, assumed to be 45\,km\,s$^{-1}$ for all spectroscopic redshifts. & phys.veloc & km\,s$^{-1}$ & double \\ 
GroupID & ID of the group the galaxy is member (value of 0 for isolated galaxies). & meta.id; meta.main &  & integer \\ 
RankIterCen & Relative ranking of galaxy by projected distance to the iteratively-calculated group centre. 1=closest galaxy ($i.e.$, central),  & stat.rank  &  & integer \\ 
&  n=n-th closest galaxy. -999 for isolated galaxies. &  &  &  \\ 
SepIterCen & Projected angular separation of the galaxy to the RankIterCen = 1 galaxy. & phys.angSize & arcsec & double \\ 
CoSepIterCen & Projected comoving distance separation of the galaxy to the RankIterCen = 1 galaxy. & phys.size.radius & Mpc\,$h$$^{-1}$ & double \\ 
AngSepIterCen & Projected angular size distance separation of the galaxy to the RankIterCen = 1 galaxy. & phys.size.radius & Mpc\,$h$$^{-1}$ & double \\ 
RankBCG & Relative ranking of galaxy by projected distance to the BCG. 1=BCG in the group, 2=closest satellite,  & stat.rank &  & integer \\ 
&  n=(n-1)-th closest galaxy. -999 for isolated galaxies. &  &  &  \\ 
SepBCG & Projected angular separation of the galaxy to the BCG. & phys.angSize & arcsec & double \\ 
CoSepBCG & Projected comoving distance separation of the galaxy to the BCG. & phys.size.radius & Mpc\,$h$$^{-1}$ & double \\ 
AngSepBCG & Projected angular size distance separation of the galaxy to the BCG. & phys.size.radius & Mpc\,$h$$^{-1}$ & double \\ 
RankCen & 1= closest galaxy to group centre, 2=closest satellite, n=n-th closest galaxy. -999 for isolated galaxies. & stat.rank &  & integer \\ 
SepCen & Projected angular separation of the galaxy to the group centre. & phys.angSize & arcsec & double \\ 
CoSepCen & Projected comoving distance separation of the galaxy to the group centre. & phys.size.radius & Mpc\,$h$$^{-1}$ & double \\ 
AngSepCen & Projected angular size distance separation of the galaxy to the group centre. & phys.size.radius & Mpc\,$h$$^{-1}$ & double \\ 
\end{tabular} 
\end{scriptsize} 
\end{center} 
\label{tab:D10GroupGals} 
\end{table*}

\begin{table*} 
\caption{\textcolor{black}{Additional c}olumns in the \texttt{devils\_dr1.D10Groups} catalogue (which are not in Table \ref{tab:baseColumns}).} 
\begin{center} 
\begin{scriptsize} 
\begin{tabular}{l l l c l } 
Name  &  Description & UCD & Unit & Type \\ 
\hline 
GroupID & Unique ID of the galaxy group. & meta.id; meta.main &  & integer \\ 
Nfof & Number of member galaxies. & stat.value  &  & integer \\ 
Zfof & Median redshift of the group members. & src.redshift &  & double \\ 
IterCenUID & UID of the RankIterCen = 1 galaxy. & meta.id &  & integer \\ 
IterCenRA & Right ascension of the RankIterCen = 1 galaxy. & pos.eq.ra & deg & double \\ 
IterCenDEC & Declination of the RankIterCen = 1 galaxy. & pos.eq.dec & deg & double \\ 
IterCenZ & Redshift of the RankIterCen = 1 galaxy. & src.redshift &  & double \\ 
BCGUID & UID of the RankBCG = 1 galaxy. & meta.id &  & integer \\ 
BCGRA & Right ascension of the RankBCG = 1 galaxy. & pos.eq.ra & deg & double \\ 
BCGDEC & Declination of the RankBCG = 1 galaxy. & pos.eq.dec & deg & double \\ 
CenRA & Right ascension of centre of luminosity of the group. & pos.eq.ra & deg & double \\ 
CenDEC & Declination of centre of luminosity of the group. & pos.eq.dec & deg & double \\ 
Rad50 & Radius of the smallest projected circle containing 50\% of the group members, centred on the IterCenUID galaxy. & phys.size.radius & Mpc\,$h$$^{-1}$ & double \\ 
Rad1Sig & Radius of the smallest projected circle containing 68\% of the group members, centred on the IterCenUID galaxy. & phys.size.radius & Mpc\,$h$$^{-1}$ & double \\ 
Rad100 & Radius of the smallest projected circle containing 100\% of the group members, centred on the IterCenUID galaxy. & phys.size.radius & Mpc\,$h$$^{-1}$ & double \\ 
RelDen & Relative galaxy space density within a cylinder of radius 1.5\,Mpc\,$h$$^{-1}$ and length 24 Mpc\,$h$$^{-1}$ centred on the IterCenUID galaxy,  &  &  & double \\ 
&  $i.e.$, ratio between measured density and expected density from the $N(z)$ fit in the cylinder. & src.density &  &  \\ 
GroupEdge & Fraction of the group within survey volume, estimated using IterCen and Rad100. &  &  & double \\ 
VelDispRaw & Raw measurement of the group velocity dispersion. & phys.veloc & km\,s$^{-1}$ & double \\ 
VelErr & Total velocity error of the group. & phys.veloc & km\,s$^{-1}$ & double \\ 
VelDisp & Group velocity dispersion, corrected for the total velocity measurement error.  & phys.veloc & km\,s$^{-1}$ & double \\ 
&  Set to 0 if total error is larger than the raw velocity ($i.e.$, VelErr$>$VelDisp). &  &  &  \\ 
GroupMass & Dynamical group mass estimated using M$_{\mathrm{halo}}$ $\propto$ (Rad$_{50}$ $\times$ VelDisp$^2$). See section 4.3 of \cite{Robotham11} for further details. & phys.mass & M$_{\odot}$\,$h$$^{-1}$ & double \\ 
GroupMassA & Dynamical group mass estimated using M$_{\mathrm{halo}}$ $\propto$ (Rad$_{50}$ $\times$ VelDisp$^2$), corrected for median biases by a scalar factor. & phys.mass & M$_{\odot}$\,$h$$^{-1}$ & double \\ 
&  See section 4.3 of \cite{Robotham11} for further details. &  &  &  \\ 
GroupMassAfunc & Dynamical group mass estimated using M$_{\mathrm{halo}}$ $\propto$ (Rad$_{50}$ $\times$ VelDisp$^2$), & phys.mass & M$_{\odot}$\,$h$$^{-1}$ & double \\ 
&  corrected for median biases by a function of both Nfof and IterCenZ. See section 4.3 of \cite{Robotham11} for further details. &  &  &  \\ 
LinkTot & Number of FoF links in the group. & stat.value &  & integer \\ 
LinkStrenProj & Mean of one minus the ratio between projected link length and projected link length limit for the FoF, & stat.correlation  &  & double \\ 
&  $i.e.$, $\overline{1-(L_\mathrm{link,rad}/L_\mathrm{limit,rad})}$. Values from 0 to 1, with larger values meaning stronger links. &  stat.correlation &  &  \\ 
LinkStrenZ & Mean of one minus the ratio between radial link length and radial link length limit for the FoF,  &  &  & double \\ 
&  $i.e.$, $\overline{1-(L_\mathrm{link,rad}/L_\mathrm{limit,rad})}$. Values from 0 to 1, with larger values meaning stronger links. &  &  &  \\ 
LinkStrenComb & The average of LinkStrenProj $\times$ LinkStrenZ. &  stat.correlation &  & double \\ 
RadKurt & Kurtosis of the projected radial distribution of galaxies, measured from IterCen. &  stat.fit.param &  & double \\ 
VelSkew & Skewness of the distribution of galaxy velocities. Value of -999 if Nfof$<$ 4. & stat.fit.param &  & double \\ 
VelKurt & Kurtosis of the distribution of galaxy velocities. Value of -999 if Nfof$<$ 4. & stat.fit.param &  & double \\ 
 Modality & Modality of the system, calculated as (1 + VelSkew$^2$) / (3 + VelKurt$^2$). & stat.fit.param &  & double \\ 
AxAng & Angle of the group major axis on the sky using the \cite{Carter80}. Value of -999 if Nfof$<$ 3. & pos.posAng & deg \\ 
AxRat & Axial ratio of the minor and major axes of the group using the \cite{Carter80} algorithm. Value of -999 if Nfof$<$3. & phys.size.axisRatio &  \\ 
SymKS & Kolmogorov-Smirnov test of the uniformity of groups, after correction for the axial-ratio. Value of -999 if Nfof$<$3. & stat.probability  &  & double \\ 
d3vol & 3D convex hull volume of the group. Value of -999 if Nfof$<$4. &  phys.volume & Mpc$^3$\,$h^{-3}$ & double \\ 
d3area & Surface area of the group hull. Value of -999 if Nfof$<$4. & phys.area & Mpc$^2$\,$h$$^{-2}$ & double \\ 
d2radec & Projected area of the group hull in the RA-Dec plane. Value of -999 if Nfof$<$3. & phys.area & Mpc$^2$\,$h$$^{-2}$ & double \\ 
d2raz & Projected area of the group hull in the RA-z plane. Value of -999 if Nfof$<$3. & phys.area & Mpc$^2$\,$h$$^{-2}$& double \\ 
d2decz & Projected area of the group hull in the Dec-z plane. Value of -999 if Nfof$<$3. & phys.area & Mpc$^2$\,$h$$^{-2}$ & double \\ 
distortcirc & Measure of the circularity of the group in projection. Value of -999 if Nfof$<$3, equal or greater than 1 otherwise (distortcirc = 1 for a circle). & src.ellipticity &  & double \\ 
distortsph & Measure of the sphericity of the group. Value of -999 if Nfof$<$4, equal or greater than 1 otherwise (distortsph = 1 for a sphere). & src.ellipticity &  & double \\ 
distortellip & Measure of the ellipsoidicity of the group. Value of -999 if Nfof$<$4, equal or greater than 1 otherwise (distortellip = 1 for a ellipsoid). & src.ellipticity &  & double \\ 
\end{tabular} 
\end{scriptsize} 
\end{center} 
\label{tab:D10Groups} 
\end{table*}

\bsp	
\label{lastpage}
\end{document}